\documentclass[aps, prb, twocolumn, superscriptaddress, longbibliography]{revtex4-1}

\usepackage{graphicx}
\usepackage{dcolumn}
\usepackage{bm}
\usepackage{color} 
\usepackage{multirow}
\usepackage{siunitx}
\usepackage[colorlinks = true, urlcolor = blue, linkcolor = blue, citecolor=blue]{hyperref}

\begin{document}

\title{Dichotomy of saddle points in energy bands of a monolayer NbSe$_2$}

\author{Sejoong Kim}
 \email{sejoong@alum.mit.edu}
 \affiliation{University of Science and Technology (UST), Gajeong-ro 217, Daejeon 34113, Korea}
 
\author{Young-Woo Son}
 \email{hand@kias.re.kr}
 \affiliation{Korea Institute for Advanced Study, Hoegiro 85, Seoul 02455, Korea}

\date{\today}

\begin{abstract}
We theoretically show that two distinctive spin textures 
manifest themselves around saddle points of energy bands in
a monolayer NbSe$_2$ under external gate potentials. 
While the density of states at all saddle points diverge logarithmically, 
ones at the zone boundaries display a windmill-shaped spin texture while the others unidirectional spin orientations. 
The disparate spin-resolved states are demonstrated to contribute an intrinsic spin Hall conductivity significantly 
while their characteristics differ from each other.
Based on a minimal but essential tight-binding approximation reproducing first-principles computation results, 
we established distinct effective Rashba Hamiltonians 
for each saddle point, realizing
the unique spin textures depending on their momentum.
Energetic positions of the saddle points in a single layer NbSe$_2$ 
are shown to be well controlled by a gate potential so that it could be a prototypical 
system to test a competition between various collective phenomena triggered by diverging density of states and their spin textures 
in low-dimension.
\end{abstract}

\maketitle

\section{\label{sec:Intro}Introduction}
The experimental demonstrations of isolating a single layer of the layered transition 
metal dichalcogenides (TMDs)~\cite{PNAS2005Novoselov} have spurred intense researches on 
their characteristic electronic properties differing 
from those of their bulk forms~\cite{NatNano2012Wang,Nature2013Geim,NatChem2013Chhowalla}.  
For example, the indirect
band gap of bulk TMDs changes to be a direct one in their monolayer~\cite{PRL2010Mak,NatNano2014Zhang}. 
Moreover, the Coulomb interaction as well as effects
of environment such as substrates on which a single layer placed become to be essential in altering low energy physics~\cite{PRB2012Cheiwchanchamnagij,PRB2012Ashwin,PRB2012Komsa,PRL2013Qiu,NatMater2014Ugeda,PhysRevB_96_155439_Kim}. 
Like the cases in semiconducting TMDs, the metallic ones also show several intriguing
changes as their thickness decreases~\cite{NatNano2015Yu,Science2015Saito,Nature2016Li,NatPhys2016Tsen}.

Among the metallic TMDs, 
niobium diselenide of 2$H$ stacking structure  (2$H$-NbSe$_2$) has long been studied 
owing to its intriguing phase diagram 
showing a charge density wave (CDW) and subsequent superconducting (SC) states
as temperature decreases~\cite{AdvPhys2001Wilson}. 
A single layer of 2$H$-NbSe$_2$ also exhibits a similar phase diagram 
with a different set of critical temperatures for the states~\cite{NatNano2015Xi,NatPhys2016Ugeda}. 
Since there is no apparent diverging susceptibility 
for the bulk and monolayer NbSe$_2$~\cite{PhysRevB_96_155439_Kim,PRB2006Johannes,PRB2008Johaness,PRB2009Calandra,PRB2012Ge}, 
the weak coupling scenario for the CDW may not work very well
and several other proposals have been put forward~\cite{JPCM2011Rossnagel,PRL2008Shen,PRL2009Borisenko,PRL2015Arguello,NatComm2020Lin}.
Among those, there is an alternative weak coupling scenario
of CDW formation originating from the nesting 
van Hove singularities (vHSs) at saddle points~\cite{Rice_Scott_PRL_35_120_1975}.
Although there is no direct evidence for the vHS-driven CDW,
we can expect other interesting low energy physics 
thanks to the logarithmically diverging local density of states 
at the saddle points~\cite{PRL2008Honerkamp,PRB2011Makogon,NatPhys2012Rahul,PRL2014Yudin,PRB2021Kiesel}.
Moreover, the low energy physical properties of monolayer metallic TMDs
can also be altered by external control knobs such as ion depositions and bottom gates~\cite{NatNano2015Yu,Science2015Saito,Nature2016Li,NatPhys2016Tsen}
so that  
the monolayer NbSe$_2$ could be an interesting material platform 
to understand the peculiar physics originated from saddle shaped electronic energy 
bands in low dimension. 

\begin{figure}[b]
\begin{center}
\includegraphics[width=1.0\columnwidth, clip=true]{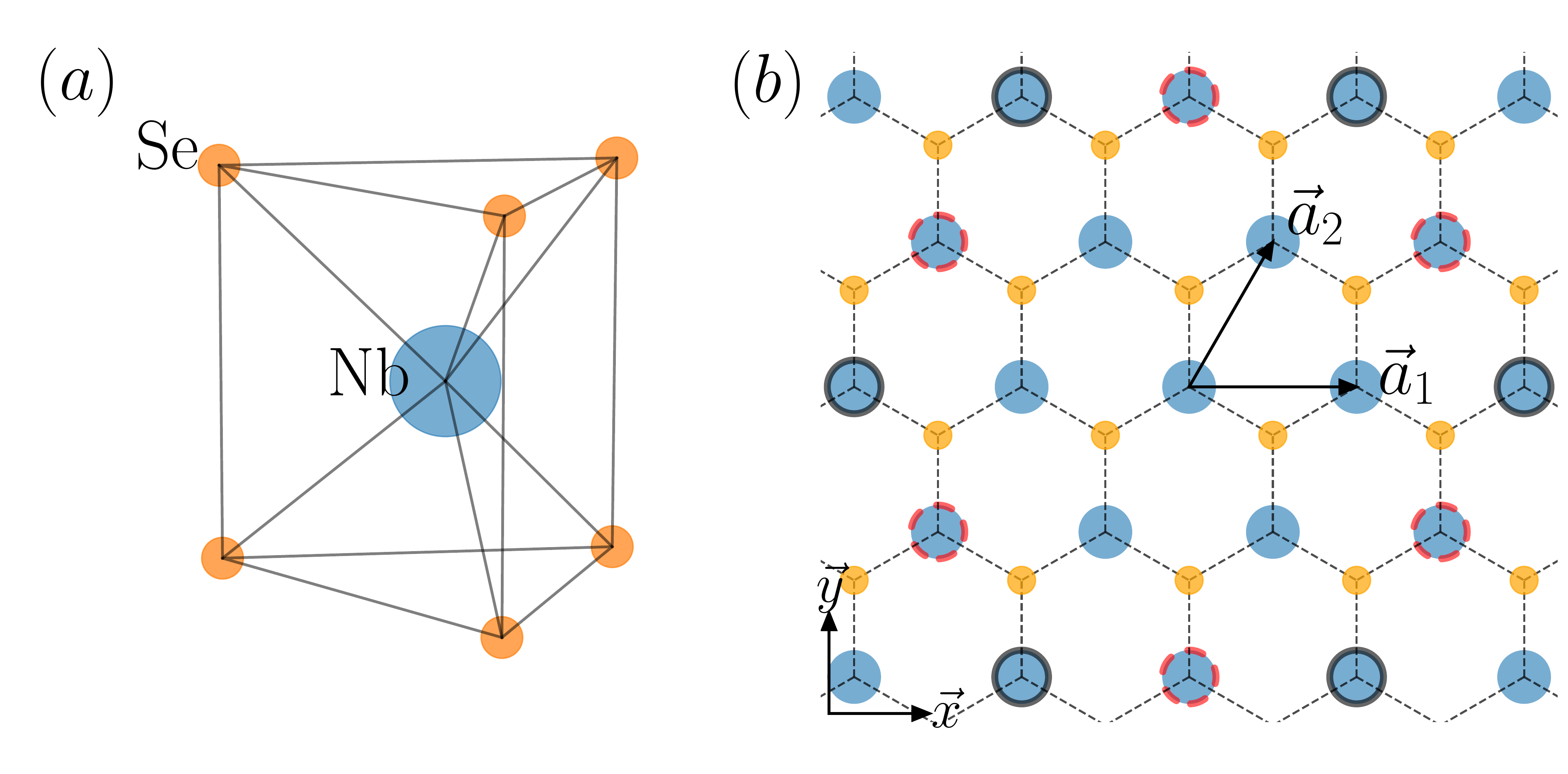} 
\end{center}
\caption{\label{Crystal_Structure} (Color online) Crystal structure of monolayer NbSe$_{2}$. (a) Nb atom is sandwiched by six Se atoms, 
A light blue (larger) sphere and orange (smaller) ones represent Nb and Se atoms, respectively. 
(b) The top down view of the lattice structure. $\vec{a}_{1}$ and $\vec{a}_{2}$ are unit vectors. 
The second and third nearest neighbors are highlighted by red dashed circles and black solid circles, respectively.} 
\end{figure}

The metallic single layer of bulk 2$H$-NbSe$_2$ has the trigonal prismatic structure where 
 the triangular lattice of transition metals sandwiched by two triangular lattice of chalcogen atoms (Fig.~\ref{Crystal_Structure}).
 The chalcogen atoms are in the mirror-reflection symmetric position 
 with respect to the transition metal layer as shown in Figs.~\ref{Crystal_Structure}(a) and (b), 
suppressing the Rashba spin-orbit interaction and allowing the Zeeman splitting only. 
The suppressed Rashba interaction can be 
revived by applying external perturbations, e.g., the electric field perpendicular to the monolayer plane
in the field effect transistor (FET) setup~\cite{Yuan_NatPhys_9_563_2013, Cheng_Nanoscale_8_17854_2016, PhysRevB_91_235145_2015_Shanavas}. 
The induced Rashba spin-orbit interaction can lead to nontrivial spin textures lying on the plane, 
which is different from the Ising-type spin orientation~\cite{NatPhys2016Xi,NatPhys2016Saito} of the mirror-symmetric structures. 
A single layer of NbSe$_2$ has a merit in that the energetic position of vHSs in NbSe$_2$ is quite close to the Fermi energy ($E_F$)~\cite{PhysRevB_96_155439_Kim}, quite contrary to the case of graphene where vHSs are very far away from 
$E_F$~\cite{PhysRevLett.104.136803,PRL2020Rosenzweig}. 
Therefore, we expect interesting spin-related physical properties from interplay between spin textures  and distinctive vHSs in the monolayer of NbSe$_2$ that can be easily accessible in experiments.

In this paper, we show that the Rashba interaction on vHSs in the monolayer NbSe$_2$
can induce two characteristic spin textures. 
One is a windmill-shaped spin texture circling around the saddle points
while the other uniform in-plane spin orientation. 
The peculiar windmill-shaped spin texture can be regarded as a projection of the spin vortex
induced by Rashba interaction onto the crossed linear lines of local Fermi surface around the saddle points.
To compute spin transport with the FET gating effectively, 
we develop a tight-binding model with a minimal but essential set of atomic orbitals 
to reproduce our {\it ab initio} computational results reliably. 
With these methods, we derive the two distinct Rashba Hamiltonians describing 
the local low energy physics around two disparate saddle points in the first Brillouin Zone (BZ), respectively,
and compute the associated intrinsic spin Hall conductivities. 
It is shown that the energetic position of vHSs can be controlled well by the FET gating regardless 
of metallicity of the monolayer. 
We expect that our complete TB approximations 
and the distinct models for the spin-orbit interactions with vHSs in the monolayer NbSe$_2$ with FET gating
will be of interest in understanding various spin-related phenomena triggered by 
diverging density of states in low dimension. 

This paper is organized as follows. In Sec.~\ref{sec:DFT}, we present band structures based 
on density functional theory (DFT) calculations with {\it ab initio} simulation of the FET gating. 
We also discuss the evolution of band structures as a function of hole doping concentrations associated with the FET gating simulation, especially focusing on positions of saddle points relative to the Fermi level. 
In Sec.~\ref{sec:TB} we construct the tight-binding model of five $d$-orbitals including atomic spin-orbit interactions, which is best fitted to DFT band structures obtained in Sec.~\ref{sec:DFT}. 
In Sec.~\ref{sec:spin_texture}, we present spin textures obtained by DFT calculations and the tight-binding model constructed in Sec.~\ref{sec:TB}. 
In Sec.~\ref{sec:ISHC} we compute and discuss the static intrinsic spin Hall conductivities.  
In Sec.~\ref{sec:effective_model} we develop effective minimal models around $\mathbf{k}$ points, where major contributions to the intrinsic spin Hall conductivity occur. 
Conclusions are in Sec.~\ref{sec:conclusions}. 
Other details of the five $d$-orbital 
TB model and the intrinsic spin Hall conductivity are provided in Appendix.

\section{\label{sec:DFT}DFT Band Structures}
\begin{figure}[t]
\begin{center}
\includegraphics[width=1.0\columnwidth, clip=true]{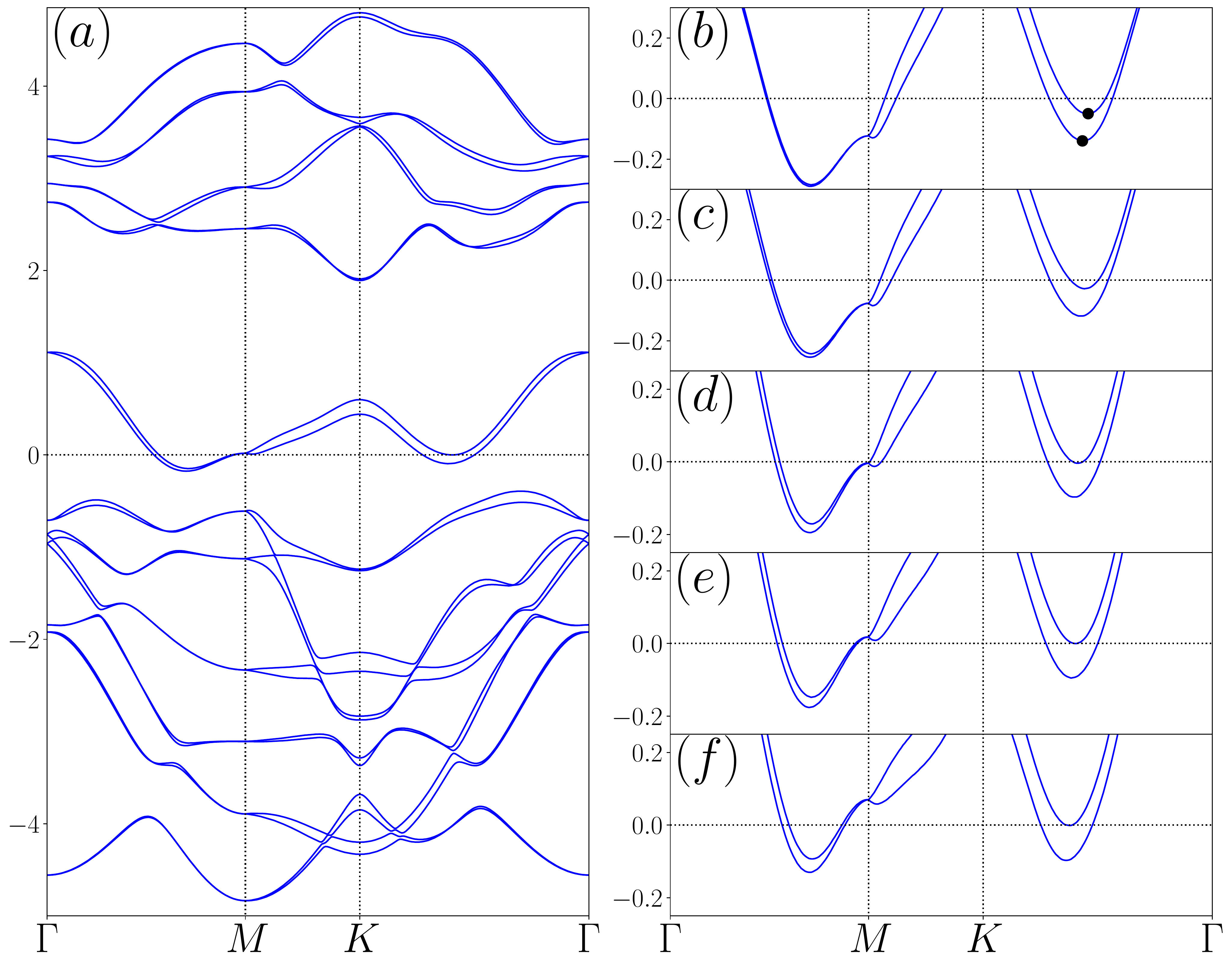} 
\end{center}
\caption{\label{FET_bands} (Color online) Band structures under the FET gating from DFT calculations. Here the Fermi level $E_{F}$ is set to be zero. Band structures with 0.4 holes per unit cell are drawn in (a). Two bands around $E_{F}$ with different hole doping concentration are shown in (b)--(f). Hole doping concentrations of (b), (c), (d), (e), and (f) are 0.100, 0.200, 0.355, 0.400, and 0.500 holes per unit cell, respectively. Two saddle points hosting van Hove singularities in the ${K\Gamma}$ line are denoted by black filled circles in (b).} 
\end{figure}
We perform the first-principles calculations based on the density functional theory (DFT) in order to obtain the reference band structures and spin textures. 
The DFT calculations are performed by using the \textsc{Quantum  Espresso}~\cite{JPhys_CM_21_395502_2009,JPhys_CM_29_465901_2017} with the plane-wave basis, the PBE exchange-correlation functional~\cite{PhysRevLett_77_3865_1996} and norm-conserving pseudopotentials~\cite{PhysRevB_88_085117_2013, CompPhysComms_196_36_2015, JChemTheoryComput_12_3523_2016}. 
We adopt $20\times20\times1$ $k$-point mesh, and the smearing temperature $0.005$ Ry with the cold smearing technique~\cite{PhysRevLett_82_3296_1999_Marzari}, and the kinetic energy cutoff $120$ Ry for the self-consistent calculation. 
We also combine the recently developed technique~\cite{PhysRevB_89_245406_2014_Brumme, PhysRevB_91_155436_2015_Brumme} to simulate the field effect transistor (FET) gating set-up, which breaks the mirror symmetry with respect to the two-dimensional plane denoted by $\mathcal{M}_{z}$. 

Figure~\ref{FET_bands} shows DFT band structures under the FET gating. 
Here we focus the evolution of two bands around the Fermi energy $E_{F}$ as a function of hole doping concentrations.
We note that the part of energy bands is split by spin-orbit interactions.   
When the hole doping concentration is changed by the FET gating, 
the saddle point of the upper band along the ${K\Gamma}$ line approaches the Fermi energy. 
At the same time, the degenerate energy bands at $M$, also a saddle point, 
shifts up as the hole doping concentration increases. 
With low hole doping concentrations of 0.1 (0.2) holes per unit cell, 
the energy of states at $M$ is not aligned with another saddle point energy along the ${K\Gamma}$ line 
[Figs.~\ref{FET_bands}(b) and \ref{FET_bands}(c)].
When the hole concentration increases to 0.355 holes per unit cell, 
the energy bands at $M$ point and the upper saddle point are aligned just at $-3.7$ meV 
below the Fermi level as shown in  Fig.~\ref{FET_bands}(d).
At the hole concentration of 0.4 holes per unit cell, 
the saddle point almost touches the $E_F$, while the band energy at $M$ is slightly higher than $E_{F}$. 
When the hole concentration is further increased to 0.5 holes per unit cell [Fig.~\ref{FET_bands}(f)], 
the band energy at $M$ is pushed away from the $E_F$, 
but the upper saddle point is still located in its vicinity.

\section{\label{sec:TB}Tight-Binding Model}
Using DFT band structures as reference, we show that the effective tight-binding (TB) model requires the five $d$-orbitals of $\textrm{Nb}$ atoms
as a minimal basis set to reproduce the first-principles results with the FET gating. 
When the mirror symmetry $\mathcal{M}_{z}$ is preserved, the effective TB model with three $d$- orbitals $d_{z^2}$, $d_{xy}$, and $d_{x^2-y^2}$ can reproduce energy bands around the $E_F$ as shown in Ref.~\onlinecite{PhysRevB_88_085433_Liu}. 
The energy bands consisting of the other two $d$-orbitals $d_{zx}$ and $d_{yz}$ are not mixed with bands with $d_{z^2}$, $d_{xy}$, and $d_{x^2-y^2}$. 
In contrast, when the mirror symmetry $\mathcal{M}_{z}$ is broken under the FET gating, all of the five $d$-orbitals are needed to consider in order to well reproduce energy bands around the Fermi energy.

\begin{figure}[b]
\begin{center}
\includegraphics[width=1.0\columnwidth, clip=true]{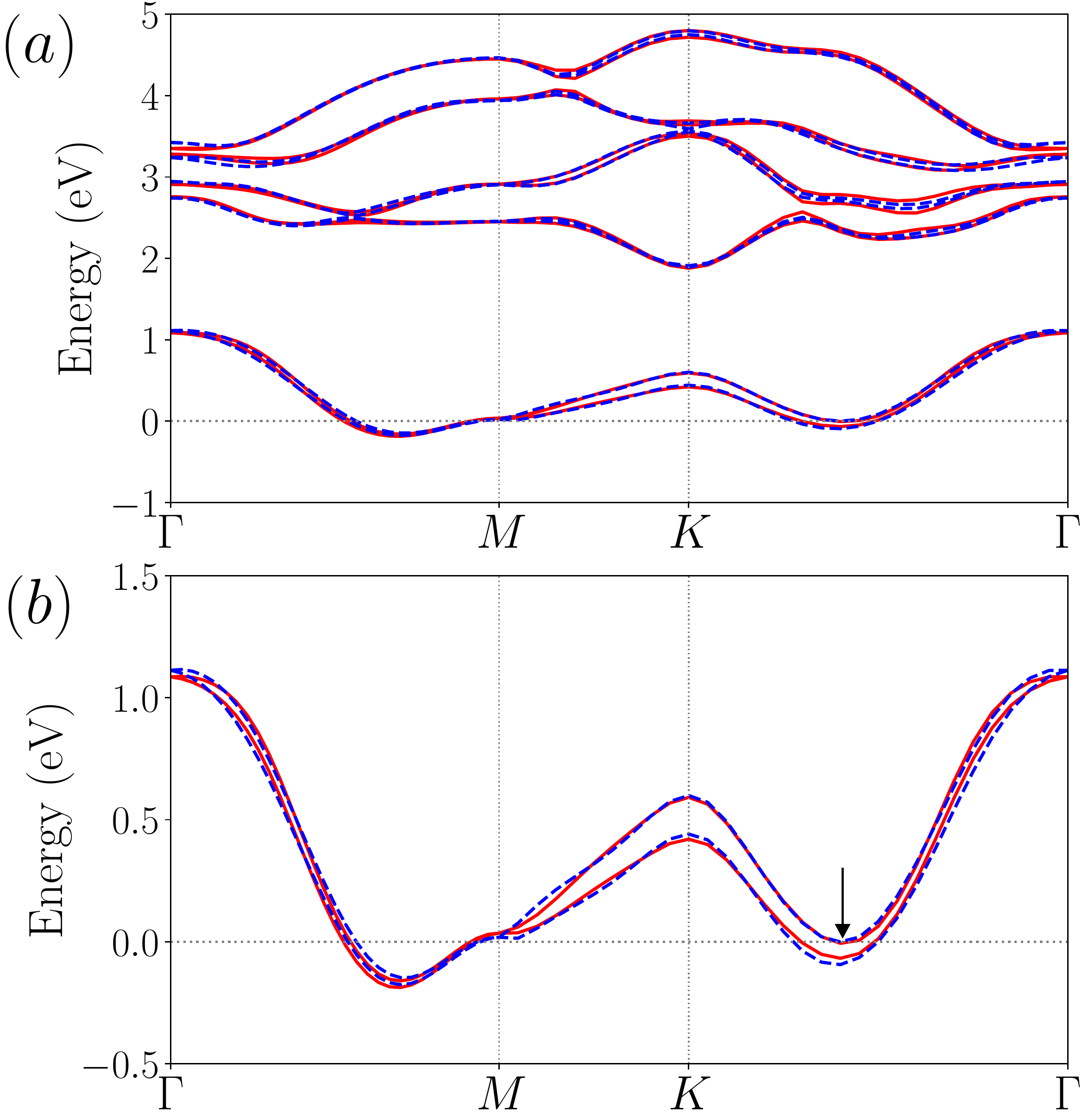} 
\end{center}
\caption{\label{bands} (Color online) Band structures from DFT calculations (blue dashed lines) and TB model (red solid lines) 
with 0.4 holes per unitcell.
Hereafter, all figures are for 0.4 holes per unitcell, otherwise noted explicitly.
Ten bands composing of $d$-orbitals are shown in (a). Two lowest bands around the Fermi energy are magnified in (b). The saddle point occurs in the middle of ${\Gamma K}$ as indicated by a black arrow in (b).} 
\end{figure}

Following the Slater-Koster scheme~\cite{PhysRev_94_1498_1954_Slater_Koster}, the TB model consists of energy integerals, 
which are defined as 
\begin{equation}
\label{energy_integral}E_{ij}(\mathbf{R}) = \langle \phi_{i}(\mathbf{0}) | \mathcal{H} | \phi_{j} (\mathbf{R}) \rangle,    
\end{equation}
where $|\phi_{i}(\mathbf{0})\rangle$ and $|\phi_{j}(\mathbf{R})\rangle$ are $i$th and $j$th orbitals located at the origin and at the lattice vector $\mathbf{R}$. 
Here we denote the five $d$-orbitals of the transition metal as 
$|\phi_{1}\rangle = |d_{z^2}\rangle$, 
$|\phi_{2}\rangle = |d_{x^2-y^2}\rangle$,  
$|\phi_{3}\rangle = |d_{xy}\rangle$,  
$|\phi_{4}\rangle = |d_{zx}\rangle$,  
and $|\phi_{5}\rangle = |d_{yz}\rangle$. 
Since energy integrals are related to one another via the lattice symmetry, the set of independent energy integrals can be determined by using the group theoretical approach~\cite{PhysRevB_88_085433_Liu, Group_Theory_Dresselhaus}. 
Note that our TB model is extended up to the third nearest-neighbor (TNN) hopping, 
following former studies that show the electronic structure of TMDC is well fitted by including the TNN hopping.  Reference~\onlinecite{PhysRevB_91_235145_2015_Shanavas} reported the TB model of monolayer 
TMDC whose mirror symmetry $\mathcal{M}_{z}$ is broken under electric field, but the TB model of Ref~\onlinecite{PhysRevB_91_235145_2015_Shanavas} is based on the two-center approximation instead of the energy integral.
Our current model can be regarded as an extension of Ref.~\onlinecite{PhysRevB_88_085433_Liu} to the five $d$-orbital case, in a sense that the TB model consists of energy integrals shown in Eq.~(\ref{energy_integral}).

Band structures can be obtained by diagonalizing the effective tight-binding Hamiltonian,
\begin{equation}
\left[\mathcal{H}_{\textrm{tot}}\right]_{i\sigma,j\sigma^\prime}(k) =\sum_{\mathbf{R}}e^{i\mathbf{k}\cdot\mathbf{R}}\langle\phi_{i}(\mathbf{0})\sigma|\mathcal{H}_{\textrm{tot}}|\phi_{j}(\mathbf{R})\sigma^\prime \rangle,    
\end{equation}
where $\sigma$ and $\sigma^\prime$ denote spin states ($\uparrow$ or $\downarrow$) and  $|\phi_{i}(\mathbf{R})\sigma\rangle \equiv |\phi_{i}(\mathbf{R})\rangle|\sigma\rangle$. Here the total Hamiltonian $\mathcal{H}_{\textrm{tot}}$ includes the electronic Hamiltonian $\mathcal{H}$ describing inter-orbital hoppings and the atomic spin-orbit coupling term $\mathcal{H}_{\textrm{soc}}$, i.e., $\mathcal{H}_{\textrm{tot}} = \mathcal{H} + \mathcal{H}_{\textrm{soc}}$. 
The detailed expression and fitting parameters of the TB model are summarized in Appendix~\ref{Appendix-TB} and Table~\ref{Table_parameters}.

Figure~\ref{bands} shows band structures obtained from DFT calculations and the TB model of this work. 
As shown in Fig.~\ref{bands}(a), band structures of the TB model well match those from DFT calculations. Figure~\ref{bands}(b) focuses on two lowest energy bands, which constitute the Fermi surface. The atomic spin-orbit coupling splits the energy level around the Fermi level to two lowest bands as shown in Fig.~\ref{bands}(b). 
The band splitting due to the spin-orbit coupling is most apparent around $K$ point. 
It is known that the atomic spin-orbit coupling does not split energy bands along ${\Gamma M}$ when the mirror symmetry $\mathcal{M}_{z}$ in the direction perpendicular to the plane is preserved. 
In contrast, energy bands along ${\Gamma M}$ are split off in the system of our interest where the mirror symmetry $\mathcal{M}_{z}$ is broken due the FET gating.

\begin{figure}[b]
\begin{center}
\includegraphics[width=1.0\columnwidth, clip=true]{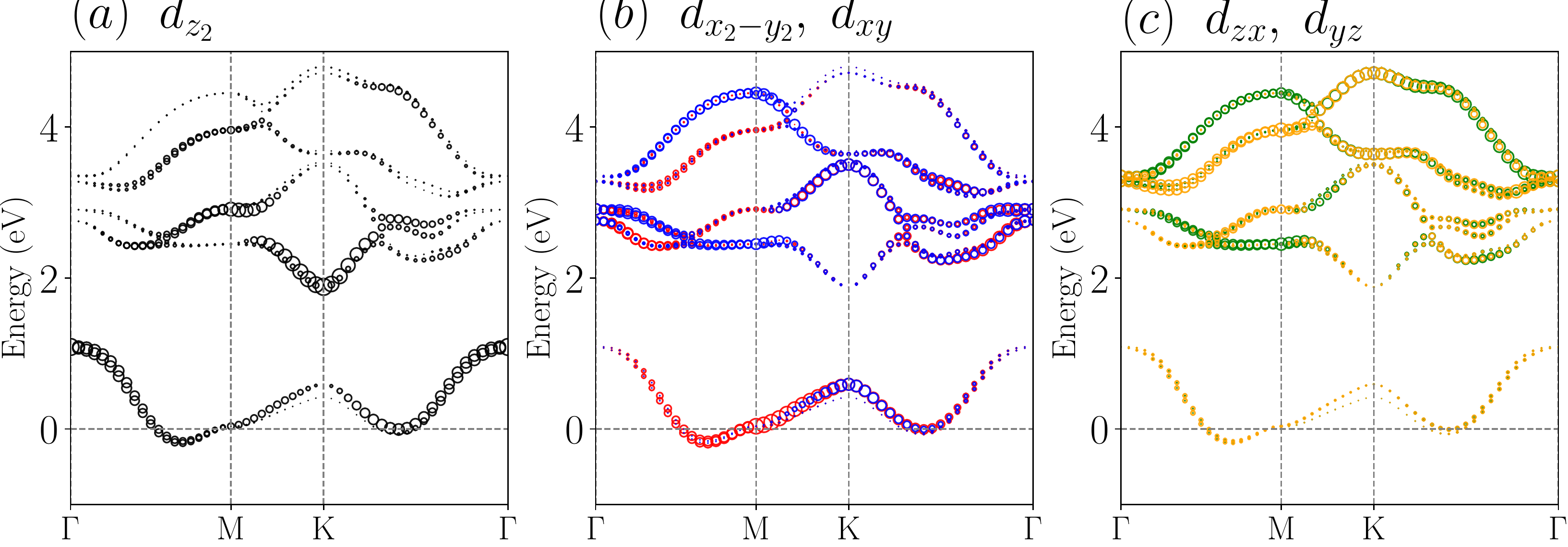} 
\end{center}
\caption{\label{orbital_projections} (Color online) Orbital-projected band structures from the TB model. $d_{z^2}$-orbital, 
$d_{xy}$ and $d_{x^2-y^2}$, and $d_{zx}$ and $d_{yz}$-orbital projections are denoted by (a) black, (b) blue and red, and (c) green and orange circles, respectively. }
\end{figure}

Another important feature of band structures is that the second lowest band almost touches the $E_F$ in the middle of ${\Gamma K}$. 
The touching point is the saddle point around which energy landscape is described by a hyperbolic surface. The hole doping level of the FET gating can be controlled as the input parameter in the DFT calculations. 
The hole doping concentration is set to be $0.4$ holes per unit cell ($7.641 \times 10^{14}$ cm$^{-2}$) in order to place the saddle point close to the Fermi level. 
When the spin-orbit interaction is turned off, the energy band around the $E_F$ hosts another saddle point at the $M$ point. Unlike the saddle point in the middle of ${\Gamma K}$ where the spin-orbit interaction splits two hyperbolic energy surfaces, the energy degeneracy at $M$ is not lifted up by the spin-orbit interaction. 
Energy surfaces with the spin-orbit interaction at $M$ minutely differs from those of the exact saddle points lying on ${\Gamma K}$. 
This feature will be discussed in detail in the section~\ref{sec:effective_model}.

Figure~\ref{orbital_projections} exhibits orbital-projected band structures. The two bands crossing $E_F$ are mainly composed of $d_{z^2}$, $d_{xy}$, and $d_{x^2-y^2}$. It is noticed that $d_{zx}$ and $d_{yz}$ orbitals also contribute to the two bands around the $E_F$ to some extent. It is also found that all of the $d$-orbitals are involved for higher bands. This mixing of two subsets of $d$-orbitals, $\{d_{z^2}, d_{xy}, d_{x^2-y^2}\}$ and $\{d_{zx}, d_{yz}\}$ is due to two factors, the atomic spin-orbit coupling and the FET gating, which breaks the mirror reflection symmetry $\mathcal{M}_{z}$. 
When the mirror reflection symmetry $\mathcal{M}_{z}$ is preserved and there is no atomic spin-orbit coupling, $d_{zx}$ and $d_{yz}$ orbitals are decoupled to $d_{z^2}$, $d_{xy}$, and $d_{x^2-y^2}$ due to symmetry consideration~\cite{PhysRevB_88_085433_Liu, PhysRevB_96_155439_Kim}. 
When the atomic spin-orbit coupling is introduced, one spin state of $d_{z^2}$, $d_{xy}$, and $d_{x^2-y^2}$ can be mixed with the opposite spin state of $d_{zx}$ and $d_{yz}$ orbitals as apparently shown in the atomic spin-orbit interaction, Eq.~(\ref{Eq:Atomic_SOC}). 
When the mirror symmetry $\mathcal{M}_{z}$ is broken, the same spin states of $\{d_{z^2}, d_{xy}, d_{x^2-y^2}\}$ and $\{d_{zx}, d_{yz}\}$ are coupled.

\begin{figure}[b]
\begin{center}
\includegraphics[width=1.0\columnwidth, clip=true]{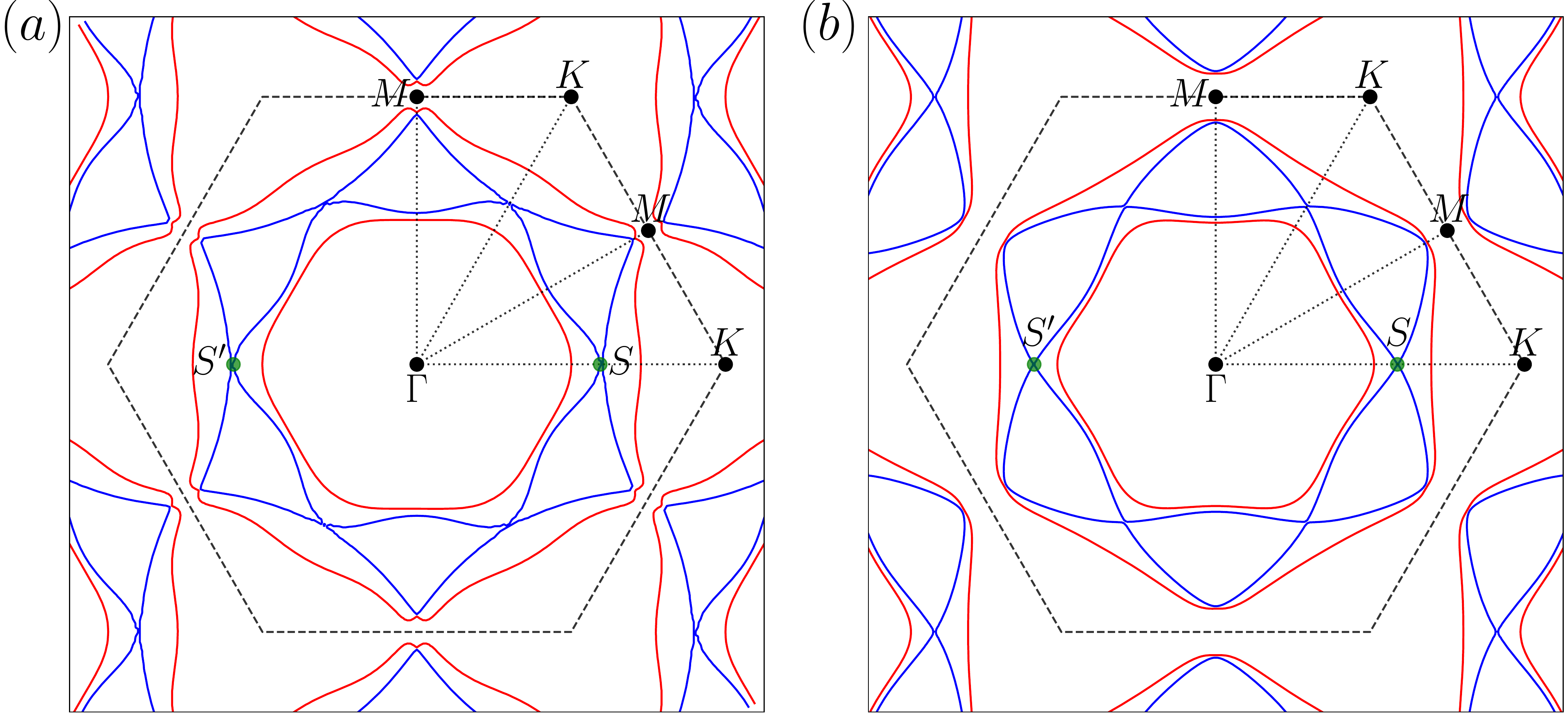} 
\end{center}
\caption{\label{Fermi_surface} (Color online) Fermi surfaces from (a) DFT calculations and (b) the TB model. Black dashed lines indicate the first Brillouin zone (1BZ), and black dotted lines form the irreducible Brillouin zone (IBZ). Symmetric points $\Gamma$, $M$, and $K$ are denoted. Green dots indicate two saddle points $S$ and $S^\prime$ located on $k_{x}$ axis. Red and blue solid lines correspond to Fermi surfaces coming from the lowest energy band and the second lowest one, respectively.}
\end{figure}
Fermi surfaces (FSs) obtained from DFT bands and TB ones are illustrated in Fig.~\ref{Fermi_surface}. Their detailed shapes such as sharpness and convexity of lines do not match perfectly, but the overall features are shared by and agreed with the two approaches very well.
Inside the first BZ, there are four contours originated from the two bands around the $E_F$. 
Two red contours and two blue contours originate from the lower energy band and the other around the $E_F$, respectively.

\section{\label{sec:spin_texture}Overall Spin Textures}

The spin-orbit interaction induced by the broken mirror symmetry leads to the change of spin textures in the monolayer NbSe$_2$. 
When the mirror-reflection symmetry $\mathcal{M}_{z}$ is respected, only the $z$-component of the atomic spin-orbit coupling survives
because planar ($x$ and $y$) components and the $z$ component of the atomic spin-orbit coupling have odd and even parity numbers under mirror reflection symmetry $\mathcal{M}_{z}$ respectively. 
As a consequence, the spin eigenstates of the $S_{z}$ operator are good quantum states of the full Hamiltonian $\mathcal{H}_{\textrm{tot}}=\mathcal{H} + \mathcal{H}_{\textrm{soc}}$, which means that spin directions induced by the atomic spin-orbit coupling are effectively the Ising-type.

\begin{figure}[b]
\begin{center}
\includegraphics[width=1.0\columnwidth, clip=true]{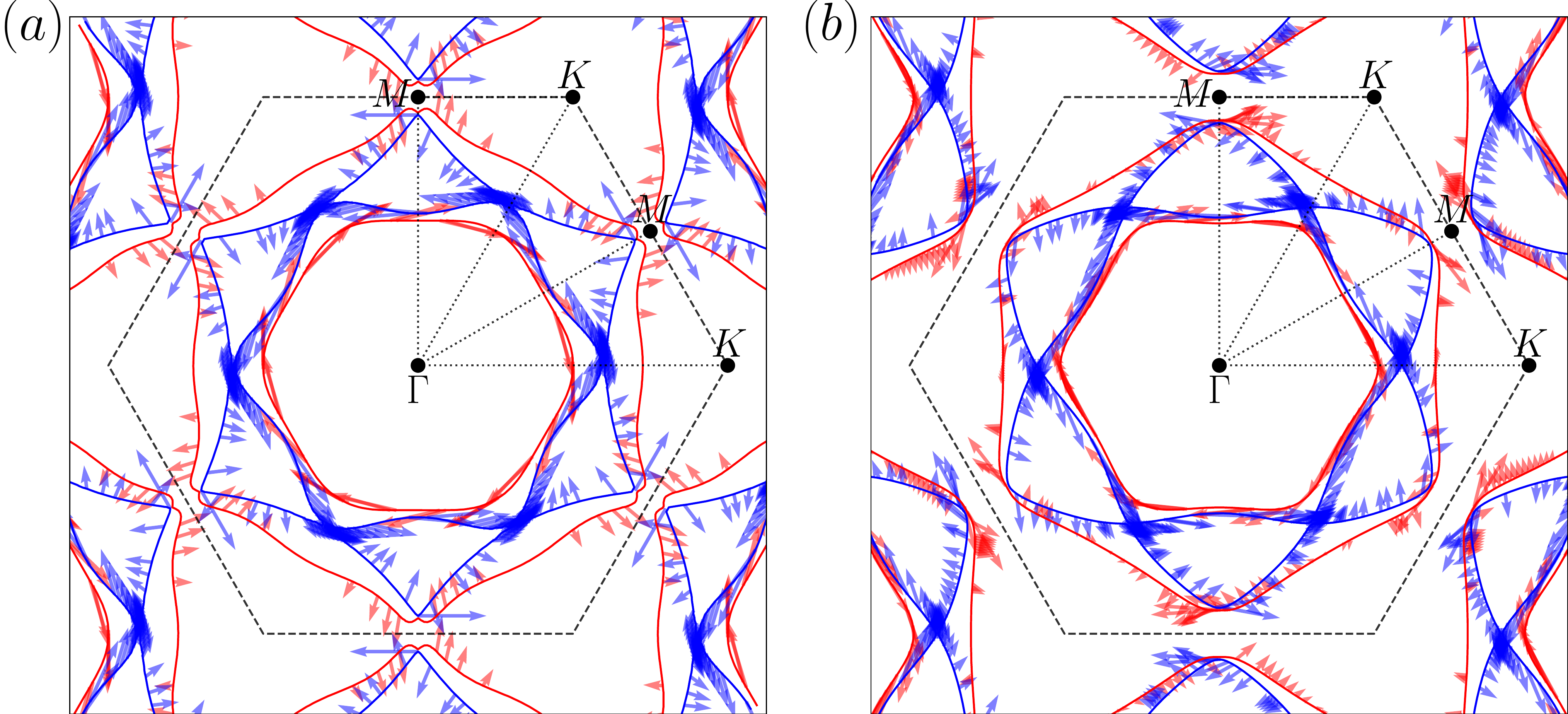} 
\end{center}
\caption{\label{Spin_texture} (Color online) Spin textures projected on Fermi surfaces from (a) DFT calculations and (b) the TB model. Spin directions are drawn by arrows whose length is arbitrarily chosen for clear illustration. Red and blue solid lines are used to indicate Fermi surfaces and spin directions of the lowest energy band and the second lowest one, respectively.}
\end{figure}

When the mirror-reflection symmetry is broken by the FET gating, the constraint on spin states discussed above no longer holds. Hence the expectation values for the planar spin orientation are not zero.   
Essentially, the in-plane spin orientations follow the helical or vortex shape circling 
around the $\Gamma$-point as discussed in a typical Rashba interaction. 
In NbSe$_2$, the helicity of spin vortex for the two energy bands near the $E_F$ is opposite to each other.

Since there is no spontaneous time-reversal symmetry breaking, 
we expect that the non-trivial spin texture around $E_F$ will have the largest contribution 
to the spin-dependent physical observables. 
So, in Fig.~\ref{Spin_texture}, we project the computed spin textures on the FSs of Fig.~\ref{Fermi_surface}.
Two innermost FSs exhibit helical spin textures. 
The innermost FS from the lower energy band has the clockwise helicity, 
while the spin texture of the second innermost FS from the other band is counter-clockwise. 
One important feature is that many spin vectors are concentrated in the vicinity of saddle points, 
reflecting a divergent density of states at the saddle point. 
This implies that spin-related physical properties could be influenced 
not only by helical patterns of planar spin textures, 
but also by vHSs at saddle points. 
Therefore, the local spin texture are shown to depend on the geometry of energy bands and their positions in the first BZ.

\section{\label{sec:effective_model}Effective Rashba Models}

As discussed above, the spin-dependent properties of metallic monolayer NbSe\textsubscript{2} 
could be mainly determined by spin states around saddle points hosting diverging density of states. 
We can develop an effective minimal theory in order to capture the essential physics around such special points. 
The minimal model can be constructed by using the $k \cdot p$ approximation. 
Using the quasi-degenerate perturbation theory based on the Schrieffer-Wolff transformation~\cite{SOC_Winkler}, 
we derive the effective $2 \times 2$ Hamiltonian in terms of ${\bf q}=(q_x,q_y)$, 
which is the small deviation from the given crystal momentum of ${\bf k}$, e.g., for $M$ point, ${\bf k}=\overrightarrow{\Gamma M}+{\bf q}$
with $|{\bf q}|\ll |{\bf k}|$.
In this section we present the minimal models around the symmetric points. 
Detailed derivation methods are in Appendix~\ref{Appendix-kp}. 

We first consider the trivial cases around $\Gamma$ and $K$-points. 
Around the $\Gamma$ point, the $k \cdot p$ approximation leads to the minimal Hamiltonian, 
\begin{equation}
\label{Eq:H_eff_G}\mathcal{H}_{\Gamma}({\bf q}) = -\alpha_{\Gamma}\left(q_{x}^{2}+q_{y}^{2}\right)\sigma_0
+\lambda_{R}\left(q_{x}\sigma_{y} - q_{y}\sigma_{x}\right),
\end{equation}
where $\alpha_{\Gamma}=\hbar^2/2m^{*}_{\Gamma}$ with the effective mass $m_{\Gamma}^{*}$ at $\Gamma$, 
and $\lambda_{R}$ is the Rashba interaction strength. 
$\sigma_0$ is an identity and $\sigma_i~(i=x,y,z)$ is the Pauli matrix.
The energy dispersion around $\Gamma$ is isotropic, and the effective spin-orbit coupling is of the Rashba type. 
The numerical values for the parameters are in the Table~\ref{parameter}.

The effective Hamiltonian in the vicinity of $K=\left(\frac{4\pi}{3a}, 0\right)$ is given by 
\begin{eqnarray}
\label{Eq:H_eff_K}\mathcal{H}_{K}({\bf q}) &=& -\alpha_K \left(q_{x}^{2}+q_{y}^{2}\right)\sigma_0
+\lambda_{R}\left(q_{y}\sigma_{x}-q_{x}\sigma_{y}\right)\nonumber\\
&&+\lambda_{Z}\sigma_{z},
\end{eqnarray}
where $\alpha_K=\hbar^2/2m^{*}_{K}$ with the effective mass $m_{K}^{*}$ at $K$, and $\lambda_Z$  the effective Zeeman term. 
As like the $\Gamma$ point, the energy dispersion around $K$ is isotropic. 
The effective Rashba spin-orbit interaction is residual 
when compared to the Zeeman term that is independent of the displacement vector ${\bf q}$. 
When the FET gating is not applied, the Rashba interaction vanishes, 
and only the Zeeman term survives, thereby leading to the Ising-type spin texture. 
The effective Hamiltonian around $K^\prime=\left(\frac{2\pi}{3a}, \frac{2\pi}{\sqrt{3}a}\right)$
can be obtained by flipping sign of $\lambda_Z$ in Eq.~\ref{Eq:H_eff_K}.

\begin{figure}[t]
\begin{center}
\includegraphics[width=1.0\columnwidth, clip=true]{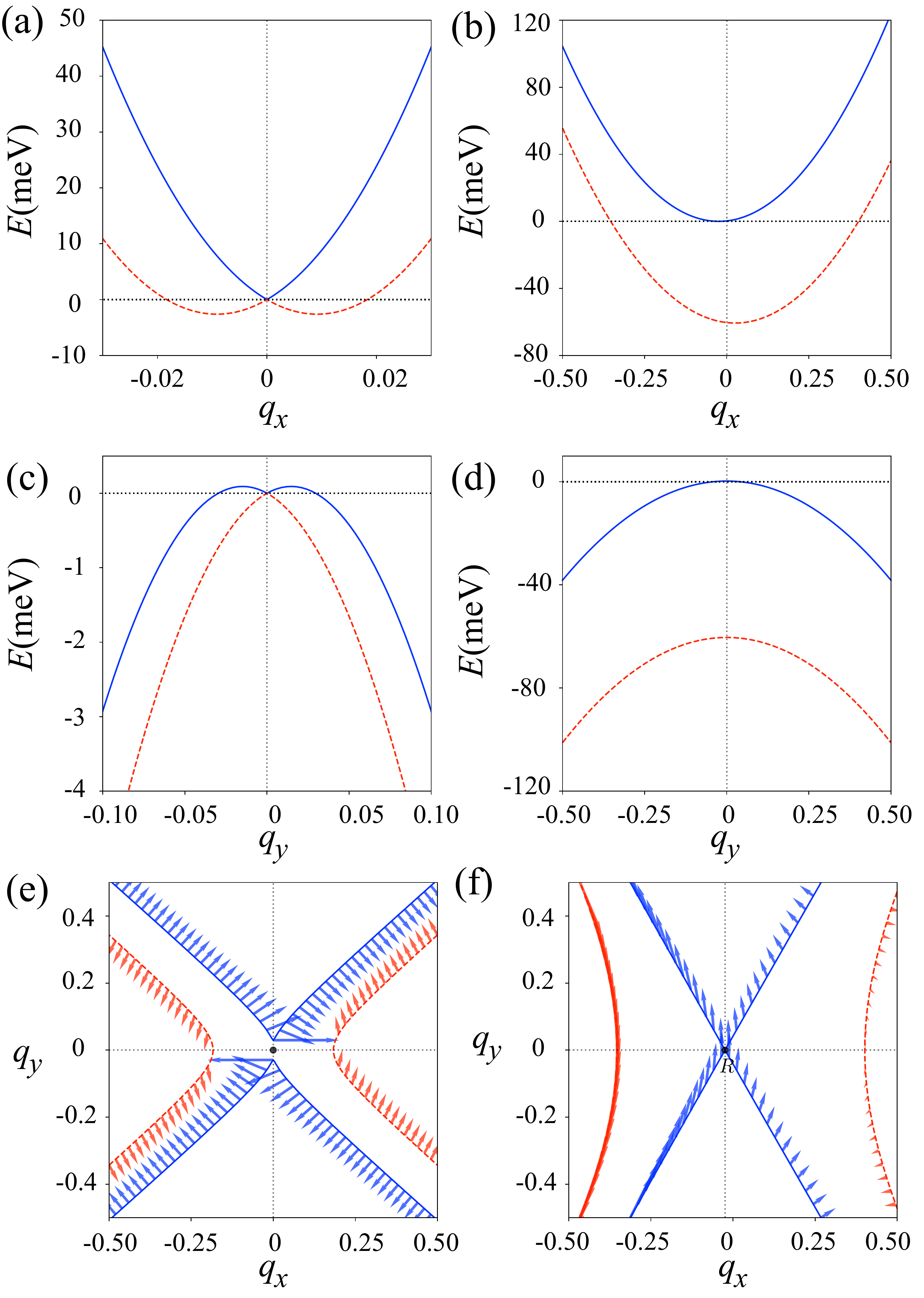}
\end{center}
\caption{\label{spin_texture_2x2} (Color online) 
Cross sections of energy bands around $M_1$ along (a) $q_{x}$ and (c) $q_{y}$ axes, respectively.
Similarly, the bands around $S$ along (b) $q_{x}$ and (d) $q_{y}$ axes, respectively.
Here blue and red colors represent upper and lower energy bands crossing the $E_F$. 
$E_F$ is set to zero.
For $M_{1}$ the $E_F$ is tuned at the band energy of $M_{1}$. For $S$, the $E_F$ is located at the saddle point of the upper energy band. 
Spin textures projected on the Fermi surfaces around (e) $M_{1}$ and (f) $S$. 
}
\end{figure}

\begin{table*}[t]
\begin{ruledtabular}
\begin{tabular}{ c | c  c  c  c  c  c  c  c  c  } 
$\mathbf{k}$-points & $\alpha_x$  & $\alpha_y$   & $\alpha_{xy}$ & $\lambda_R$ & $\zeta_R$ & $\lambda_D$ & $\zeta_D$ & $\lambda_Z$ & $\zeta_Z$ \\
\hline
$\Gamma$           & $0.204a^2$ & $0.204a^2$ &         0              & $0.043a$       & 0               &           0            &     0      &          0          &      0      \\
$K$                      &  $0.538a^2$ & $0.538a^2$ &        0              & $0.086a$       & 0                &          0            &     0      &        0.086     &      0      \\
$M_1$                  & $0.312a^2$ & $0.415a^2$ &        0               & $0.012a$       & 1.220        &           0           &     0       &     $0.055a$   &      0      \\
$M_2$                  & $0.234a^2$ & $0.130a^2$ & $0.629a^2$     & $0.014a$       &  0.901       &      $0.001a$   & 1.009    & $-0.028a$        &$-1.732$ \\
$S$                       & $0.440a^2$ & $0.159a^2$ &  0                    & $-0.017a$      & $-1.161$   &           0           &      0      & $-0.031a$      &      0 \\
\end{tabular}
\end{ruledtabular}
\caption{\label{parameter} Calculated parameters for the effective Hamiltonians around the symmetric points 
in Eqs.~\ref{Eq:H_eff_G},~\ref{Eq:H_eff_K}, ~\ref{Eq:H_eff_M1},~\ref{Eq:H_eff_M2}
and~\ref{Eq:H_eff_S1}. Here $a=\left|\vec{a}_{1}\right|=\left|\vec{a}_{2}\right|$ is the lattice constant.}
\end{table*}

Now, we turn to the effective models around the saddle points where vHSs are most significant. 
The effective Hamiltonian around $M_{1}=\left({\textstyle 0,\frac{2\pi}{\sqrt{3}a}}\right)$ reads
\begin{eqnarray}
\label{Eq:H_eff_M1}\mathcal{H}_{M_1}({\bf q})&=&\left( \alpha_x q_{x}^{2}-\alpha_y q_{y}^{2}\right)\sigma_0
+\lambda_R \left(q_{y}\sigma_{x}-\zeta_R q_{x}\sigma_{y}\right) \nonumber\\
& &+\lambda_Z q_{x}\sigma_{z},
\end{eqnarray}
where 
$\alpha_i=\hbar^{2}/2m_{i}~(i=x,y)$,
$m_{i}$ effective mass along $k_{i}$ direction, $\zeta_R$ anisotropy for the $\lambda_R$, 
and $\lambda_Z$ the momentum dependent effective Zeeman term. 
Numerical values for the parameters based on the TB model 
are summarized in the Table~\ref{parameter}. 
The effective Hamiltonian constitutes the hyperbolic energy surface, the Rashba-like spin-orbit interaction, and the effective Zeeman term proportional to $\sigma_{z}$. Note that the Rashba-like interaction is anisotropic unlike those of $\Gamma$ and $K(K^\prime)$ points. 

At the $M_{2}={\textstyle (\frac{\pi}{a},\frac{\pi}{\sqrt{3}a})}$ point, the minimal model is 
\begin{eqnarray}
\label{Eq:H_eff_M2}\mathcal{H}_{M_2}({\bf q}) &=&\left(-\alpha_x q_{x}^{2}+\alpha_y q_{y}^{2}-\alpha_{xy} q_{x} q_{y}\right)\sigma_0 \nonumber \\
&&+\lambda_R\left(q_{y}\sigma_{x}-\zeta_{R}q_x \sigma_{y}\right)
+\lambda_D \left(q_{x}\sigma_{x}-\zeta_D q_{y}\sigma_{y}\right) \nonumber\\
& &+\lambda_Z \left(q_{x}+\zeta_{Z}q_{y}\right)\sigma_{z},
\end{eqnarray}
where $\alpha_{xy}=\hbar^{2}/m_{2xy}$, $\lambda_D$ the effective Dresselhaus interaction, $\zeta_{D}$ the anisotropy for $\lambda_D$, 
and $\zeta_{Z}$ the anisotropy for $\lambda_Z$. 
Compared with $M_{1}$, the hyperbolic energy surface at $M_{2}$ is rotated by ${\pi}/{3}$, 
thereby including the term proportional to $q_{x} q_{y}$. 
The numerical values for the parameters are in the Table~\ref{parameter}.
Such a relative rotation also leads to both Rashba-like and Dresselhaus-like spin-orbit interaction, 
while the induced spin-orbit interactions in $\mathcal{H}_{\textrm{eff}}(M_{1})$ is purely the Rashba-type.

For the saddle points of $S$($S'$) at ${\bf k}_{S(S')}=\left(\pm2\xi_{0}/a, 0\right)$
where $\xi_{0}\simeq 1.247 \simeq 2\pi/5$, the effective $2 \times 2$ Hamiltonian is 
\begin{eqnarray}
\label{Eq:H_eff_S1}\mathcal{H}_{S(S')}({\bf q})
&=&\left(\alpha_x q_{x}^{2}-\alpha_y q_{y}^{2}\right)\sigma_0
+\lambda_R\left(q_{y}\sigma_{x}-\zeta_{R}q_{x}\sigma_{y}\right) \nonumber \\ 
&&+\lambda_Z q_{x}\sigma_{z}+{\bf B}_\textrm{eff}\cdot{\bm\sigma},
\end{eqnarray}
where ${\bf B}_\textrm{eff} =\pm(0, B_y, B_z)$ and ${\bm\sigma}=(\sigma_x,\sigma_y,\sigma_z)$. 
The first term of Eq.~(\ref{Eq:H_eff_S1}) describes the hyperbolic energy surface around the saddle point of $S$ as expected. 
The second and third terms in Eq.~(\ref{Eq:H_eff_S1}) are the Rashba-type spin-orbit interaction and the Zeeman term, respectively, 
both of which depend on the displacement $\left(q_{x}, q_{y} \right)$ from the saddle point $\mathbf{k}_{S}$. 
The last term is the constant spin-orbit interaction at the saddle point, playing an effective constant magnetic field. 
From numerical calculations based on the TB model, $B_y = 0.011/a$ and $B_z = -0.028/a$, 
and other values are in the Table~\ref{parameter}. 

The difference between the effective models for $M$ and $S$ is that 
Eq.~(\ref{Eq:H_eff_S1}) includes the constant background term of ${\bf B}_\textrm{eff}\cdot{\bm\sigma}$, 
which does not depend on $\mathbf{q}$. 
This constant term can lead to differences in energy and spin profiles around $M_{1}$ and $S$ 
as shown in Fig.~\ref{spin_texture_2x2}. 
Without spin-orbit interactions, the two points $M$ and $S$ are saddle points of hyperbolic energy surfaces. 
When the spin-orbit interaction is turned on, the energy surfaces at $M_{1}$ and $S$ 
evolve to two energy surfaces in the different fashion. 
The spin-orbit interaction for $M_{1}$ 
depends only on $\mathbf{q}$ as shown in Eq.~\ref{Eq:H_eff_M1}.
So, the energy degeneracy at $M_{1}$ cannot not lifted up by the spin-orbit interaction. 
The cross section of energy bands in the $q_{y}=0$ plane shown in Fig.~\ref{spin_texture_2x2}(a) 
looks like the conventional Rashba bands. 
The cross section in the $q_{x}=0$ plane in Fig.~\ref{spin_texture_2x2}(c) also resembles 
that of the conventional Rasbha model, but it is upside down with respect to that in Fig.~\ref{spin_texture_2x2}(b). 
Such peculiar split bands are of consequence from both Rashba-type spin-orbit interactions
and vHS of the hyperbolic shaped energy bands. 
In contrast, the constant spin-orbit interaction in Eq.~(\ref{Eq:H_eff_S1}) for $S$ 
opens a finite gap between two hyperbolic energy surfaces breaking degeneracies along all $k$ points as shown 
in Figs.~\ref{spin_texture_2x2}(b) and \ref{spin_texture_2x2}(d).  

The difference in the spin-orbit interaction for $M_{1}$ and $S$ also affects the spin texture in the vicinity of the points.
Figure~\ref{spin_texture_2x2}(e) shows the spin texture projected on the Fermi surface when the $E_F$ is located at the degeneracy point at $M_{1}$. 
The spin texture around $M_{1}$ exhibits a helical behavior to rotate with respect to the $M_{1}$ point due to $\lambda_R$ term in Eq.~\ref{Eq:H_eff_M1}. 
When projected on the FSs, the resulting spin texture looks like the fan of windmill rotating with respect to the saddle point $M_1$. 
In contrast, the background term ${\bf B}_\textrm{eff}$ in Eq.~(\ref{Eq:H_eff_S1}) forces spin texture in the vicinity of $S$ almost be aligned in one direction
as shown in Fig.~\ref{spin_texture_2x2}(f). 
This explains that spin textures concentrated on $S$ are aligned to the same direction as illustrated in Fig.~\ref{Spin_texture}.

\begin{figure}[b]
\begin{center}
\includegraphics[width=1.0\columnwidth, clip=true]{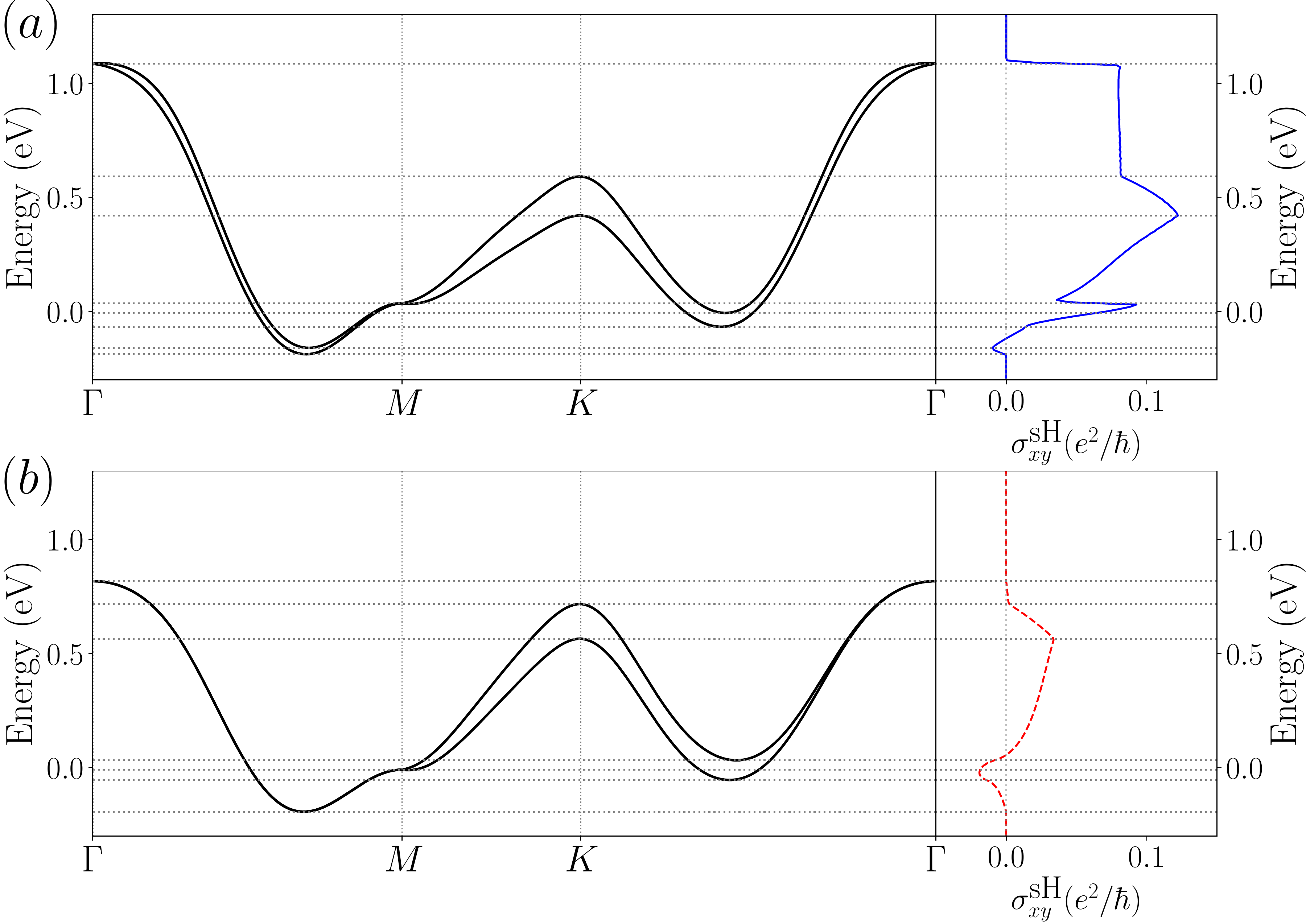} 
\end{center}
\caption{\label{sHc} (Color online) Intrinsic spin Hall conductivities as a function of the Fermi energy. Intrinsic spin Hall conductivities with and without FET gating are denoted by (a) blue solid lines and (b) red dashed ones, respectively. Two energy bands (black solid lines) crossing the $E_F$
 are drawn in the left panels. Gray dotted lines are guidelines to indicate band energies at points $\Gamma$, $M$, and $K$, and local energy minimums, 
 respectively.}
\end{figure}

\begin{figure*}[t]
\begin{center}
\includegraphics[width=0.7\textwidth]{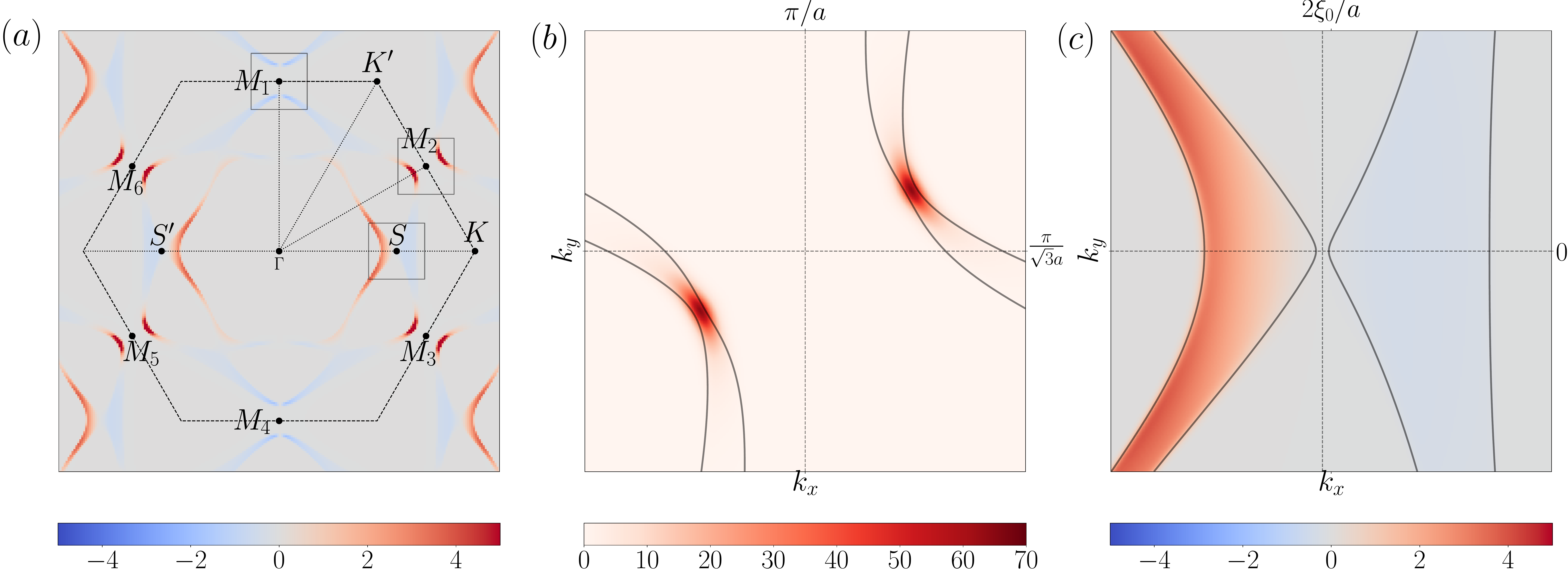}
\end{center}
\caption{\label{sHc_kpts} (Color online) The $k$-resolved spin Hall conductivity $\sigma_{xy}^{\textrm{sH}}(\mathbf{k})$ 
at the $E_F$ under the FET gating 
corresponding to 0.4 holes per unicell. 
The conductivity is plotted in a unit of $e^2/\hbar$. (a) The $k$-resolved spin Hall conductivity in the 1BZ. The value of conductivity is set to be in the range from $-2$ and $2$ in order to clearly display the $k$-resolved conductivity over the 1BZ. Note that the  $k$-resolved conductivity exceeds $2$ around four equivalent points $M_2$, $M_3$, $M_5$, and $M_6$. The $k$-resolved spin Hall conductivity in the vicinity of $M_2$ and $S$ marked by gray boxes are shown in (b) and (c), respectively. 
}
\end{figure*}

\section{\label{sec:ISHC}Intrinsic Spin Hall Conductivity}

We calculate the intrinsic spin Hall conductivity~\cite{RevModPhys_87_1213_2015_Sinova}
in order to investigate the implication of the spin texture change and saddle points on spin transport properties. The intrinsic spni Hall conductivity in the static limit can be derived from the Kubo formula~\cite{PhysRevLett_92_126603_2004_Sinova, PhysRevLett_94_226601_2005_Guo,PhysRevLett_95_156601_2005_Yao,PhysRevLett_100_096401_2008_Guo,PhysRevB_94_085410_2016_Matthes,RevModPhys_87_1213_2015_Sinova,PhysRevB_86_165108_2012_Feng,PhysRevB_99_060408_2019_Zhou, PhysRevB_99_235113_2019_Ryoo} as follows:
\begin{eqnarray}
\label{Eq:sHc}\sigma_{xy}^{\textrm{sH}} (E_F)= -\frac{e\hbar}{N_\mathbf{k} V}&&\sum_{\mathbf{k}} \sum_{n}\sum_{n^\prime \neq n}  \left(f_{n\mathbf{k}}-f_{n^\prime \mathbf{k}} \right) \nonumber \\
&&\times \frac{\textrm{Im}\langle nk|\hat{j}_{x,\mathbf{k}}^{z}|n^{\prime}\mathbf{k}\rangle\langle n^{\prime}\mathbf{k}|\hat{v}_{y,\mathbf{k}}|n\mathbf{k}\rangle}{\left(E_{n\mathbf{k}}-E_{n^\prime \mathbf{k}}\right)^{2}},
\end{eqnarray}
where $e$ is the electron charge, $V$ is the volume of the primitive unit cell, $N_{k}$ is the total number of $k$-grid points, 
and $f_{n\mathbf{k}}\equiv f_\textrm{FD}(E_{n\mathbf{k}}-E_F)$ is
 the Fermi-Dirac distribution for the band energy $E_{n\mathbf{k}}$. 
 The velocity operator is defined as
$\hat{v}_{y,\mathbf{k}}=\frac{1}{\hbar}\frac{\partial}{\partial k_{y}} \mathcal{H}_{\textrm{tot}} (\mathbf{k})$ 
and the spin current operator $\hat{j}_{x,\mathbf{k}}^{z}= \frac{\hbar}{4}\left\{\sigma_{z},\hat{v}_{x,\mathbf{k}}\right\}$. 
 By multiplying $\frac{e}{\hbar}$ to Eq.~(\ref{Eq:sHc}), 
 the intrinsic spin Hall conductivity is calculated in the unit of $\frac{e^2}{\hbar}$.

Figure~\ref{sHc} shows the static intrinsic spin Hall conductivity of monolayer $\textrm{NbSe}_{2}$ 
by tuning the $E_{F}$  together with the energy band structures. 
In this calculation, we only consider two cases; the monolayer with the FET gating 
corresponding to 0.4 holes per unitcell
and without one.
For each case, the TB model parameters are fixed respectively and the  $E_{F}$ varies in Eq.~(\ref{Eq:sHc})
to compute energy dependent spin Hall conductivity. 
So, for the larger $E_F$ away from zero in each case, the TB parameters may not be correct in reflecting realistic situations.
As discussed in Sec.~\ref{sec:DFT}, 
when the system is doped, the energy bands are slightly shifted. 
Notably, the energetic positions of saddle points can be aligned 
with higher doping as shown in Fig.~\ref{FET_bands}.
So, the following computed spin Hall conductivities are precise at the zero energy for each doping and the contributions 
from the two disparate local spin textures at different saddle points can be compared and analyzed. 

In Fig.~\ref{sHc}, it is immediately noticeable that the formation of the in-plane 
spin texture due to the FET gating 
enhances the intrinsic spin Hall conductivity $\sigma_{xy}^{\textrm{sH}}$. 
In particular, the intrinsic spin Hall conductivity changes to $0.072e^2/\hbar$ at $E_F=0$
under the FET gating, while the conductivity without the FET gating is $-0.018e^2/\hbar$.

We calculate the $\mathbf{k}$-resolved 
spin Hall conductivity $\sigma_{xy}^{\textrm{sH}}(\mathbf{k})$ 
defined as $\sigma_{xy}^{\textrm{sH}} = \sum_{\mathbf{k}} \sigma_{xy}^{\textrm{sH}}(\mathbf{k})$ 
in order to investigate the contribution from saddle points to the intrinsic spin Hall conductivity. 
Figure~\ref{sHc_kpts} shows the $\mathbf{k}$-resolved spin Hall conductivity at the Fermi energy. 
Due to the Fermi factor $\left(f_{n\mathbf{k}} - f_{n^\prime \mathbf{k}}\right)$ in Eq.~(\ref{sHc}), 
nonzero values of $\sigma_{xy}^{\textrm{sH}}(\mathbf{k})$ are distributed around the Fermi surface. 
Major contributions to the spin Hall conductivity are concentrated on two regions centered at saddle points $M$ and $S$, respectively. 
Note that six equivalent symmetric points $M$ in the 1BZ do not give equal contributions to the spin Hall conductivity
thanks to the Hall measurement setup of the charge current along the $x$-direction. 
Two large symmetric peaks emerge in the vicinity of four $M$ points, $M_{2}$, $M_{3}$, $M_{5}$, and $M_{6}$ in Fig.~\ref{sHc_kpts}(a). 
The maximum value of the peaks reaches almost $60$ as shown in Fig.~\ref{sHc_kpts}(b). 
Around the other two $M$ points, $M_{1}$ and $M_{4}$ located on the $k_{y}$ axis, $\sigma_{xy}^{\textrm{sH}}(\mathbf{k})$ is negative, 
but much smaller than the four $M$ points aforementioned. 

Another contributions originate from two saddle points $S$ and $S^\prime$ located on the $k_{x}$ axis. 
In comparison with $M$ points, the landscape of $\sigma_{\textrm{xy}}^{\textrm{sH}}(\mathbf{k})$ around $S$ and $S^\prime$ 
disperses over the wider range as shown in Fig.~\ref{sHc_kpts}(c), 
but its maximum height reaches about $4$, smaller than peak heights of $M$ points. 
The $\mathbf{k}$-resolved conductivity around $S$ is asymmetric; 
On the side close to $\Gamma$ with respect to $S$, the $\mathbf{k}$-resolved conductivity gives positive contributions, 
which are almost constantly $4$. 
On the opposite side close to $K$, the $\mathbf{k}$-resolved conductivity shows very small negative values.

The effective models, Eqs.~(\ref{Eq:H_eff_M2}) and (\ref{Eq:H_eff_S1}), 
can be analytically solved to obtain the $\mathbf{k}$-resolved intrinsic spin Hall conductivity. 
See detailed calculations in Appendix~\ref{Appendix-2by2-model}. 
We find that the $\mathbf{k}$-resolved conductivities $\sigma_{xy}^{\textrm{sH}}(\mathbf{k})$ 
in the neighborhood of $M_{2}$ and $S$ from the minimal $2\times2$ Hamiltonians [not shown here] 
are in excellent agreement with $\sigma_{xy}^{\textrm{sH}}(\mathbf{k})$ 
from the five $d$-orbital tight-binding model, implying that the minimal models around the saddle points
can be used to describe spin transport properties as well as other local spin related physical properties.

\begin{figure}[t]
\begin{center}
\includegraphics[width=1.0\columnwidth, clip=true]{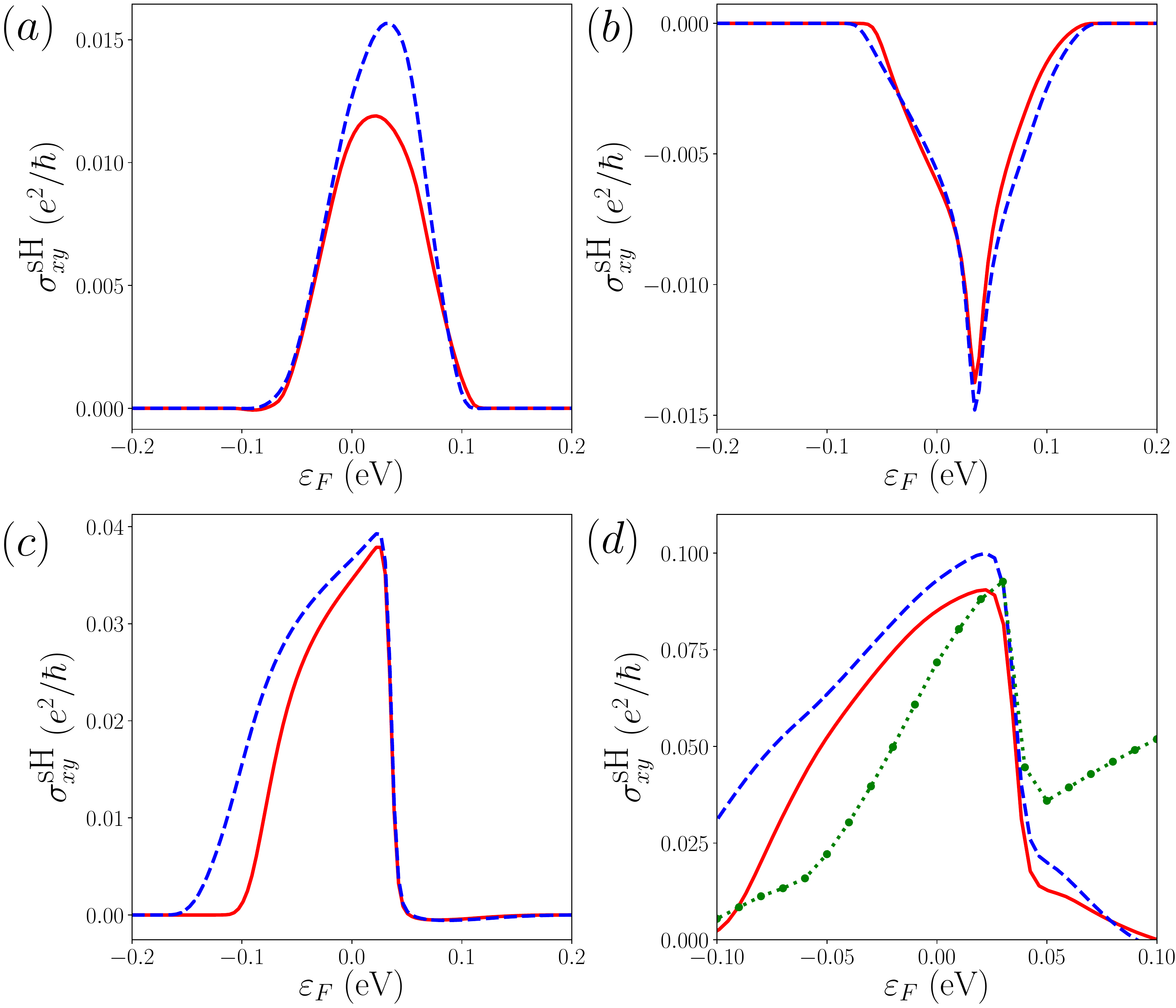}
\end{center}
\caption{\label{fig:sHc_3pts} (Color online) 
Intrinsic momentum-resolved spin Hall conductivity over the vicinity of (a) $S$, (b) $M_{1}$, and (c) $M_{2}$
for the case of 0.4 holes per unit cell. Here the momentum-resolved local spin Hall conductivities for each momentum
are computed within small rectangular boxes shown in Fig.~\ref{sHc_kpts}(a).
Blue dashed and red solid lines correspond to the full TB model and the effective $2 \times 2$ model calculations, respectively. 
(d) Comparison between sum of local saddle point contributions shown in  (a), (b), and (c) 
and the intrinsic spin Hall conductivity (green dotted line) including the whole first BZ contributions.}
\end{figure}

We also compare the sum of three major contributions at $S$, $M_{1}$, and $M_{2}$ 
with the intrinsic spin Hall conductivity Eq.~(\ref{Eq:sHc}) calculated by the full TB model.
Considering symmetry in the presence of charge current along $x$-direction in the spin Hall setup, 
two $S$ points, one single $M_{1}$ point, and two $M_{2}$ points mainly contribute to the total intrinsic spin Hall conductivity over the 1BZ. 
Figure~\ref{fig:sHc_3pts}(d) shows total contributions of five points calculated by 
effective $2 \times 2$ models and the total intrinsic spin Hall conductivity Eq.~(\ref{Eq:sHc}) of the full TB model. 
This comparison shows that the summed contributions of $S$, $M_{1}$, and $M_{2}$ points 
reproduce the sharp drop observed in $\sigma_{xy}^{\textrm{sH}}$ around 0.035 eV very well. 
From these, we can point out that the sharp drop comes from the contribution of the $M_{2}$ 
point to the intrinsic spin Hall conductivity as shown in Fig.~\ref{fig:sHc_3pts}(c).

\section{\label{sec:conclusions}Conclusions}
We investigate the effect of the Rashba spin-orbit 
interaction induced by mirror-reflection symmetry breaking on a single layer of 2$H$-NbSe$_2$. 
We develop the minimal tight-binding model for its electronic structures under the FET gating, 
reproducing the first-principles computational results very well. 
The spin-orbit interaction induced by the broken mirror symmetry 
leads to interesting in-plane spin textures differing from the Ising-type one of the mirror-symmetric system. 
Such differences in spin states are highlighted in spin transport properties, e.g., the intrinsic spin Hall conductivity. 
It is shown that major contributions to the intrinsic spin Hall conductivity take place around energy saddle points 
hosting divergent density of states. 
Unique spin textures depending on the momentum of the van Hove singularities are also shown. 
Since the current system has the singularities quite close to the charge neutral point, 
the well-controlled energetic positions of energy bands in layered materials~\cite{NatNano2015Yu,Science2015Saito,Nature2016Li,NatPhys2016Tsen}
will provide an excellent platform to study relations between unique spin textures and several 
intriguing collective phenomena triggered by the saddle points of hexagonal two-dimensional
crystals~\cite{PRL2008Honerkamp,PRB2011Makogon,NatPhys2012Rahul,PRL2014Yudin,PRB2021Kiesel}.

\section*{Acknowledgement}
We thank Eun-Gook Moon and SangEun Han for fruitful discussions.
S. K. was supported by the National Research Foundation of Korea (NRF) grant funded by the Korea government(MSIT) (Grant No. 2018R1C1B6007233) and by the Open KIAS Center at Korea Institute for Advanced Study.
Y.-W.S. was supported by NRF of Korea (Grant No. 2017R1A5A1014862, SRC program: vdWMRC center) and KIAS individual Grant No. (CG031509). Computations were supported by the CAC of KIAS.

\appendix

\section{\label{Appendix-TB}The Five d-orbital TB model}
Here we provide the full description of the five $d$-orbital TB model, which are constructed by using the Slater-Koster method based on energy integrals Eq.~(\ref{energy_integral}). Independent parameters of energy integrals are determined by irreducible representations and symmetric operations~\cite{Group_Theory_Dresselhaus, PhysRevB_88_085433_Liu}. 

\begin{eqnarray}
\mathcal{H}_{11}&=&\varepsilon_{1} + t_{1}\left[2\cos2\xi+4\cos\xi\cos\eta\right] \nonumber\\
                          &&+ r_{1}\left[4\cos3\xi\cos\eta+2\cos2\eta\right] \nonumber \\
                           &&+u_{1}\left[2\cos4\xi+4\cos2\xi\cos2\eta\right]  
\end{eqnarray}
\begin{eqnarray}
\mathcal{H}_{12}&=&2t_{12}\cos2\xi-2t_{12}\cos\xi\cos\eta+i2\sqrt{3}t_{13}\cos\xi\sin\eta \nonumber\\
&&+2\left(r_{12}+r_{21}\right)\left(\cos3\xi\cos\eta-\cos2\eta\right) \nonumber \\
&&+i2\left(r_{12}-r_{21}\right)\left(\cos3\xi\sin\eta+\sin2\eta\right) \nonumber\\
&&+2u_{12}\cos4\xi-2u_{12}\cos2\xi\cos2\eta \nonumber \\
&&+i2\sqrt{3}u_{13}\cos2\xi\sin2\eta
\end{eqnarray}
\begin{eqnarray}
\mathcal{H}_{13}&=&2it_{13}\sin2\xi-2\sqrt{3}t_{12}\sin\xi\sin\eta+i2t_{13}\sin\xi\cos\eta \nonumber \\
&&-2\sqrt{3}\left(r_{12}+r_{21}\right)\sin3\xi\sin\eta \nonumber \\
&&+i2\sqrt{3}\left(r_{12}-r_{21}\right)\sin3\xi\cos\eta \nonumber\\
&&+2iu_{13}\sin4\xi-2\sqrt{3}u_{12}\sin2\xi\sin2\eta \nonumber \\
&&+i2u_{13}\sin2\xi\cos2\eta
\end{eqnarray}
\begin{eqnarray}
\mathcal{H}_{14}&=&2it_{14}\sin2\xi+2it_{14}\sin\xi\cos\eta-2\sqrt{3}t_{15}\sin\xi\sin\eta \nonumber\\
&&-2\sqrt{3}\left(r_{15}+r_{51}\right)\sin3\xi\sin\eta \nonumber \\
&&+i2\sqrt{3}\left(r_{15}-r_{51}\right)\sin3\xi\cos\eta +2iu_{14}\sin4\xi\nonumber \\
&&+2iu_{14}\sin2\xi\cos2\eta-2\sqrt{3}u_{15}\sin2\xi\sin2\eta
\end{eqnarray}
\begin{eqnarray}
\mathcal{H}_{15}&=&2t_{15}\cos2\xi+i2\sqrt{3}t_{14}\cos\xi\sin\eta-2t_{15}\cos\xi\cos\eta \nonumber\\
&&+2\left(r_{15}+r_{51}\right)\left(\cos3\xi\cos\eta-\cos2\eta\right) \nonumber \\
&&+i2\left(r_{15}-r_{51}\right)\left(\cos3\xi\sin\eta+\sin2\eta\right)+2u_{15}\cos4\xi \nonumber\\
&&+i2\sqrt{3}u_{14}\cos2\xi\sin2\eta-2u_{15}\cos2\xi\cos2\eta
\end{eqnarray}

\begin{eqnarray}
\mathcal{H}_{22}&=& \varepsilon_{2} + 2t_{2}\cos2\xi+\left[t_{2}+3t_{3}\right]\cos\xi\cos\eta \nonumber\\
&& +4r_{2}\cos3\xi\cos\eta+2\left(r_{2}+\sqrt{3}r_{23}\right)\cos2\eta \nonumber\\ 
&&+2u_{2}\cos4\xi+\left[u_{2}+3u_{3}\right]\cos2\xi\cos2\eta
\end{eqnarray}

\begin{eqnarray}
\mathcal{H}_{23}&=&\sqrt{3}\left[t_{2}-t_{3}\right]\sin\xi\sin\eta+i4t_{23}\sin\xi\left(\cos\xi-\cos\eta\right)\nonumber\\
&&-4r_{23}\sin3\xi\sin\eta+\sqrt{3}\left[u_{2}-u_{3}\right]\sin2\xi\sin2\eta \nonumber \\
&&+i4u_{23}\sin2\xi\left(\cos2\xi-\cos2\eta\right)
\end{eqnarray}

\begin{eqnarray}
\mathcal{H}_{24}&=&4it_{24}\sin\xi\cos\xi-i\left[t_{24}-3t_{35}\right]\sin\xi\cos\eta\nonumber\\
&&-\sqrt{3}\left[t_{34}-t_{25}\right]\sin\xi\sin\eta \nonumber \\
&&+i\sqrt{3}\left(r_{34}-r_{43}-r_{25}+r_{52}\right)\sin3\xi\cos\eta \nonumber\\
&&-\sqrt{3}\left(r_{34}+r_{43}-r_{25}-r_{52}\right)\sin3\xi\sin\eta \nonumber \\
&&+i4u_{24}\sin2\xi\cos2x-i\left[u_{24}-3u_{35}\right]\sin2\xi\cos2\eta \nonumber\\
&&-\sqrt{3}\left[u_{34}-u_{25}\right]\sin2\xi\sin2\eta
\end{eqnarray}

\begin{eqnarray}
\mathcal{H}_{25}&=&2t_{25}\cos2\xi-i\sqrt{3}\left[t_{24}+t_{35}\right]\cos\xi\sin\eta \nonumber\\
&&+\left[3t_{34}+t_{25}\right]\cos\xi\cos\eta \nonumber \\
&&+2r_{25}e^{i\eta}\cos3\xi+2r_{52}e^{-i\eta}\cos3\xi \nonumber\\
&&+\left(\frac{3}{2}r_{43}-\frac{1}{2}r_{52}\right)e^{i2\eta}+\left(\frac{3}{2}r_{34}-\frac{1}{2}r_{25}\right)e^{-i2\eta}\nonumber\\
&&+2u_{25}\cos4\xi-i\sqrt{3}\left(u_{24}+u_{35}\right)\cos2\xi\sin2\eta\nonumber\\
&&+\left(3u_{34}+u_{25}\right)\cos2\xi\cos2\eta
\end{eqnarray}

\begin{eqnarray}
\mathcal{H}_{33}&=&\varepsilon_{2} + 2t_{3}\cos2\xi+\left[3t_{2}+t_{3}\right]\cos\xi\cos\eta\nonumber\\
&&+4\left(r_{2}+\frac{2}{\sqrt{3}}r_{23}\right)\cos3\xi\cos\eta \nonumber\\
&&+2\left(r_{2}-\frac{\sqrt{3}}{3}r_{23}\right)\cos2\eta\nonumber\\
&&+2u_{3}\cos4\xi+\left[3u_{2}+u_{3}\right]\cos2\xi\cos2\eta
\end{eqnarray}

\begin{eqnarray}
\mathcal{H}_{34}&=&2t_{34}\cos2\xi+i\sqrt{3}\left[t_{24}+t_{35}\right]\cos\xi\sin\eta \nonumber\\
&&+\left[t_{34}+3t_{25}\right]\cos\xi\cos\eta\nonumber\\
&&+2r_{34}e^{i\eta}\cos3\xi+2r_{43}e^{-i\eta}\cos3\xi\nonumber\\
&&+\left(-\frac{1}{2}r_{43}+\frac{3}{2}r_{52}\right)e^{i2\eta}+\left(\frac{3}{2}r_{25}-\frac{1}{2}r_{34}\right)e^{-i2\eta}\nonumber\\
&&+2u_{34}\cos4\xi+i\sqrt{3}\left[u_{24}+u_{35}\right]\cos2\xi\sin2\eta\nonumber\\
&&+\left[u_{34}+3u_{25}\right]\cos2\xi\cos2\eta
\end{eqnarray}

\begin{eqnarray}
\mathcal{H}_{35}&=&i4t_{35}\sin\xi\cos\xi+i\left[3t_{24}-t_{35}\right]\sin\xi\cos\eta \nonumber\\
&&-\sqrt{3}\left[t_{34}-t_{25}\right]\sin\xi\sin\eta\nonumber\\
&&+i\sqrt{3}\left(r_{34}-r_{43}-r_{25}+r_{52}\right)\sin3\xi\cos\eta \nonumber\\
&&-\sqrt{3}\left(r_{34}+r_{43}-r_{25}-r_{52}\right)\sin3\xi\sin\eta\nonumber\\
&&+i4u_{35}\sin2\xi\cos2\xi+i\left[3u_{24}-u_{35}\right]\sin2\xi\cos2\eta \nonumber\\
&&-\sqrt{3}\left[u_{34}-u_{25}\right]\sin2\xi\sin2\eta
\end{eqnarray}

\begin{eqnarray}
\mathcal{H}_{44}&=&\varepsilon_{3} + 2t_{4}\cos2\xi+\left[t_{4}+3t_{5}\right]\cos\xi\cos\eta\nonumber\\
&&+4r_{4}\cos3\xi\cos\eta+2\left(r_{4}-\sqrt{3}r_{45}\right)\cos2\eta\nonumber\\
&&+2u_{4}\cos4\xi+\left[u_{4}+3u_{5}\right]\cos2\xi\cos2\eta
\end{eqnarray}

\begin{eqnarray}
\mathcal{H}_{45}&=&\sqrt{3}\left[t_{5}-t_{4}\right]\sin\xi\sin\eta+i4t_{45}\sin\xi\left(\cos\xi-\cos\eta\right)\nonumber \\
&&-4r_{45}\sin3\xi\sin\eta
+\sqrt{3}\left(u_{5}-u_{4}\right)\sin2\xi\sin2\eta\nonumber\\
&&+i4u_{45}\sin2\xi\left(\cos2\xi-\cos2\eta\right)
\end{eqnarray}

\begin{eqnarray}
\mathcal{H}_{55}&=&\varepsilon_{3} + 2t_{5}\cos2\xi+\left[3t_{4}+t_{5}\right]\cos\xi\cos\eta\nonumber\\
&&+4\left(r_{4}-\frac{2}{\sqrt{3}}r_{45}\right)\cos3\xi\cos\eta \nonumber \\
&&+2\left(r_{4}+\frac{\sqrt{3}}{3}r_{45}\right)\cos2\eta\nonumber\\
&&+2u_{5}\cos4\xi+\left[3u_{4}+u_{5}\right]\cos2\xi\cos2\eta,
\end{eqnarray}
where $\xi \equiv \frac{1}{2}k_{x}a$ and $\eta=\frac{\sqrt{3}}{2}k_{y}a$. Here $\varepsilon_{1}$, $\varepsilon_{2}$, and $\varepsilon_{3}$ are on-site energies for $\{d_{z^2}\}$, $\{d_{xy}, d_{x^2-y^2}\}$, and $\{d_{zx}, d_{yz}\}$, respectively. 
The first-, second- and third-nearest neighbor energy integrals are given by $t_{ij}=E_{ij}(R_1)$, $r_{ij}=E_{ij}(\tilde{R}_{1})$
and $u_{ij}=E_{ij}(2R_1)$, respectively where $i,j = 1,\cdots, 5~ (i\ge j)$, $t_{i}\equiv t_{ii}$, $u_{i}\equiv u_{ii}$, $r_{i}\equiv r_{ii}$,
$R_1=|\vec{a}_{1}|=|\vec{a}_2|$ and $\tilde{R}_1 =|\vec{a}_1+\vec{a}_2|$.

\begin{table*}[t]
\centering
\begin{center}
\begin{tabular}{ c  c  c  c  c  c  c  c  c  c  c  c  c  c  c } 
\hline\hline
\multicolumn{15}{c}{On-site energies} \\
\hline
& & & & & & $\varepsilon_{1}$ & $\varepsilon_{2}$ & $\varepsilon_{3}$ & & & & & \\
& & & & & & 1.859 & 2.303 & 3.381 & & & & & \\
\hline
\multicolumn{15}{c}{Nearest neighbor energy integrals} \\
\hline
$t_{1}$ & $t_{2}$ & $t_{3}$ & $t_{4}$ & $t_{5}$ & $t_{12}$ & $t_{13}$ & $t_{14}$ & $t_{15}$ & $t_{23}$ & $t_{24}$ & $t_{25}$ & $t_{34}$ & $t_{35}$ & $t_{45}$\\
-0.1554 & -0.2531 & 0.4279 & -0.0541 & -0.1490 & 0.2097 & 0.3873 & -0.1422 & 0.1796 & -0.2459 & 0.0756 & 0.2188 & -0.2071 & -0.0857 & -0.1384\\
\hline
\multicolumn{15}{c}{Second nearest neighbor energy integrals} \\
\hline
& $r_{1}$ & $r_{2}$ & $r_{4}$ & $r_{12}$ & $r_{21}$ & $r_{15}$ & $r_{51}$ & $r_{23}$ & $r_{25}$ & $r_{52}$ & $r_{34}$ & $r_{43}$ & $r_{45}$ & \\
& -0.0398 & 0.0174 &  0.1243 & -0.0853 & 0.0038 & 0.0578 &  0.0693 & -0.0235 & -0.0039 & 0.0460 & 0.0259 & -0.0015 &  0.0743 & \\
\hline
\multicolumn{15}{c}{Third nearest neighbor energy integrals} \\
\hline
$u_{1}$ & $u_{2}$ & $u_{3}$ & $u_{4}$ & $u_{5}$ & $u_{12}$ & $u_{13}$ & $u_{14}$ & $u_{15}$ & $u_{23}$ & $u_{24}$ & $u_{25}$ & $u_{34}$ & $u_{35}$ & $u_{45}$\\

0.0671 & -0.0397 & 0.0377 & 0.0051 & 0.0083 & -0.0551 & -0.0755 & -0.0236 & -0.0311 & 0.0494 & 0.0394 & 0.0467 & -0.0881 & -0.0521 & -0.0280\\
\hline\hline
\end{tabular}
\end{center}
\caption{\label{Table_parameters} Fitting parameters in a unit of $eV$.}
\end{table*}
\begin{widetext}
We can also add the atomic spin-orbit coupling term to the tight-binding model. For the five $d$-orbitals of the transition metal atoms, the atomic spin-orbit interaction reads
\begin{equation}
\label{Eq:Atomic_SOC}
\mathcal{H}_{\textrm{SOC}} = \lambda_{\textrm{TM}}\hat{S}\cdot\hat{L}_{m}\dot{=}\lambda_{\textrm{TM}}\left[\begin{array}{cccccccccc}
0 & 0 & 0 & 0 & 0 & 0 & 0 & 0 & -\frac{\sqrt{3}}{2} & i\frac{\sqrt{3}}{2}\\
0 & 0 & -i & 0 & 0 & 0 & 0 & 0 & \frac{1}{2} & \frac{i}{2}\\
0 & i & 0 & 0 & 0 & 0 & 0 & 0 & -\frac{i}{2} & \frac{1}{2}\\
0 & 0 & 0 & 0 & -\frac{i}{2} & \frac{\sqrt{3}}{2} & -\frac{1}{2} & \frac{i}{2} & 0 & 0\\
0 & 0 & 0 & \frac{i}{2} & 0 & -i\frac{\sqrt{3}}{2} & -\frac{i}{2} & -\frac{1}{2} & 0 & 0\\
0 & 0 & 0 & \frac{\sqrt{3}}{2} & i\frac{\sqrt{3}}{2} & 0 & 0 & 0 & 0 & 0\\
0 & 0 & 0 & -\frac{1}{2} & \frac{i}{2} & 0 & 0 & i & 0 & 0\\
0 & 0 & 0 & -\frac{i}{2} & -\frac{1}{2} & 0 & -i & 0 & 0 & 0\\
-\frac{\sqrt{3}}{2} & \frac{1}{2} & \frac{i}{2} & 0 & 0 & 0 & 0 & 0 & 0 & \frac{i}{2}\\
-i\frac{\sqrt{3}}{2} & -\frac{i}{2} & \frac{1}{2} & 0 & 0 & 0 & 0 & 0 & -\frac{i}{2} & 0
\end{array}\right], 
\end{equation}
which is written with the basis $|z^{2},\uparrow\rangle$, $|x^{2}-y^{2},\uparrow\rangle$, $|xy,\uparrow\rangle$, $|zx,\uparrow\rangle$, $|yz,\uparrow\rangle$, $|z^{2},\downarrow\rangle$, $|x^{2}-y^{2},\downarrow\rangle$, $|xy,\downarrow\rangle$, $|zx,\downarrow\rangle$, and $|yz,\downarrow\rangle$. Here $|\uparrow\rangle$ and $|\downarrow\rangle$ are spin eigenstates of the $\hat{S}_{z}$ operator. $\lambda_{\textrm{TM}}$ is the atomic spin-orbit coupling constant of transition metal atoms, which is determined by fitting the TB model to DFT energy bands. Table~\ref{Table_parameters} summarizes a set of TB parameters obtained by performing the least-squared fitting procedure. 
\end{widetext}

\section{\label{Appendix-kp}Derivation: The Effective Minimal Model}
Here we adopt the $k \cdot p$ approximation in order to obtain the minimal model in the vicinity of the saddle point $k_{S}=\left(2\xi_{0}/a, 0\right)$. Since the lowest energy band of the effective five orbital TB model is of our interest, we first diagonalize the TB Hamiltonian at $k_{S}$ such that 
\begin{equation}
\mathcal{U}^{\dagger}(k_{S})\mathcal{H}(k_{S})\mathcal{U}(k_{S}) = \mathcal{D}(k_{S}).
\end{equation}
Here $\mathcal{D}(k_{S})$ and $\mathcal{U}(k_{S})$ are matrices of energy eigenvalues and corresponding eigenvectors:
\begin{equation}
\mathcal{D}(k_{S}) = \left[\begin{array}{ccccc}
E_{1} & 0 & 0 & 0 & 0\\
0 & E_{2} & 0 & 0 & 0\\
0 & 0 & E_{3} & 0 & 0\\
0 & 0 & 0 & E_{4} & 0\\
0 & 0 & 0 & 0 & E_{5}
\end{array}\right],
\end{equation}
and
\begin{equation}
\mathcal{U}(k_{S})=\left[\begin{array}{ccccc}
| & | & | & | & |\\
\vec{u}_{1} & \vec{u}_{2} & \vec{u}_{3} & \vec{u}_{4} & \vec{u}_{5}\\
| & | & | & | & |
\end{array}\right],
\end{equation}
where $E_{i}$ are the $i$th band energy at $k_{S}$, and $\vec{u}_{i}$ is the corresponding eigenvector. 

The TB Hamiltonian $\mathcal{H}$ can be expanded in terms of very small displacement $\delta k_{x}$ and $\delta k_{y}$ from the saddle point $k_{S}$:
\begin{equation}
\mathcal{H}\left(\mathbf{k}_{S} + \delta \mathbf{k} \right) \approx \mathcal{H}(\mathbf{k}_{S}) + \left. \delta \mathcal{H} \right|_{\mathbf{k}_{S}}
\end{equation}
When the Hamiltonian is denoted by $\bar{\mathcal{H}}$ under the unitary transformation $\mathcal{U} \equiv \mathcal{U}(\mathbf{k}_{S})$, the expanded Hamiltonian reads
\begin{equation}
\bar{\mathcal{H}}\left(\mathbf{k}_{S} + \delta \mathbf{k} \right) \approx \bar{\mathcal{H}}(\mathbf{k}_{S}) + \left. \delta \bar{\mathcal{H}} \right|_{\mathbf{k}_{S}} = \mathcal{D}(\mathbf{k}_{S}) + \left. \delta \bar{\mathcal{H}} \right|_{\mathbf{k}_{S}},
\end{equation}
where 
\begin{eqnarray}
\delta \bar{\mathcal{H}} &=& 
\frac{\partial \bar{\mathcal{H}}}{\partial k_{x}} \delta k_{x} + \frac{\partial \bar{\mathcal{H}}}{\partial k_{y}} \delta k_{y} 
+ \frac{1}{2}\frac{\partial^{2} \bar{\mathcal{H}}}{\partial k_{x}^{2}} \delta k_{x}^{2} 
+ \frac{\partial^{2} \bar{\mathcal{H}}}{\partial k_{x} \partial k_{y}} \delta k_{x} \delta k_{y}\nonumber\\
&& + \frac{1}{2}\frac{\partial^{2} \bar{\mathcal{H}}}{\partial k_{y}^{2}} \delta k_{y}^{2} + \mathcal{O}(\delta \mathbf{k}^3)
\end{eqnarray}
Note that the expansion can be performed up to the second order of the displacement $\delta \mathbf{k}$ in order to obtain the quadratic form of the minimal model. 

When the atomic-spin orbit coupling is included, the total Hamiltonian $\mathcal{H}_{\textrm{tot}}$, which is ten-dimensional, is expressed in a form of the $2\times 2$ block matrix, each of which is five-dimensional, as follows:
as follows:

\begin{eqnarray}
\mathcal{H}_{\textrm{tot}} &=& \left[\begin{array}{cc}
\mathcal{H}_{\textrm{tot}}^{\uparrow\uparrow} & \mathcal{H}_{\textrm{tot}}^{\uparrow\downarrow} \\
\mathcal{H}_{\textrm{tot}}^{\downarrow\uparrow} & \mathcal{H}_{\textrm{tot}}^{\downarrow\downarrow}
\end{array} \right] = \left[\begin{array}{cc}
\mathcal{H} & 0 \\
0 & \mathcal{H}
\end{array} \right] + \mathcal{H}_{\textrm{soc}} \nonumber \\
&=& \left[\begin{array}{cc}
\mathcal{H} & 0 \\
0 & \mathcal{H}
\end{array} \right] + \left[\begin{array}{cc}
\mathcal{H}_{\textrm{soc}}^{\uparrow\uparrow} & \mathcal{H}_{\textrm{soc}}^{\uparrow\downarrow} \\
\mathcal{H}_{\textrm{soc}}^{\downarrow\uparrow} & \mathcal{H}_{\textrm{soc}}^{\downarrow\downarrow}
\end{array} \right],
\end{eqnarray}
where superscripts $\uparrow$ and $\downarrow$ indicate spin components, and $\mathcal{H}$ is the spinless TB model. For simplicity the $\mathbf{k}$ vector $\mathbf{k}_{S}$ is not explicitly written from now on. It means that Hamiltonians are assumed to be defined at \textbf{}. If Hamiltonian components defined not at $\mathbf{k}_{S}$ are needed to consider, the $\mathbf{k}$ vector associated with the components will be explicitly specified. 

The total Hamiltonian $\mathcal{H}_{\textrm{tot}}$ can be unitarily transformed by using $\mathcal{U}$
such as 
$
\bar{\mathcal{H}}_{\textrm{tot}} 
= [\mathcal{U}^\dagger\otimes\mathcal{I}_{2\times 2}] \mathcal{H}_\textrm{tot}[\mathcal{U}\otimes\mathcal{I}_{2\times 2}]
=\bar{\mathcal{H}}_\textrm{tot}^{0} +\delta \bar{\mathcal{H}}_\textrm{tot} +
\bar{\mathcal{H}}_\textrm{SOC},
$
where $\mathcal{I}_{2\times2}$ is a 2$\times$2 identical matrix.

Here we define that 
\begin{eqnarray}
\bar{\mathcal{H}}_\textrm{tot}^{0} &=& \left[\begin{array}{cc}
\bar{\mathcal{H}}^{0} & 0 \\
0 & \bar{\mathcal{H}}^{0} 
\end{array} \right] \\
\delta \bar{\mathcal{H}}_\textrm{tot} &=& \left[\begin{array}{cc}
\delta \bar{\mathcal{H}} & 0 \\
0 & \delta \bar{\mathcal{H}}
\end{array} \right] \\
\bar{\mathcal{H}}_\textrm{SOC} &=& \left[\begin{array}{cc}
\mathcal{U}^{\dagger}\mathcal{H}_{\textrm{soc}}^{\uparrow\uparrow}\mathcal{U} & \mathcal{U}^{\dagger}\mathcal{H}_{\textrm{soc}}^{\uparrow\downarrow}\mathcal{U} \\
\mathcal{U}^{\dagger}\mathcal{H}_{\textrm{soc}}^{\downarrow\uparrow}\mathcal{U} & \mathcal{U}^{\dagger}\mathcal{H}_{\textrm{soc}}^{\downarrow\downarrow}\mathcal{U}
\end{array} \right]
\end{eqnarray}

Here $\delta \bar{\mathcal{H}}_\textrm{tot} + \bar{\mathcal{H}}_\textrm{soc}$ are treated as a perturbation to $\bar{\mathcal{H}}_\textrm{tot}^0$.
Now we apply the quasi-degenerate perturbation theory to this decomposition of the total Hamiltonian. We can consider two subspaces $A$ and $B$: $A=\{|\psi_{1 \mathbf{k}_{S}}\rangle|\uparrow\rangle, |\psi_{1 \mathbf{k}_{S}}\rangle|\downarrow\rangle \}$ where $|\psi_{1\mathbf{k}_{S}}\rangle$ is the lowest energy state of $\bar{\mathcal{H}}^{0}$ at the saddle point $\mathbf{k}_{S}$, and $B$ is the subspace consisting of remaining energy levels, whose dimension is eight. We can further decompose the perturbation $\delta \bar{\mathcal{H}}_{\textrm{tot}} + \bar{\mathcal{H}}_{\textrm{soc}}$ into two parts such that 
\begin{equation}
\delta \bar{\mathcal{H}}_{\textrm{tot}} + \bar{\mathcal{H}}_{\textrm{soc}} = \bar{\mathcal{H}}_{\textrm{tot}}^{1} + \bar{\mathcal{H}}_{\textrm{tot}}^{2},
\end{equation}
where $\bar{\mathcal{H}}_{\textrm{tot}}^{1}$ is the perturbation Hamiltonian describing interactions only between states within $A$ and $B$ subspaces, and $\bar{\mathcal{H}}_{\textrm{tot}}^{2}$ is the Hamiltonian part interacting only between $A$ and $B$ subspaces. 

By using the Schrieffer-Wolff transformation where the unitary transformation $e^{-S}$ is applied to to $\bar{\mathcal{H}}_{\textrm{tot}}$,
\begin{equation}
\tilde{\mathcal{H}}_{\textrm{tot}} = e^{-S} \bar{\mathcal{H}}_{\textrm{tot}} e^{S},
\end{equation}
one can find out the generator $S$ such that there is no interaction matrix between $A$ and $B$ in the unitarily transformed Hamiltonian $\tilde{\mathcal{H}}_{\textrm{tot}}$. By expanding the generator $S$ in terms of the perturbation $\delta \bar{\mathcal{H}}_\textrm{tot} + \bar{\mathcal{H}}_\textrm{soc}$, one can obtain the perturbative expansion approximation to $\tilde{\mathcal{H}}_{\textrm{tot}}$,
\begin{equation}
\tilde{\mathcal{H}}_{\textrm{tot}} = \tilde{\mathcal{H}}^{(0)}_{\textrm{tot}} + \tilde{\mathcal{H}}^{(1)}_{\textrm{tot}} + \tilde{\mathcal{H}}^{(2)}_{\textrm{tot}} + \cdots,
\end{equation}
where the superscript $(n)$ stands for the perturbation order. 
Using the notation that $\mathcal{H}_{m\sigma,n\sigma^\prime} = \langle \psi_{m \mathbf{k}_{S}\sigma} | \mathcal{H} | \psi_{n \mathbf{k}_{S}\sigma^\prime}\rangle$ for Bloch states $|\psi_{m \mathbf{k}_{S}} \sigma \rangle$ at the saddle point $k_{S}$, the perturbative expansion terms for the $A$ subspace up to the second order reads
\begin{eqnarray}
\label{Htot0}\left[\tilde{\mathcal{H}}^{(0)}_{\textrm{tot}}\right]_{1\sigma, 1\sigma^\prime} &=& \bar{\mathcal{H}}^{0}_{11} \delta_{\sigma\sigma^\prime} \\
\label{Htot1}\left[\tilde{\mathcal{H}}^{(1)}_{\textrm{tot}}\right]_{1\sigma, 1\sigma^\prime} &=& 
\left[\bar{\mathcal{H}}^{1}_{\textrm{tot}}\right]_{1\sigma,1\sigma^{\prime}} =\left[\delta \bar{\mathcal{H}}\right]_{11}\delta_{\sigma \sigma^{\prime}} + \left[\bar{\mathcal{H}}_{\textrm{soc}}\right]_{1\sigma, 1\sigma^\prime} \nonumber \\ \\
\label{Htot2}\left[\tilde{\mathcal{H}}^{(2)}_{\textrm{tot}}\right]_{1\sigma, 1\sigma^\prime} &=& \sum_{m=2}^{5}\sum_{\sigma^{\prime\prime}=\uparrow,\downarrow} \frac{\left[\bar{\mathcal{H}}^{2}_{\textrm{tot}}\right]_{1\sigma, m\sigma^{\prime\prime}}\left[\bar{\mathcal{H}}^{2}_{\textrm{tot}}\right]_{ m\sigma^{\prime\prime},1\sigma^{\prime}}}{E_{1}-E_{m}} \nonumber \\
\end{eqnarray}

\begin{widetext}
Using the fact that
\begin{eqnarray}
\left[\bar{\mathcal{H}}^{2}_{\textrm{tot}}\right]_{1\sigma, m\sigma^{\prime\prime}} &=& \delta \bar{\mathcal{H}}_{1m}\delta_{\sigma\sigma^{\prime\prime}} + \left[ \bar{\mathcal{H}}_{\textrm{soc}} \right]_{1\sigma, m \sigma^{\prime\prime}} \\
\left[\bar{\mathcal{H}}^{2}_{\textrm{tot}}\right]_{m\sigma^{\prime\prime}, 1\sigma^{\prime}} &=& \delta \bar{\mathcal{H}}_{m1}\delta_{\sigma^{\prime\prime}\sigma^{\prime}} + \left[ \bar{\mathcal{H}}_{\textrm{soc}} \right]_{m \sigma^{\prime\prime}, 1\sigma^{\prime}}, 
\end{eqnarray}
the second-order term can be further decomposed into 
\begin{equation}
\left[\tilde{\mathcal{H}}^{(2)}_{\textrm{tot}}\right]_{1\sigma, 1\sigma^\prime} = \left[\tilde{\mathcal{H}}^{(2)}_{\textrm{tot},0}\right]_{1\sigma, 1\sigma^\prime} + 
\left[\tilde{\mathcal{H}}^{(2)}_{\textrm{tot},1}\right]_{1\sigma, 1\sigma^\prime} + 
\left[\tilde{\mathcal{H}}^{(2)}_{\textrm{tot},2}\right]_{1\sigma, 1\sigma^\prime},    
\end{equation}
where
\begin{eqnarray}
\label{Htot20}\left[\tilde{\mathcal{H}}_{\textrm{tot},0}^{(2)}\right]_{1\sigma,1\sigma^{\prime}} &=&	\sum_{m=2}^{5}\sum_{\sigma^{\prime\prime}=\uparrow,\downarrow}\frac{\left[\bar{\mathcal{H}}_{\textrm{soc}}\right]_{1\sigma,m\sigma^{\prime\prime}}\left[\bar{\mathcal{H}}_{\textrm{soc}}\right]_{m\sigma^{\prime\prime},1\sigma^{\prime}}}{E_{1}-E_{m}} \nonumber\\ \\
\label{Htot21}\left[\tilde{\mathcal{H}}_{\textrm{tot},1}^{(2)}\right]_{1\sigma,1\sigma^{\prime}} &=&	\sum_{m=2}^{5}\frac{1}{E_{1}-E_{m}}\left(\left[\delta\bar{\mathcal{H}}\right]_{1m}\left[\bar{\mathcal{H}}_{\textrm{soc}}\right]_{m\sigma,1\sigma^{\prime}}+\left[\bar{\mathcal{H}}_{\textrm{soc}}\right]_{1\sigma,m\sigma^{\prime}}\left[\delta\bar{\mathcal{H}}\right]_{m1}\right) \\
\label{Htot22}\left[\tilde{\mathcal{H}}_{\textrm{tot},2}^{(2)}\right]_{1\sigma,1\sigma^{\prime}} &=&	\sum_{m=2}^{5}\frac{\left[\delta\bar{\mathcal{H}}\right]_{1m}\left[\delta\bar{\mathcal{H}}\right]_{m1}}{E_{1}-E_{m}}\delta_{\sigma\sigma^{\prime}}
\end{eqnarray}
\end{widetext}

The resulting $k \cdot p$ Hamiltonian in the vicinity of the saddle point $\mathbf{k}_{S}$ can be expressed in terms of $\sigma_i~(i=0,x,y,z)$. 
To be specific, Eq.~(\ref{Htot0}), the first term of Eq.~(\ref{Htot1}), and Eq.~(\ref{Htot22}) contribute to the Hamiltonian part proportional to $\sigma_0$. 
The second term of Eq.~(\ref{Htot1}), Eq.~(\ref{Htot20}), and Eq.~(\ref{Htot21}), which involves the atomic spin-orbit coupling $\mathcal{H}_\textrm{soc}$, lead to the Hamiltonian expressed in terms of $\sigma_i~(i=x,y,z)$, which corresponds to the effective spin-orbit interaction.

\section{\label{Appendix-2by2-model}Spin Expectations and the Intrinsic Spin Hall Conductivity of the Minimal Model}
Let us consider a generic $2\times2$ Hermitian matrix, which is generally written as
\begin{eqnarray}
\label{Eq:H_generic}
\mathcal{H}(\mathbf{k})&=& H_{0}(\mathbf{k})\sigma_0 + \alpha(\mathbf{k})\sigma_{x}+\beta(\mathbf{k})\sigma_{y}+\gamma(\mathbf{k})\sigma_{z} \\
&=&\left[\begin{array}{cc}
\gamma & \alpha-i\beta\\
\alpha+i\beta & -\gamma
\end{array}\right],
\end{eqnarray}
where $H_{0}(\mathbf{k})$, $\alpha(\mathbf{k})$, $\beta(\mathbf{k})$, and $\gamma(\mathbf{k})$ are real functions of $\mathbf{k}$.
Eigenvalues $\varepsilon_{\pm}$ can be obtained by solving the characteristic equation, 
\begin{equation}
\varepsilon_{\pm}=H_{0}\pm\lambda,
\end{equation}
where $\lambda\equiv\sqrt{\alpha^{2}+\beta^{2}+\gamma^{2}}$.
The corresponding eigenvectors are 
\begin{equation}
|v_{\pm}\rangle = \left[\begin{array}{c}
x_{\pm}\\
y_{\pm}\end{array}\right]=\frac{1}{\sqrt{2\lambda^{2}\mp2\lambda\gamma}}\left[\begin{array}{c}
\alpha-i\beta\\
\pm\lambda-\gamma
\end{array}\right],
\end{equation}
when $\lambda \neq 0$. If $\lambda \neq 0$, i.e., $\alpha=\beta=\gamma=0$, eigenvalues are degenerate, so eigenvectors can be chosen to be 
\begin{equation}
|v_{+}\rangle = \left[\begin{array}{c}
1\\
0\end{array}\right],\;
|v_{-}\rangle = \left[\begin{array}{c}
0\\
1\end{array}\right].
\end{equation}
Using these results, one can calculate spin expectation values as follows:
\begin{eqnarray}
\langle S_{x}\rangle_{\pm} &=& \frac{\hbar}{2}\left(x_{\pm}y_{\pm}^{*}+x_{\pm}^{*}y_{\pm}\right) = \pm \frac{\hbar}{2}\frac{\alpha}{\lambda} \\
\langle S_{y}\rangle_{\pm} &=& i\frac{\hbar}{2}\left(x_{\pm}y_{\pm}^{*}-x_{\pm}^{*}y_{\pm}\right)= \pm\frac{\hbar}{2}\frac{\beta}{\lambda} \\
\langle S_{z}\rangle_{\pm} &=& \frac{\hbar}{2}\left(\left|x_{\pm}\right|^{2}-\left|y_{\pm}\right|^{2}\right) = \pm\frac{\hbar}{2}\frac{\gamma}{\lambda} \\
\langle\vec{S}\rangle_{\pm} &=& \pm\frac{\hbar}{2\lambda}\left(\alpha\hat{x}+\beta\hat{y}+\gamma\hat{z}\right)=\pm\frac{\hbar}{2\lambda}\left[\begin{array}{c}
\alpha\\
\beta\\
\gamma
\end{array}\right],
\end{eqnarray}
where $\lambda \neq 0$.
The intrnsic spin Hall conductivity derived from the Kubo formula is 
\begin{eqnarray}
\sigma_{xy}^{\textrm{sH}} &=& -\frac{e^2}{N_\mathbf{k} V}
\sum_{\mathbf{k}} \sum_{n}\sum_{n^\prime \neq n}  
\left(f_{n\mathbf{k}}-f_{n^\prime \mathbf{k}} \right) \nonumber\\
&& \times\frac{\textrm{Im}\langle nk|\hat{j}_{x,\mathbf{k}}^{z}|n^{\prime}\mathbf{k}\rangle\langle n^{\prime}\mathbf{k}|\hat{v}_{y,\mathbf{k}}|n\mathbf{k}\rangle}{\left(E_{n\mathbf{k}}-E_{n^\prime \mathbf{k}}\right)^{2}}.
\end{eqnarray}
Here the spin current operator $\hat{j}_{x}^{z}=\frac{\hbar}{4}\left\{ \sigma_{z},v_{x}\right\} $and the velocity operator $\hat{v}_{y}$ are 
\begin{eqnarray}
\hbar\hat{v}_{y}&=&\partial_{y}H_{0}\mathbf{I}+\partial_{y}\alpha\sigma_{x}+\partial_{y}\beta\sigma_{y}+\partial_{y}\gamma\sigma_{z} \\
\hat{j}_{x}^{z}&=&\left(\frac{1}{2}\partial_{x}H_{0}\sigma_{z}+\frac{1}{2}\partial_{x}\gamma\right) \mathbf{I},
\end{eqnarray}
where $\partial_{i} \equiv \frac{\partial}{\partial k_{i}}$. 
Using these results, it can be calculated as 
\begin{eqnarray}
\langle+|\hbar\hat{v}_{y}|-\rangle&=&\partial_{y}\alpha\frac{-\alpha\gamma-i\beta\lambda}{\lambda\sqrt{\alpha^{2}+\beta^{2}}}+\partial_{y}\beta\frac{i\alpha\lambda-\beta\gamma}{\lambda\sqrt{\alpha^{2}+\beta^{2}}}\nonumber\\
&&+\partial_{y}\gamma\frac{\alpha^{2}+\beta^{2}}{\lambda\sqrt{\alpha^{2}+\beta^{2}}} \\
\langle+|\hat{j}_{x}^{z}|-\rangle&=&\frac{1}{2}\partial_{x}H_{0}\frac{\sqrt{\alpha^{2}+\beta^{2}}}{\sqrt{\alpha^{2}+\beta^{2}+\gamma^{2}}}.
\end{eqnarray}
Using these results, 
one can compute 
$ \sigma_{xy}^{\textrm{sH}}=\sum_{\mathbf{k}}\sigma_{xy}^{\textrm{sH}}(\mathbf{k})$. 
The momentum resolved $\sigma_{xy}^{\textrm{sH}}(\mathbf{k})$ within the effective model approach
can be written as
\begin{eqnarray}\label{Eq:k_sigma}
\sigma_{xy}^{\textrm{sH}}(\mathbf{k})&=&\frac{e^2}{\hbar}\frac{1}{V}\left(f_{+}-f_{-}\right)\nonumber \\
&&\times\partial_{x}H_{0}\frac{\alpha\partial_{y}\beta-\beta\partial_{y}\alpha}{4\lambda^{3}},
\end{eqnarray}
where $f_\pm\equiv f(\varepsilon_\pm-E_F)$ is the Fermi-Dirac distribution function for bands of $\varepsilon_\pm$.

By inspecting the effective equations in Eqs.~\ref{Eq:H_eff_M1},~\ref{Eq:H_eff_M2},~\ref{Eq:H_eff_S1},~\ref{Eq:H_eff_G}
and~\ref{Eq:H_eff_K}, all can be expressed as the generic spin Hamiltonian shown in Eq.~\ref{Eq:H_generic}.
So, it is straightforward to compute local $\sigma_{xy}^\textrm{sH}$ around each $k$-point using Eq.~\ref{Eq:k_sigma}.


\begin{thebibliography}{63}%
\makeatletter
\providecommand \@ifxundefined [1]{%
 \@ifx{#1\undefined}
}%
\providecommand \@ifnum [1]{%
 \ifnum #1\expandafter \@firstoftwo
 \else \expandafter \@secondoftwo
 \fi
}%
\providecommand \@ifx [1]{%
 \ifx #1\expandafter \@firstoftwo
 \else \expandafter \@secondoftwo
 \fi
}%
\providecommand \natexlab [1]{#1}%
\providecommand \enquote  [1]{``#1''}%
\providecommand \bibnamefont  [1]{#1}%
\providecommand \bibfnamefont [1]{#1}%
\providecommand \citenamefont [1]{#1}%
\providecommand \href@noop [0]{\@secondoftwo}%
\providecommand \href [0]{\begingroup \@sanitize@url \@href}%
\providecommand \@href[1]{\@@startlink{#1}\@@href}%
\providecommand \@@href[1]{\endgroup#1\@@endlink}%
\providecommand \@sanitize@url [0]{\catcode `\\12\catcode `\$12\catcode
  `\&12\catcode `\#12\catcode `\^12\catcode `\_12\catcode `\%12\relax}%
\providecommand \@@startlink[1]{}%
\providecommand \@@endlink[0]{}%
\providecommand \url  [0]{\begingroup\@sanitize@url \@url }%
\providecommand \@url [1]{\endgroup\@href {#1}{\urlprefix }}%
\providecommand \urlprefix  [0]{URL }%
\providecommand \Eprint [0]{\href }%
\providecommand \doibase [0]{http://dx.doi.org/}%
\providecommand \selectlanguage [0]{\@gobble}%
\providecommand \bibinfo  [0]{\@secondoftwo}%
\providecommand \bibfield  [0]{\@secondoftwo}%
\providecommand \translation [1]{[#1]}%
\providecommand \BibitemOpen [0]{}%
\providecommand \bibitemStop [0]{}%
\providecommand \bibitemNoStop [0]{.\EOS\space}%
\providecommand \EOS [0]{\spacefactor3000\relax}%
\providecommand \BibitemShut  [1]{\csname bibitem#1\endcsname}%
\let\auto@bib@innerbib\@empty
\bibitem [{\citenamefont {Novoselov}\ \emph {et~al.}(2005)\citenamefont
  {Novoselov}, \citenamefont {Jiang}, \citenamefont {Schedin}, \citenamefont
  {Booth}, \citenamefont {Khotkevich}, \citenamefont {Morozov},\ and\
  \citenamefont {Geim}}]{PNAS2005Novoselov}%
  \BibitemOpen
  \bibfield  {author} {\bibinfo {author} {\bibfnamefont {K.~S.}\ \bibnamefont
  {Novoselov}}, \bibinfo {author} {\bibfnamefont {D.}~\bibnamefont {Jiang}},
  \bibinfo {author} {\bibfnamefont {F.}~\bibnamefont {Schedin}}, \bibinfo
  {author} {\bibfnamefont {T.~J.}\ \bibnamefont {Booth}}, \bibinfo {author}
  {\bibfnamefont {V.~V.}\ \bibnamefont {Khotkevich}}, \bibinfo {author}
  {\bibfnamefont {S.~V.}\ \bibnamefont {Morozov}}, \ and\ \bibinfo {author}
  {\bibfnamefont {A.~K.}\ \bibnamefont {Geim}},\ }\bibfield  {title} {\enquote
  {\bibinfo {title} {Two-dimensional atomic crystals},}\ }\href@noop {}
  {\bibfield  {journal} {\bibinfo  {journal} {Proc. Natl. Acad. Sci. USA}\
  }\textbf {\bibinfo {volume} {102}},\ \bibinfo {pages} {10451--10453}
  (\bibinfo {year} {2005})}\BibitemShut {NoStop}%
\bibitem [{\citenamefont {Wang}\ \emph {et~al.}(2012)\citenamefont {Wang},
  \citenamefont {Kalantar-Zadeh}, \citenamefont {Kis}, \citenamefont
  {Coleman},\ and\ \citenamefont {Strano}}]{NatNano2012Wang}%
  \BibitemOpen
  \bibfield  {author} {\bibinfo {author} {\bibfnamefont {Q.~H.}\ \bibnamefont
  {Wang}}, \bibinfo {author} {\bibfnamefont {K.}~\bibnamefont
  {Kalantar-Zadeh}}, \bibinfo {author} {\bibfnamefont {A.}~\bibnamefont {Kis}},
  \bibinfo {author} {\bibfnamefont {J.~N.}\ \bibnamefont {Coleman}}, \ and\
  \bibinfo {author} {\bibfnamefont {M.~S.}\ \bibnamefont {Strano}},\ }\bibfield
   {title} {\enquote {\bibinfo {title} {Electronics and optoelectronics of
  two-dimensional transition metal dichalcogenides},}\ }\href@noop {}
  {\bibfield  {journal} {\bibinfo  {journal} {Nat. Nano.}\ }\textbf {\bibinfo
  {volume} {7}},\ \bibinfo {pages} {699--712} (\bibinfo {year}
  {2012})}\BibitemShut {NoStop}%
\bibitem [{\citenamefont {Geim}\ and\ \citenamefont
  {Grigorieva}(2013)}]{Nature2013Geim}%
  \BibitemOpen
  \bibfield  {author} {\bibinfo {author} {\bibfnamefont {A.~K.}\ \bibnamefont
  {Geim}}\ and\ \bibinfo {author} {\bibfnamefont {I.~V.}\ \bibnamefont
  {Grigorieva}},\ }\bibfield  {title} {\enquote {\bibinfo {title} {Van der
  $\text{Waals}$ heterostructures},}\ }\href@noop {} {\bibfield  {journal}
  {\bibinfo  {journal} {Nature}\ }\textbf {\bibinfo {volume} {499}},\ \bibinfo
  {pages} {419--425} (\bibinfo {year} {2013})}\BibitemShut {NoStop}%
\bibitem [{\citenamefont {Chhowalla}\ \emph {et~al.}(2013)\citenamefont
  {Chhowalla}, \citenamefont {Shin}, \citenamefont {Eda}, \citenamefont {Li},
  \citenamefont {Loh},\ and\ \citenamefont {Zhang}}]{NatChem2013Chhowalla}%
  \BibitemOpen
  \bibfield  {author} {\bibinfo {author} {\bibfnamefont {M.}~\bibnamefont
  {Chhowalla}}, \bibinfo {author} {\bibfnamefont {H.~S.}\ \bibnamefont {Shin}},
  \bibinfo {author} {\bibfnamefont {G.}~\bibnamefont {Eda}}, \bibinfo {author}
  {\bibfnamefont {L.-J.}\ \bibnamefont {Li}}, \bibinfo {author} {\bibfnamefont
  {K.~P.}\ \bibnamefont {Loh}}, \ and\ \bibinfo {author} {\bibfnamefont
  {H.}~\bibnamefont {Zhang}},\ }\bibfield  {title} {\enquote {\bibinfo {title}
  {The chemistry of two-dimensional layered transition metal dichalcogenide
  nanosheets},}\ }\href@noop {} {\bibfield  {journal} {\bibinfo  {journal}
  {Nat. Chem.}\ }\textbf {\bibinfo {volume} {5}},\ \bibinfo {pages} {263--275}
  (\bibinfo {year} {2013})}\BibitemShut {NoStop}%
\bibitem [{\citenamefont {Mak}\ \emph {et~al.}(2010)\citenamefont {Mak},
  \citenamefont {Lee}, \citenamefont {Hone}, \citenamefont {Shan},\ and\
  \citenamefont {Heinz}}]{PRL2010Mak}%
  \BibitemOpen
  \bibfield  {author} {\bibinfo {author} {\bibfnamefont {Kin~Fai}\ \bibnamefont
  {Mak}}, \bibinfo {author} {\bibfnamefont {Changgu}\ \bibnamefont {Lee}},
  \bibinfo {author} {\bibfnamefont {James}\ \bibnamefont {Hone}}, \bibinfo
  {author} {\bibfnamefont {Jie}\ \bibnamefont {Shan}}, \ and\ \bibinfo {author}
  {\bibfnamefont {Tony~F.}\ \bibnamefont {Heinz}},\ }\bibfield  {title}
  {\enquote {\bibinfo {title} {Atomically thin $\text{MoS}_2$: A new direct-gap
  semiconductor},}\ }\href@noop {} {\bibfield  {journal} {\bibinfo  {journal}
  {Phys. Rev. Lett.}\ }\textbf {\bibinfo {volume} {105}},\ \bibinfo {pages}
  {136805} (\bibinfo {year} {2010})}\BibitemShut {NoStop}%
\bibitem [{\citenamefont {Zhang}\ \emph {et~al.}(2014)\citenamefont {Zhang},
  \citenamefont {Chang}, \citenamefont {Zhou}, \citenamefont {Cui},
  \citenamefont {Yan}, \citenamefont {Liu}, \citenamefont {Schmitt},
  \citenamefont {Lee}, \citenamefont {Moore}, \citenamefont {Chen},
  \citenamefont {Lin}, \citenamefont {Jeng}, \citenamefont {Mo}, \citenamefont
  {Hussain}, \citenamefont {Bansil},\ and\ \citenamefont
  {Shen}}]{NatNano2014Zhang}%
  \BibitemOpen
  \bibfield  {author} {\bibinfo {author} {\bibfnamefont {Yi}~\bibnamefont
  {Zhang}}, \bibinfo {author} {\bibfnamefont {Tay-Rong}\ \bibnamefont {Chang}},
  \bibinfo {author} {\bibfnamefont {Bo}~\bibnamefont {Zhou}}, \bibinfo {author}
  {\bibfnamefont {Yong-Tao}\ \bibnamefont {Cui}}, \bibinfo {author}
  {\bibfnamefont {Hao}\ \bibnamefont {Yan}}, \bibinfo {author} {\bibfnamefont
  {Zhongkai}\ \bibnamefont {Liu}}, \bibinfo {author} {\bibfnamefont {Felix}\
  \bibnamefont {Schmitt}}, \bibinfo {author} {\bibfnamefont {James}\
  \bibnamefont {Lee}}, \bibinfo {author} {\bibfnamefont {Rob}\ \bibnamefont
  {Moore}}, \bibinfo {author} {\bibfnamefont {Yulin}\ \bibnamefont {Chen}},
  \bibinfo {author} {\bibfnamefont {Hsin}\ \bibnamefont {Lin}}, \bibinfo
  {author} {\bibfnamefont {Horng-Tay}\ \bibnamefont {Jeng}}, \bibinfo {author}
  {\bibfnamefont {Sung-Kwan}\ \bibnamefont {Mo}}, \bibinfo {author}
  {\bibfnamefont {Zahid}\ \bibnamefont {Hussain}}, \bibinfo {author}
  {\bibfnamefont {Arun}\ \bibnamefont {Bansil}}, \ and\ \bibinfo {author}
  {\bibfnamefont {Zhi-Xun}\ \bibnamefont {Shen}},\ }\bibfield  {title}
  {\enquote {\bibinfo {title} {Direct observation of the transition from
  indirect to direct bandgap in atomically thin epitaxial $\text{MoSe}_2$},}\
  }\href@noop {} {\bibfield  {journal} {\bibinfo  {journal} {Nat.
  Nanotechnol.}\ }\textbf {\bibinfo {volume} {9}},\ \bibinfo {pages} {111--115}
  (\bibinfo {year} {2014})}\BibitemShut {NoStop}%
\bibitem [{\citenamefont {Cheiwchanchamnangij}\ and\ \citenamefont
  {Lambrecht}(2012)}]{PRB2012Cheiwchanchamnagij}%
  \BibitemOpen
  \bibfield  {author} {\bibinfo {author} {\bibfnamefont {T.}~\bibnamefont
  {Cheiwchanchamnangij}}\ and\ \bibinfo {author} {\bibfnamefont {W.~R.~L.}\
  \bibnamefont {Lambrecht}},\ }\bibfield  {title} {\enquote {\bibinfo {title}
  {Quasiparticle band structure calculation of monolayer, bilayer, and bulk
  $\text{MoS}_{2}$},}\ }\href@noop {} {\bibfield  {journal} {\bibinfo
  {journal} {Phys. Rev. B}\ }\textbf {\bibinfo {volume} {85}},\ \bibinfo
  {pages} {205302} (\bibinfo {year} {2012})}\BibitemShut {NoStop}%
\bibitem [{\citenamefont {Ramasubramaniam}(2012)}]{PRB2012Ashwin}%
  \BibitemOpen
  \bibfield  {author} {\bibinfo {author} {\bibfnamefont {A.}~\bibnamefont
  {Ramasubramaniam}},\ }\bibfield  {title} {\enquote {\bibinfo {title} {Large
  excitonic effects in monolayers of molybdenum and tungsten
  dichalcogenides},}\ }\href@noop {} {\bibfield  {journal} {\bibinfo  {journal}
  {Phys. Rev. B}\ }\textbf {\bibinfo {volume} {86}},\ \bibinfo {pages} {115409}
  (\bibinfo {year} {2012})}\BibitemShut {NoStop}%
\bibitem [{\citenamefont {Komsa}\ and\ \citenamefont
  {Krasheninnikov}(2012)}]{PRB2012Komsa}%
  \BibitemOpen
  \bibfield  {author} {\bibinfo {author} {\bibfnamefont {H.-P.}\ \bibnamefont
  {Komsa}}\ and\ \bibinfo {author} {\bibfnamefont {A.~V.}\ \bibnamefont
  {Krasheninnikov}},\ }\bibfield  {title} {\enquote {\bibinfo {title} {Effects
  of confinement and environment on the electronic structure and exciton
  binding energy of $\text{MoS}_{2}$ from first principles},}\ }\href@noop {}
  {\bibfield  {journal} {\bibinfo  {journal} {Phys. Rev. B}\ }\textbf {\bibinfo
  {volume} {86}},\ \bibinfo {pages} {241201} (\bibinfo {year}
  {2012})}\BibitemShut {NoStop}%
\bibitem [{\citenamefont {Qiu}\ \emph {et~al.}(2013)\citenamefont {Qiu},
  \citenamefont {da~Jornada},\ and\ \citenamefont {Louie}}]{PRL2013Qiu}%
  \BibitemOpen
  \bibfield  {author} {\bibinfo {author} {\bibfnamefont {D.~Y.}\ \bibnamefont
  {Qiu}}, \bibinfo {author} {\bibfnamefont {F.~H.}\ \bibnamefont {da~Jornada}},
  \ and\ \bibinfo {author} {\bibfnamefont {S.~G.}\ \bibnamefont {Louie}},\
  }\bibfield  {title} {\enquote {\bibinfo {title} {Optical spectrum of
  $\text{MoS}_{2}$: Many-body effects and diversity of exciton states},}\
  }\href@noop {} {\bibfield  {journal} {\bibinfo  {journal} {Phys. Rev. Lett.}\
  }\textbf {\bibinfo {volume} {111}},\ \bibinfo {pages} {216805} (\bibinfo
  {year} {2013})}\BibitemShut {NoStop}%
\bibitem [{\citenamefont {Ugeda}\ \emph {et~al.}(2014)\citenamefont {Ugeda},
  \citenamefont {Bradley}, \citenamefont {Shi}, \citenamefont {da~Jornada},
  \citenamefont {Zhang}, \citenamefont {Qiu}, \citenamefont {Ruan},
  \citenamefont {Mo}, \citenamefont {Hussain}, \citenamefont {Shen},
  \citenamefont {Wang}, \citenamefont {Louie},\ and\ \citenamefont
  {Crommie}}]{NatMater2014Ugeda}%
  \BibitemOpen
  \bibfield  {author} {\bibinfo {author} {\bibfnamefont {M.~M.}\ \bibnamefont
  {Ugeda}}, \bibinfo {author} {\bibfnamefont {A.~J.}\ \bibnamefont {Bradley}},
  \bibinfo {author} {\bibfnamefont {S.-F.}\ \bibnamefont {Shi}}, \bibinfo
  {author} {\bibfnamefont {F.~H.}\ \bibnamefont {da~Jornada}}, \bibinfo
  {author} {\bibfnamefont {Y.}~\bibnamefont {Zhang}}, \bibinfo {author}
  {\bibfnamefont {D.~Y.}\ \bibnamefont {Qiu}}, \bibinfo {author} {\bibfnamefont
  {W.}~\bibnamefont {Ruan}}, \bibinfo {author} {\bibfnamefont {S.-K.}\
  \bibnamefont {Mo}}, \bibinfo {author} {\bibfnamefont {Z.}~\bibnamefont
  {Hussain}}, \bibinfo {author} {\bibfnamefont {Z.-X.}\ \bibnamefont {Shen}},
  \bibinfo {author} {\bibfnamefont {F.}~\bibnamefont {Wang}}, \bibinfo {author}
  {\bibfnamefont {S.~G.}\ \bibnamefont {Louie}}, \ and\ \bibinfo {author}
  {\bibfnamefont {M.~F.}\ \bibnamefont {Crommie}},\ }\bibfield  {title}
  {\enquote {\bibinfo {title} {Giant bandgap renormalization and excitonic
  effects in a monolayer transition metal dichalcogenide semiconductor},}\
  }\href@noop {} {\bibfield  {journal} {\bibinfo  {journal} {Nat. Mater.}\
  }\textbf {\bibinfo {volume} {13}},\ \bibinfo {pages} {1091--1095} (\bibinfo
  {year} {2014})}\BibitemShut {NoStop}%
\bibitem [{\citenamefont {Kim}\ and\ \citenamefont
  {Son}(2017)}]{PhysRevB_96_155439_Kim}%
  \BibitemOpen
  \bibfield  {author} {\bibinfo {author} {\bibfnamefont {Sejoong}\ \bibnamefont
  {Kim}}\ and\ \bibinfo {author} {\bibfnamefont {Young-Woo}\ \bibnamefont
  {Son}},\ }\bibfield  {title} {\enquote {\bibinfo {title} {Quasiparticle
  energy bands and fermi surfaces of monolayer $\textrm{NbSe}_2$},}\
  }\href@noop {} {\bibfield  {journal} {\bibinfo  {journal} {Phys. Rev. B}\
  }\textbf {\bibinfo {volume} {96}},\ \bibinfo {pages} {155439} (\bibinfo
  {year} {2017})}\BibitemShut {NoStop}%
\bibitem [{\citenamefont {Yu}\ \emph {et~al.}(2015)\citenamefont {Yu},
  \citenamefont {Yang}, \citenamefont {Lu}, \citenamefont {Yan}, \citenamefont
  {Cho}, \citenamefont {Ma}, \citenamefont {Niu}, \citenamefont {Kim},
  \citenamefont {Son}, \citenamefont {Feng}, \citenamefont {Li}, \citenamefont
  {Cheong}, \citenamefont {Chen},\ and\ \citenamefont {Zhang}}]{NatNano2015Yu}%
  \BibitemOpen
  \bibfield  {author} {\bibinfo {author} {\bibfnamefont {Y.}~\bibnamefont
  {Yu}}, \bibinfo {author} {\bibfnamefont {F.}~\bibnamefont {Yang}}, \bibinfo
  {author} {\bibfnamefont {X.~F.}\ \bibnamefont {Lu}}, \bibinfo {author}
  {\bibfnamefont {Y.~J.}\ \bibnamefont {Yan}}, \bibinfo {author} {\bibfnamefont
  {Y.-H.}\ \bibnamefont {Cho}}, \bibinfo {author} {\bibfnamefont
  {L.}~\bibnamefont {Ma}}, \bibinfo {author} {\bibfnamefont {X.}~\bibnamefont
  {Niu}}, \bibinfo {author} {\bibfnamefont {S.}~\bibnamefont {Kim}}, \bibinfo
  {author} {\bibfnamefont {Y.-W.}\ \bibnamefont {Son}}, \bibinfo {author}
  {\bibfnamefont {D.}~\bibnamefont {Feng}}, \bibinfo {author} {\bibfnamefont
  {S.}~\bibnamefont {Li}}, \bibinfo {author} {\bibfnamefont {S.-W.}\
  \bibnamefont {Cheong}}, \bibinfo {author} {\bibfnamefont {X.~H.}\
  \bibnamefont {Chen}}, \ and\ \bibinfo {author} {\bibfnamefont
  {Y.}~\bibnamefont {Zhang}},\ }\bibfield  {title} {\enquote {\bibinfo {title}
  {Gate-tunable phase transitions in thin flakes of $\text{1T-TaS}_2$},}\
  }\href@noop {} {\bibfield  {journal} {\bibinfo  {journal} {Nat. Nano.}\
  }\textbf {\bibinfo {volume} {10}},\ \bibinfo {pages} {270--276} (\bibinfo
  {year} {2015})}\BibitemShut {NoStop}%
\bibitem [{\citenamefont {Saito}\ \emph {et~al.}(2015)\citenamefont {Saito},
  \citenamefont {Kasahara}, \citenamefont {Ye}, \citenamefont {Iwasa},\ and\
  \citenamefont {Nojima}}]{Science2015Saito}%
  \BibitemOpen
  \bibfield  {author} {\bibinfo {author} {\bibfnamefont {Y.}~\bibnamefont
  {Saito}}, \bibinfo {author} {\bibfnamefont {Y.}~\bibnamefont {Kasahara}},
  \bibinfo {author} {\bibfnamefont {J.~T.}\ \bibnamefont {Ye}}, \bibinfo
  {author} {\bibfnamefont {Y.}~\bibnamefont {Iwasa}}, \ and\ \bibinfo {author}
  {\bibfnamefont {T.}~\bibnamefont {Nojima}},\ }\bibfield  {title} {\enquote
  {\bibinfo {title} {Metallic ground state in an ion-gated two-dimensional
  superconductor},}\ }\href@noop {} {\bibfield  {journal} {\bibinfo  {journal}
  {Science}\ }\textbf {\bibinfo {volume} {350}},\ \bibinfo {pages} {409--413}
  (\bibinfo {year} {2015})}\BibitemShut {NoStop}%
\bibitem [{\citenamefont {Li}\ \emph {et~al.}(2016)\citenamefont {Li},
  \citenamefont {O'Farrell}, \citenamefont {Loh}, \citenamefont {Eda},
  \citenamefont {\"Ozyilmaz},\ and\ \citenamefont
  {Castro-Neto}}]{Nature2016Li}%
  \BibitemOpen
  \bibfield  {author} {\bibinfo {author} {\bibfnamefont {L.~J.}\ \bibnamefont
  {Li}}, \bibinfo {author} {\bibfnamefont {E.~C.~T.}\ \bibnamefont
  {O'Farrell}}, \bibinfo {author} {\bibfnamefont {K.~P.}\ \bibnamefont {Loh}},
  \bibinfo {author} {\bibfnamefont {G.}~\bibnamefont {Eda}}, \bibinfo {author}
  {\bibfnamefont {B.}~\bibnamefont {\"Ozyilmaz}}, \ and\ \bibinfo {author}
  {\bibfnamefont {A.~H.}\ \bibnamefont {Castro-Neto}},\ }\bibfield  {title}
  {\enquote {\bibinfo {title} {Controlling many-body states by the
  electric-field effect in a two-dimensional material},}\ }\href@noop {}
  {\bibfield  {journal} {\bibinfo  {journal} {Nature}\ }\textbf {\bibinfo
  {volume} {529}},\ \bibinfo {pages} {185--189} (\bibinfo {year}
  {2016})}\BibitemShut {NoStop}%
\bibitem [{\citenamefont {Tsen}\ \emph {et~al.}(2016)\citenamefont {Tsen},
  \citenamefont {Hunt}, \citenamefont {Kim}, \citenamefont {Yuan},
  \citenamefont {Jia}, \citenamefont {Cava}, \citenamefont {Hone},
  \citenamefont {Kim}, \citenamefont {Dean},\ and\ \citenamefont
  {Pasupathy}}]{NatPhys2016Tsen}%
  \BibitemOpen
  \bibfield  {author} {\bibinfo {author} {\bibfnamefont {A.~W.}\ \bibnamefont
  {Tsen}}, \bibinfo {author} {\bibfnamefont {B.}~\bibnamefont {Hunt}}, \bibinfo
  {author} {\bibfnamefont {Y.~D.}\ \bibnamefont {Kim}}, \bibinfo {author}
  {\bibfnamefont {Z.~J.}\ \bibnamefont {Yuan}}, \bibinfo {author}
  {\bibfnamefont {S.}~\bibnamefont {Jia}}, \bibinfo {author} {\bibfnamefont
  {R.~J.}\ \bibnamefont {Cava}}, \bibinfo {author} {\bibfnamefont
  {J.}~\bibnamefont {Hone}}, \bibinfo {author} {\bibfnamefont {P.}~\bibnamefont
  {Kim}}, \bibinfo {author} {\bibfnamefont {C.~R.}\ \bibnamefont {Dean}}, \
  and\ \bibinfo {author} {\bibfnamefont {A.~N.}\ \bibnamefont {Pasupathy}},\
  }\bibfield  {title} {\enquote {\bibinfo {title} {Nature of the quantum metal
  in a two-dimensional crystalline superconductor},}\ }\href@noop {} {\bibfield
   {journal} {\bibinfo  {journal} {Nat. Phys.}\ }\textbf {\bibinfo {volume}
  {12}},\ \bibinfo {pages} {208--212} (\bibinfo {year} {2016})}\BibitemShut
  {NoStop}%
\bibitem [{\citenamefont {Wilson}\ \emph {et~al.}(2001)\citenamefont {Wilson},
  \citenamefont {Di~Salvo},\ and\ \citenamefont {Mahajan}}]{AdvPhys2001Wilson}%
  \BibitemOpen
  \bibfield  {author} {\bibinfo {author} {\bibfnamefont {J.~A.}\ \bibnamefont
  {Wilson}}, \bibinfo {author} {\bibfnamefont {F.~J.}\ \bibnamefont
  {Di~Salvo}}, \ and\ \bibinfo {author} {\bibfnamefont {S.}~\bibnamefont
  {Mahajan}},\ }\bibfield  {title} {\enquote {\bibinfo {title} {Charge-density
  waves and superlattices in the metallic layered transition metal
  dichalcogenides},}\ }\href@noop {} {\bibfield  {journal} {\bibinfo  {journal}
  {Adv. Phys.}\ }\textbf {\bibinfo {volume} {50}},\ \bibinfo {pages}
  {1171--1248} (\bibinfo {year} {2001})}\BibitemShut {NoStop}%
\bibitem [{\citenamefont {Xi}\ \emph {et~al.}(2015)\citenamefont {Xi},
  \citenamefont {Zhao}, \citenamefont {Wang}, \citenamefont {Berger},
  \citenamefont {Forr{\'o}}, \citenamefont {Shan},\ and\ \citenamefont
  {Mak}}]{NatNano2015Xi}%
  \BibitemOpen
  \bibfield  {author} {\bibinfo {author} {\bibfnamefont {X.}~\bibnamefont
  {Xi}}, \bibinfo {author} {\bibfnamefont {L.}~\bibnamefont {Zhao}}, \bibinfo
  {author} {\bibfnamefont {Z.}~\bibnamefont {Wang}}, \bibinfo {author}
  {\bibfnamefont {H.}~\bibnamefont {Berger}}, \bibinfo {author} {\bibfnamefont
  {L.}~\bibnamefont {Forr{\'o}}}, \bibinfo {author} {\bibfnamefont
  {J.}~\bibnamefont {Shan}}, \ and\ \bibinfo {author} {\bibfnamefont {K.~F.}\
  \bibnamefont {Mak}},\ }\bibfield  {title} {\enquote {\bibinfo {title}
  {Strongly enhanced charge-density-wave order in monolayer $\text{NbSe}_2$},}\
  }\href@noop {} {\bibfield  {journal} {\bibinfo  {journal} {Nat.
  Nanotechnol.}\ }\textbf {\bibinfo {volume} {10}},\ \bibinfo {pages}
  {765--769} (\bibinfo {year} {2015})}\BibitemShut {NoStop}%
\bibitem [{\citenamefont {Ugeda}\ \emph {et~al.}(2016)\citenamefont {Ugeda},
  \citenamefont {Bradley}, \citenamefont {Zhang}, \citenamefont {Onishi},
  \citenamefont {Chen}, \citenamefont {Ruan}, \citenamefont
  {Ojeda-Aristizabal}, \citenamefont {Ryu}, \citenamefont {Edmonds},
  \citenamefont {Tsai}, \citenamefont {Riss}, \citenamefont {Mo}, \citenamefont
  {Lee}, \citenamefont {Zettl}, \citenamefont {Hussain}, \citenamefont {Shen},\
  and\ \citenamefont {Crommie}}]{NatPhys2016Ugeda}%
  \BibitemOpen
  \bibfield  {author} {\bibinfo {author} {\bibfnamefont {M.~M.}\ \bibnamefont
  {Ugeda}}, \bibinfo {author} {\bibfnamefont {A.~J.}\ \bibnamefont {Bradley}},
  \bibinfo {author} {\bibfnamefont {Y.}~\bibnamefont {Zhang}}, \bibinfo
  {author} {\bibfnamefont {S.}~\bibnamefont {Onishi}}, \bibinfo {author}
  {\bibfnamefont {Y.}~\bibnamefont {Chen}}, \bibinfo {author} {\bibfnamefont
  {W.}~\bibnamefont {Ruan}}, \bibinfo {author} {\bibfnamefont {C.}~\bibnamefont
  {Ojeda-Aristizabal}}, \bibinfo {author} {\bibfnamefont {H.}~\bibnamefont
  {Ryu}}, \bibinfo {author} {\bibfnamefont {M.~T.}\ \bibnamefont {Edmonds}},
  \bibinfo {author} {\bibfnamefont {H.-Z.}\ \bibnamefont {Tsai}}, \bibinfo
  {author} {\bibfnamefont {A.}~\bibnamefont {Riss}}, \bibinfo {author}
  {\bibfnamefont {S.-K.}\ \bibnamefont {Mo}}, \bibinfo {author} {\bibfnamefont
  {D.}~\bibnamefont {Lee}}, \bibinfo {author} {\bibfnamefont {A.}~\bibnamefont
  {Zettl}}, \bibinfo {author} {\bibfnamefont {Z.}~\bibnamefont {Hussain}},
  \bibinfo {author} {\bibfnamefont {Z.-X.}\ \bibnamefont {Shen}}, \ and\
  \bibinfo {author} {\bibfnamefont {M.~F.}\ \bibnamefont {Crommie}},\
  }\bibfield  {title} {\enquote {\bibinfo {title} {Characterization of
  collective ground states in single-layer $\text{NbSe}_2$},}\ }\href@noop {}
  {\bibfield  {journal} {\bibinfo  {journal} {Nat. Phys.}\ }\textbf {\bibinfo
  {volume} {12}},\ \bibinfo {pages} {92--97} (\bibinfo {year}
  {2016})}\BibitemShut {NoStop}%
\bibitem [{\citenamefont {Johannes}\ \emph {et~al.}(2006)\citenamefont
  {Johannes}, \citenamefont {Mazin},\ and\ \citenamefont
  {Howells}}]{PRB2006Johannes}%
  \BibitemOpen
  \bibfield  {author} {\bibinfo {author} {\bibfnamefont {M.~D.}\ \bibnamefont
  {Johannes}}, \bibinfo {author} {\bibfnamefont {I.~I.}\ \bibnamefont {Mazin}},
  \ and\ \bibinfo {author} {\bibfnamefont {C.~A.}\ \bibnamefont {Howells}},\
  }\bibfield  {title} {\enquote {\bibinfo {title} {Fermi-surface nesting and
  the origin of the charge-density wave in $\text{NbSe}_{2}$},}\ }\href@noop {}
  {\bibfield  {journal} {\bibinfo  {journal} {Phys. Rev. B}\ }\textbf {\bibinfo
  {volume} {73}},\ \bibinfo {pages} {205102} (\bibinfo {year}
  {2006})}\BibitemShut {NoStop}%
\bibitem [{\citenamefont {Johannes}\ and\ \citenamefont
  {Mazin}(2008)}]{PRB2008Johaness}%
  \BibitemOpen
  \bibfield  {author} {\bibinfo {author} {\bibfnamefont {M.~D.}\ \bibnamefont
  {Johannes}}\ and\ \bibinfo {author} {\bibfnamefont {I.~I.}\ \bibnamefont
  {Mazin}},\ }\bibfield  {title} {\enquote {\bibinfo {title} {Fermi surface
  nesting and the origin of charge density waves in metals},}\ }\href@noop {}
  {\bibfield  {journal} {\bibinfo  {journal} {Phys. Rev. B}\ }\textbf {\bibinfo
  {volume} {77}},\ \bibinfo {pages} {165135} (\bibinfo {year}
  {2008})}\BibitemShut {NoStop}%
\bibitem [{\citenamefont {Calandra}\ \emph {et~al.}(2009)\citenamefont
  {Calandra}, \citenamefont {Mazin},\ and\ \citenamefont
  {Mauri}}]{PRB2009Calandra}%
  \BibitemOpen
  \bibfield  {author} {\bibinfo {author} {\bibfnamefont {M.}~\bibnamefont
  {Calandra}}, \bibinfo {author} {\bibfnamefont {I.~I.}\ \bibnamefont {Mazin}},
  \ and\ \bibinfo {author} {\bibfnamefont {F.}~\bibnamefont {Mauri}},\
  }\bibfield  {title} {\enquote {\bibinfo {title} {Effect of dimensionality on
  the charge-density wave in few-layer $\text{2H-NbSe}_{2}$},}\ }\href@noop {}
  {\bibfield  {journal} {\bibinfo  {journal} {Phys. Rev. B}\ }\textbf {\bibinfo
  {volume} {80}},\ \bibinfo {pages} {241108} (\bibinfo {year}
  {2009})}\BibitemShut {NoStop}%
\bibitem [{\citenamefont {Ge}\ and\ \citenamefont {Liu}(2012)}]{PRB2012Ge}%
  \BibitemOpen
  \bibfield  {author} {\bibinfo {author} {\bibfnamefont {Y.}~\bibnamefont
  {Ge}}\ and\ \bibinfo {author} {\bibfnamefont {A.~Y.}\ \bibnamefont {Liu}},\
  }\bibfield  {title} {\enquote {\bibinfo {title} {Effect of dimensionality and
  spin-orbit coupling on charge-density-wave transition in
  2$\text{H-TaSe}_{2}$},}\ }\href@noop {} {\bibfield  {journal} {\bibinfo
  {journal} {Phys. Rev. B}\ }\textbf {\bibinfo {volume} {86}},\ \bibinfo
  {pages} {104101} (\bibinfo {year} {2012})}\BibitemShut {NoStop}%
\bibitem [{\citenamefont {Rossnagel}(2011)}]{JPCM2011Rossnagel}%
  \BibitemOpen
  \bibfield  {author} {\bibinfo {author} {\bibfnamefont {K.}~\bibnamefont
  {Rossnagel}},\ }\bibfield  {title} {\enquote {\bibinfo {title} {On the origin
  of charge-density waves in select layered transition-metal
  dichalcogenides},}\ }\href@noop {} {\bibfield  {journal} {\bibinfo  {journal}
  {J. Phys.: Cond. Matter}\ }\textbf {\bibinfo {volume} {23}},\ \bibinfo
  {pages} {213001} (\bibinfo {year} {2011})}\BibitemShut {NoStop}%
\bibitem [{\citenamefont {Shen}\ \emph {et~al.}(2008)\citenamefont {Shen},
  \citenamefont {Zhang}, \citenamefont {Yang}, \citenamefont {Wei},
  \citenamefont {Ou}, \citenamefont {Dong}, \citenamefont {Xie}, \citenamefont
  {He}, \citenamefont {Zhao}, \citenamefont {Zhou}, \citenamefont {Arita},
  \citenamefont {Shimada}, \citenamefont {Namatame}, \citenamefont {Taniguchi},
  \citenamefont {Shi},\ and\ \citenamefont {Feng}}]{PRL2008Shen}%
  \BibitemOpen
  \bibfield  {author} {\bibinfo {author} {\bibfnamefont {D.~W.}\ \bibnamefont
  {Shen}}, \bibinfo {author} {\bibfnamefont {Y.}~\bibnamefont {Zhang}},
  \bibinfo {author} {\bibfnamefont {L.~X.}\ \bibnamefont {Yang}}, \bibinfo
  {author} {\bibfnamefont {J.}~\bibnamefont {Wei}}, \bibinfo {author}
  {\bibfnamefont {H.~W.}\ \bibnamefont {Ou}}, \bibinfo {author} {\bibfnamefont
  {J.~K.}\ \bibnamefont {Dong}}, \bibinfo {author} {\bibfnamefont {B.~P.}\
  \bibnamefont {Xie}}, \bibinfo {author} {\bibfnamefont {C.}~\bibnamefont
  {He}}, \bibinfo {author} {\bibfnamefont {J.~F.}\ \bibnamefont {Zhao}},
  \bibinfo {author} {\bibfnamefont {B.}~\bibnamefont {Zhou}}, \bibinfo {author}
  {\bibfnamefont {M.}~\bibnamefont {Arita}}, \bibinfo {author} {\bibfnamefont
  {K.}~\bibnamefont {Shimada}}, \bibinfo {author} {\bibfnamefont
  {H.}~\bibnamefont {Namatame}}, \bibinfo {author} {\bibfnamefont
  {M.}~\bibnamefont {Taniguchi}}, \bibinfo {author} {\bibfnamefont
  {J.}~\bibnamefont {Shi}}, \ and\ \bibinfo {author} {\bibfnamefont {D.~L.}\
  \bibnamefont {Feng}},\ }\bibfield  {title} {\enquote {\bibinfo {title}
  {Primary role of the barely occupied states in the charge density wave
  formation of $\text{NbSe}_{2}$},}\ }\href@noop {} {\bibfield  {journal}
  {\bibinfo  {journal} {Phys. Rev. Lett.}\ }\textbf {\bibinfo {volume} {101}},\
  \bibinfo {pages} {226406} (\bibinfo {year} {2008})}\BibitemShut {NoStop}%
\bibitem [{\citenamefont {Borisenko}\ \emph {et~al.}(2009)\citenamefont
  {Borisenko}, \citenamefont {Kordyuk}, \citenamefont {Zabolotnyy},
  \citenamefont {Inosov}, \citenamefont {Evtushinsky}, \citenamefont
  {B\"uchner}, \citenamefont {Yaresko}, \citenamefont {Varykhalov},
  \citenamefont {Follath}, \citenamefont {Eberhardt}, \citenamefont {Patthey},\
  and\ \citenamefont {Berger}}]{PRL2009Borisenko}%
  \BibitemOpen
  \bibfield  {author} {\bibinfo {author} {\bibfnamefont {S.~V.}\ \bibnamefont
  {Borisenko}}, \bibinfo {author} {\bibfnamefont {A.~A.}\ \bibnamefont
  {Kordyuk}}, \bibinfo {author} {\bibfnamefont {V.~B.}\ \bibnamefont
  {Zabolotnyy}}, \bibinfo {author} {\bibfnamefont {D.~S.}\ \bibnamefont
  {Inosov}}, \bibinfo {author} {\bibfnamefont {D.}~\bibnamefont {Evtushinsky}},
  \bibinfo {author} {\bibfnamefont {B.}~\bibnamefont {B\"uchner}}, \bibinfo
  {author} {\bibfnamefont {A.~N.}\ \bibnamefont {Yaresko}}, \bibinfo {author}
  {\bibfnamefont {A.}~\bibnamefont {Varykhalov}}, \bibinfo {author}
  {\bibfnamefont {R.}~\bibnamefont {Follath}}, \bibinfo {author} {\bibfnamefont
  {W.}~\bibnamefont {Eberhardt}}, \bibinfo {author} {\bibfnamefont
  {L.}~\bibnamefont {Patthey}}, \ and\ \bibinfo {author} {\bibfnamefont
  {H.}~\bibnamefont {Berger}},\ }\bibfield  {title} {\enquote {\bibinfo {title}
  {Two energy gaps and fermi-surface ``arcs'' in $\text{NbSe}_{2}$},}\
  }\href@noop {} {\bibfield  {journal} {\bibinfo  {journal} {Phys. Rev. Lett.}\
  }\textbf {\bibinfo {volume} {102}},\ \bibinfo {pages} {166402} (\bibinfo
  {year} {2009})}\BibitemShut {NoStop}%
\bibitem [{\citenamefont {Arguello}\ \emph {et~al.}(2015)\citenamefont
  {Arguello}, \citenamefont {Rosenthal}, \citenamefont {Andrade}, \citenamefont
  {Jin}, \citenamefont {Yeh}, \citenamefont {Zaki}, \citenamefont {Jia},
  \citenamefont {Cava}, \citenamefont {Fernandes}, \citenamefont {Millis},
  \citenamefont {Valla}, \citenamefont {Osgood},\ and\ \citenamefont
  {Pasupathy}}]{PRL2015Arguello}%
  \BibitemOpen
  \bibfield  {author} {\bibinfo {author} {\bibfnamefont {C.~J.}\ \bibnamefont
  {Arguello}}, \bibinfo {author} {\bibfnamefont {E.~P.}\ \bibnamefont
  {Rosenthal}}, \bibinfo {author} {\bibfnamefont {E.~F.}\ \bibnamefont
  {Andrade}}, \bibinfo {author} {\bibfnamefont {W.}~\bibnamefont {Jin}},
  \bibinfo {author} {\bibfnamefont {P.~C.}\ \bibnamefont {Yeh}}, \bibinfo
  {author} {\bibfnamefont {N.}~\bibnamefont {Zaki}}, \bibinfo {author}
  {\bibfnamefont {S.}~\bibnamefont {Jia}}, \bibinfo {author} {\bibfnamefont
  {R.~J.}\ \bibnamefont {Cava}}, \bibinfo {author} {\bibfnamefont {R.~M.}\
  \bibnamefont {Fernandes}}, \bibinfo {author} {\bibfnamefont {A.~J.}\
  \bibnamefont {Millis}}, \bibinfo {author} {\bibfnamefont {T.}~\bibnamefont
  {Valla}}, \bibinfo {author} {\bibfnamefont {R.~M.}\ \bibnamefont {Osgood}}, \
  and\ \bibinfo {author} {\bibfnamefont {A.~N.}\ \bibnamefont {Pasupathy}},\
  }\bibfield  {title} {\enquote {\bibinfo {title} {Quasiparticle interference,
  quasiparticle interactions, and the origin of the charge density wave in
  $\text{2H-NbSe}_{2}$},}\ }\href@noop {} {\bibfield  {journal} {\bibinfo
  {journal} {Phys. Rev. Lett.}\ }\textbf {\bibinfo {volume} {114}},\ \bibinfo
  {pages} {037001} (\bibinfo {year} {2015})}\BibitemShut {NoStop}%
\bibitem [{\citenamefont {Lin}\ \emph {et~al.}(2020)\citenamefont {Lin},
  \citenamefont {Li}, \citenamefont {Wen}, \citenamefont {Berger},
  \citenamefont {Forr{\'o}}, \citenamefont {Zhou}, \citenamefont {Jia},
  \citenamefont {Taniguchi}, \citenamefont {Watanabe}, \citenamefont {Xi},\
  and\ \citenamefont {Bahramy}}]{NatComm2020Lin}%
  \BibitemOpen
  \bibfield  {author} {\bibinfo {author} {\bibfnamefont {Dongjing}\
  \bibnamefont {Lin}}, \bibinfo {author} {\bibfnamefont {Shichao}\ \bibnamefont
  {Li}}, \bibinfo {author} {\bibfnamefont {Jinsheng}\ \bibnamefont {Wen}},
  \bibinfo {author} {\bibfnamefont {Helmuth}\ \bibnamefont {Berger}}, \bibinfo
  {author} {\bibfnamefont {L{\'a}szl{\'o}}\ \bibnamefont {Forr{\'o}}}, \bibinfo
  {author} {\bibfnamefont {Huibin}\ \bibnamefont {Zhou}}, \bibinfo {author}
  {\bibfnamefont {Shuang}\ \bibnamefont {Jia}}, \bibinfo {author}
  {\bibfnamefont {Takashi}\ \bibnamefont {Taniguchi}}, \bibinfo {author}
  {\bibfnamefont {Kenji}\ \bibnamefont {Watanabe}}, \bibinfo {author}
  {\bibfnamefont {Xiaoxiang}\ \bibnamefont {Xi}}, \ and\ \bibinfo {author}
  {\bibfnamefont {Mohammad~Saeed}\ \bibnamefont {Bahramy}},\ }\bibfield
  {title} {\enquote {\bibinfo {title} {Patterns and driving forces of
  dimensionality-dependent charge density waves in 2$\text{H}$-type transition
  metal dichalcogenides},}\ }\href@noop {} {\bibfield  {journal} {\bibinfo
  {journal} {Nature Comm.}\ }\textbf {\bibinfo {volume} {11}},\ \bibinfo
  {pages} {2406} (\bibinfo {year} {2020})}\BibitemShut {NoStop}%
\bibitem [{\citenamefont {Rice}\ and\ \citenamefont
  {Scott}(1975)}]{Rice_Scott_PRL_35_120_1975}%
  \BibitemOpen
  \bibfield  {author} {\bibinfo {author} {\bibfnamefont {T.~M.}\ \bibnamefont
  {Rice}}\ and\ \bibinfo {author} {\bibfnamefont {G.~K.}\ \bibnamefont
  {Scott}},\ }\bibfield  {title} {\enquote {\bibinfo {title} {New mechanism for
  a charge-density-wave instability},}\ }\href@noop {} {\bibfield  {journal}
  {\bibinfo  {journal} {Phys. Rev. Lett.}\ }\textbf {\bibinfo {volume} {35}},\
  \bibinfo {pages} {120} (\bibinfo {year} {1975})}\BibitemShut {NoStop}%
\bibitem [{\citenamefont {Honerkamp}(2008)}]{PRL2008Honerkamp}%
  \BibitemOpen
  \bibfield  {author} {\bibinfo {author} {\bibfnamefont {Carsten}\ \bibnamefont
  {Honerkamp}},\ }\bibfield  {title} {\enquote {\bibinfo {title} {Density waves
  and cooper pairing on the honeycomb lattice},}\ }\href@noop {} {\bibfield
  {journal} {\bibinfo  {journal} {Phys. Rev. Lett.}\ }\textbf {\bibinfo
  {volume} {100}},\ \bibinfo {pages} {146404} (\bibinfo {year}
  {2008})}\BibitemShut {NoStop}%
\bibitem [{\citenamefont {Makogon}\ \emph {et~al.}(2011)\citenamefont
  {Makogon}, \citenamefont {van Gelderen}, \citenamefont {Rold\'an},\ and\
  \citenamefont {Smith}}]{PRB2011Makogon}%
  \BibitemOpen
  \bibfield  {author} {\bibinfo {author} {\bibfnamefont {D.}~\bibnamefont
  {Makogon}}, \bibinfo {author} {\bibfnamefont {R.}~\bibnamefont {van
  Gelderen}}, \bibinfo {author} {\bibfnamefont {R.}~\bibnamefont {Rold\'an}}, \
  and\ \bibinfo {author} {\bibfnamefont {C.~Morais}\ \bibnamefont {Smith}},\
  }\bibfield  {title} {\enquote {\bibinfo {title} {Spin-density-wave
  instability in graphene doped near the van $\text{Hove}$ singularity},}\
  }\href@noop {} {\bibfield  {journal} {\bibinfo  {journal} {Phys. Rev. B}\
  }\textbf {\bibinfo {volume} {84}},\ \bibinfo {pages} {125404} (\bibinfo
  {year} {2011})}\BibitemShut {NoStop}%
\bibitem [{\citenamefont {Nandkishore}\ \emph {et~al.}(2012)\citenamefont
  {Nandkishore}, \citenamefont {Levitov},\ and\ \citenamefont
  {Chubukov}}]{NatPhys2012Rahul}%
  \BibitemOpen
  \bibfield  {author} {\bibinfo {author} {\bibfnamefont {Rahul}\ \bibnamefont
  {Nandkishore}}, \bibinfo {author} {\bibfnamefont {L.~S.}\ \bibnamefont
  {Levitov}}, \ and\ \bibinfo {author} {\bibfnamefont {A.~V.}\ \bibnamefont
  {Chubukov}},\ }\bibfield  {title} {\enquote {\bibinfo {title} {Chiral
  superconductivity from repulsive interactions in doped graphene},}\
  }\href@noop {} {\bibfield  {journal} {\bibinfo  {journal} {Nat. Phys.}\
  }\textbf {\bibinfo {volume} {8}},\ \bibinfo {pages} {158--163} (\bibinfo
  {year} {2012})}\BibitemShut {NoStop}%
\bibitem [{\citenamefont {Yudin}\ \emph {et~al.}(2014)\citenamefont {Yudin},
  \citenamefont {Hirschmeier}, \citenamefont {Hafermann}, \citenamefont
  {Eriksson}, \citenamefont {Lichtenstein},\ and\ \citenamefont
  {Katsnelson}}]{PRL2014Yudin}%
  \BibitemOpen
  \bibfield  {author} {\bibinfo {author} {\bibfnamefont {Dmitry}\ \bibnamefont
  {Yudin}}, \bibinfo {author} {\bibfnamefont {Daniel}\ \bibnamefont
  {Hirschmeier}}, \bibinfo {author} {\bibfnamefont {Hartmut}\ \bibnamefont
  {Hafermann}}, \bibinfo {author} {\bibfnamefont {Olle}\ \bibnamefont
  {Eriksson}}, \bibinfo {author} {\bibfnamefont {Alexander~I.}\ \bibnamefont
  {Lichtenstein}}, \ and\ \bibinfo {author} {\bibfnamefont {Mikhail~I.}\
  \bibnamefont {Katsnelson}},\ }\bibfield  {title} {\enquote {\bibinfo {title}
  {Fermi condensation near van $\text{Hove}$ singularities within the
  $\text{Hubbard}$ model on the triangular lattice},}\ }\href@noop {}
  {\bibfield  {journal} {\bibinfo  {journal} {Phys. Rev. Lett.}\ }\textbf
  {\bibinfo {volume} {112}},\ \bibinfo {pages} {070403} (\bibinfo {year}
  {2014})}\BibitemShut {NoStop}%
\bibitem [{\citenamefont {Kiesel}\ \emph {et~al.}(2012)\citenamefont {Kiesel},
  \citenamefont {Platt}, \citenamefont {Hanke}, \citenamefont {Abanin},\ and\
  \citenamefont {Thomale}}]{PRB2021Kiesel}%
  \BibitemOpen
  \bibfield  {author} {\bibinfo {author} {\bibfnamefont {Maximilian~L.}\
  \bibnamefont {Kiesel}}, \bibinfo {author} {\bibfnamefont {Christian}\
  \bibnamefont {Platt}}, \bibinfo {author} {\bibfnamefont {Werner}\
  \bibnamefont {Hanke}}, \bibinfo {author} {\bibfnamefont {Dmitry~A.}\
  \bibnamefont {Abanin}}, \ and\ \bibinfo {author} {\bibfnamefont {Ronny}\
  \bibnamefont {Thomale}},\ }\bibfield  {title} {\enquote {\bibinfo {title}
  {Competing many-body instabilities and unconventional superconductivity in
  graphene},}\ }\href@noop {} {\bibfield  {journal} {\bibinfo  {journal} {Phys.
  Rev. B}\ }\textbf {\bibinfo {volume} {86}},\ \bibinfo {pages} {020507}
  (\bibinfo {year} {2012})}\BibitemShut {NoStop}%
\bibitem [{\citenamefont {Yuan}\ \emph {et~al.}(2013)\citenamefont {Yuan},
  \citenamefont {Bahramy}, \citenamefont {Morimoto}, \citenamefont {Wu},
  \citenamefont {Nomura}, \citenamefont {Yang}, \citenamefont {Shimotani},
  \citenamefont {Suzuki}, \citenamefont {Toh}, \citenamefont {Kloc},
  \citenamefont {Xu}, \citenamefont {Arita}, \citenamefont {Nagaosa},\ and\
  \citenamefont {Iwasa}}]{Yuan_NatPhys_9_563_2013}%
  \BibitemOpen
  \bibfield  {author} {\bibinfo {author} {\bibfnamefont {H.}~\bibnamefont
  {Yuan}}, \bibinfo {author} {\bibfnamefont {M.~S.}\ \bibnamefont {Bahramy}},
  \bibinfo {author} {\bibfnamefont {K.}~\bibnamefont {Morimoto}}, \bibinfo
  {author} {\bibfnamefont {S.}~\bibnamefont {Wu}}, \bibinfo {author}
  {\bibfnamefont {K.}~\bibnamefont {Nomura}}, \bibinfo {author} {\bibfnamefont
  {B.-J.}\ \bibnamefont {Yang}}, \bibinfo {author} {\bibfnamefont
  {H.}~\bibnamefont {Shimotani}}, \bibinfo {author} {\bibfnamefont
  {R.}~\bibnamefont {Suzuki}}, \bibinfo {author} {\bibfnamefont
  {M.}~\bibnamefont {Toh}}, \bibinfo {author} {\bibfnamefont {C.}~\bibnamefont
  {Kloc}}, \bibinfo {author} {\bibfnamefont {X.}~\bibnamefont {Xu}}, \bibinfo
  {author} {\bibfnamefont {R.}~\bibnamefont {Arita}}, \bibinfo {author}
  {\bibfnamefont {N.}~\bibnamefont {Nagaosa}}, \ and\ \bibinfo {author}
  {\bibfnamefont {Y.}~\bibnamefont {Iwasa}},\ }\bibfield  {title} {\enquote
  {\bibinfo {title} {Zeeman-type spin splitting controlled by an electric
  field},}\ }\href@noop {} {\bibfield  {journal} {\bibinfo  {journal} {Nat.
  Phys.}\ }\textbf {\bibinfo {volume} {9}},\ \bibinfo {pages} {563} (\bibinfo
  {year} {2013})}\BibitemShut {NoStop}%
\bibitem [{\citenamefont {Cheng}\ \emph {et~al.}(2016)\citenamefont {Cheng},
  \citenamefont {Sun}, \citenamefont {Chen}, \citenamefont {Fu},\ and\
  \citenamefont {Meng}}]{Cheng_Nanoscale_8_17854_2016}%
  \BibitemOpen
  \bibfield  {author} {\bibinfo {author} {\bibfnamefont {C.}~\bibnamefont
  {Cheng}}, \bibinfo {author} {\bibfnamefont {J.-T.}\ \bibnamefont {Sun}},
  \bibinfo {author} {\bibfnamefont {X.-R.}\ \bibnamefont {Chen}}, \bibinfo
  {author} {\bibfnamefont {H.-X.}\ \bibnamefont {Fu}}, \ and\ \bibinfo {author}
  {\bibfnamefont {S.}~\bibnamefont {Meng}},\ }\bibfield  {title} {\enquote
  {\bibinfo {title} {Nonlinear $\text{Rashba}$ spin splitting in transition
  metal dichalcogenide monolayers},}\ }\href@noop {} {\bibfield  {journal}
  {\bibinfo  {journal} {Nanoscale}\ }\textbf {\bibinfo {volume} {8}},\ \bibinfo
  {pages} {17854} (\bibinfo {year} {2016})}\BibitemShut {NoStop}%
\bibitem [{\citenamefont {Shanavas}\ and\ \citenamefont
  {Satpathy}(2015)}]{PhysRevB_91_235145_2015_Shanavas}%
  \BibitemOpen
  \bibfield  {author} {\bibinfo {author} {\bibfnamefont {K.~V.}\ \bibnamefont
  {Shanavas}}\ and\ \bibinfo {author} {\bibfnamefont {S.}~\bibnamefont
  {Satpathy}},\ }\bibfield  {title} {\enquote {\bibinfo {title} {Effective
  tight-binding model for $\textrm{MX}_{2}$ under electric and magnetic
  fields},}\ }\href@noop {} {\bibfield  {journal} {\bibinfo  {journal} {Phys.
  Rev. B}\ }\textbf {\bibinfo {volume} {91}},\ \bibinfo {pages} {235145}
  (\bibinfo {year} {2015})}\BibitemShut {NoStop}%
\bibitem [{\citenamefont {Xi}\ \emph {et~al.}(2016)\citenamefont {Xi},
  \citenamefont {Wang}, \citenamefont {Zhao}, \citenamefont {Park},
  \citenamefont {Law}, \citenamefont {Berger}, \citenamefont {Forr{\'o}},
  \citenamefont {Shan},\ and\ \citenamefont {Mak}}]{NatPhys2016Xi}%
  \BibitemOpen
  \bibfield  {author} {\bibinfo {author} {\bibfnamefont {Xiaoxiang}\
  \bibnamefont {Xi}}, \bibinfo {author} {\bibfnamefont {Zefang}\ \bibnamefont
  {Wang}}, \bibinfo {author} {\bibfnamefont {Weiwei}\ \bibnamefont {Zhao}},
  \bibinfo {author} {\bibfnamefont {Ju-Hyun}\ \bibnamefont {Park}}, \bibinfo
  {author} {\bibfnamefont {Kam~Tuen}\ \bibnamefont {Law}}, \bibinfo {author}
  {\bibfnamefont {Helmuth}\ \bibnamefont {Berger}}, \bibinfo {author}
  {\bibfnamefont {L{\'a}szl{\'o}}\ \bibnamefont {Forr{\'o}}}, \bibinfo {author}
  {\bibfnamefont {Jie}\ \bibnamefont {Shan}}, \ and\ \bibinfo {author}
  {\bibfnamefont {Kin~Fai}\ \bibnamefont {Mak}},\ }\bibfield  {title} {\enquote
  {\bibinfo {title} {Ising pairing in superconducting $\text{NbSe}_2$
  atomic layers},}\ }\href@noop {} {\bibfield  {journal} {\bibinfo  {journal}
  {Nat. Phys.}\ }\textbf {\bibinfo {volume} {12}},\ \bibinfo {pages} {139--143}
  (\bibinfo {year} {2016})}\BibitemShut {NoStop}%
\bibitem [{\citenamefont {Saito}\ \emph {et~al.}(2016)\citenamefont {Saito},
  \citenamefont {Nakamura}, \citenamefont {Bahramy}, \citenamefont {Kohama},
  \citenamefont {Ye}, \citenamefont {Kasahara}, \citenamefont {Nakagawa},
  \citenamefont {Onga}, \citenamefont {Tokunaga}, \citenamefont {Nojima},
  \citenamefont {Yanase},\ and\ \citenamefont {Iwasa}}]{NatPhys2016Saito}%
  \BibitemOpen
  \bibfield  {author} {\bibinfo {author} {\bibfnamefont {Yu}~\bibnamefont
  {Saito}}, \bibinfo {author} {\bibfnamefont {Yasuharu}\ \bibnamefont
  {Nakamura}}, \bibinfo {author} {\bibfnamefont {Mohammad~Saeed}\ \bibnamefont
  {Bahramy}}, \bibinfo {author} {\bibfnamefont {Yoshimitsu}\ \bibnamefont
  {Kohama}}, \bibinfo {author} {\bibfnamefont {Jianting}\ \bibnamefont {Ye}},
  \bibinfo {author} {\bibfnamefont {Yuichi}\ \bibnamefont {Kasahara}}, \bibinfo
  {author} {\bibfnamefont {Yuji}\ \bibnamefont {Nakagawa}}, \bibinfo {author}
  {\bibfnamefont {Masaru}\ \bibnamefont {Onga}}, \bibinfo {author}
  {\bibfnamefont {Masashi}\ \bibnamefont {Tokunaga}}, \bibinfo {author}
  {\bibfnamefont {Tsutomu}\ \bibnamefont {Nojima}}, \bibinfo {author}
  {\bibfnamefont {Youichi}\ \bibnamefont {Yanase}}, \ and\ \bibinfo {author}
  {\bibfnamefont {Yoshihiro}\ \bibnamefont {Iwasa}},\ }\bibfield  {title}
  {\enquote {\bibinfo {title} {Superconductivity protected by spin--valley
  locking in ion-gated $\text{MoS}_2$},}\ }\href@noop {} {\bibfield  {journal}
  {\bibinfo  {journal} {Nat. Phys.}\ }\textbf {\bibinfo {volume} {12}},\
  \bibinfo {pages} {144--149} (\bibinfo {year} {2016})}\BibitemShut {NoStop}%
\bibitem [{\citenamefont {McChesney}\ \emph {et~al.}(2010)\citenamefont
  {McChesney}, \citenamefont {Bostwick}, \citenamefont {Ohta}, \citenamefont
  {Seyller}, \citenamefont {Horn}, \citenamefont {Gonz\'alez},\ and\
  \citenamefont {Rotenberg}}]{PhysRevLett.104.136803}%
  \BibitemOpen
  \bibfield  {author} {\bibinfo {author} {\bibfnamefont {J.~L.}\ \bibnamefont
  {McChesney}}, \bibinfo {author} {\bibfnamefont {Aaron}\ \bibnamefont
  {Bostwick}}, \bibinfo {author} {\bibfnamefont {Taisuke}\ \bibnamefont
  {Ohta}}, \bibinfo {author} {\bibfnamefont {Thomas}\ \bibnamefont {Seyller}},
  \bibinfo {author} {\bibfnamefont {Karsten}\ \bibnamefont {Horn}}, \bibinfo
  {author} {\bibfnamefont {J.}~\bibnamefont {Gonz\'alez}}, \ and\ \bibinfo
  {author} {\bibfnamefont {Eli}\ \bibnamefont {Rotenberg}},\ }\bibfield
  {title} {\enquote {\bibinfo {title} {Extended van $\text{Hove}$ singularity
  and superconducting instability in doped graphene},}\ }\href@noop {}
  {\bibfield  {journal} {\bibinfo  {journal} {Phys. Rev. Lett.}\ }\textbf
  {\bibinfo {volume} {104}},\ \bibinfo {pages} {136803} (\bibinfo {year}
  {2010})}\BibitemShut {NoStop}%
\bibitem [{\citenamefont {Rosenzweig}\ \emph {et~al.}(2020)\citenamefont
  {Rosenzweig}, \citenamefont {Karakachian}, \citenamefont {Marchenko},
  \citenamefont {K\"uster},\ and\ \citenamefont {Starke}}]{PRL2020Rosenzweig}%
  \BibitemOpen
  \bibfield  {author} {\bibinfo {author} {\bibfnamefont {Philipp}\ \bibnamefont
  {Rosenzweig}}, \bibinfo {author} {\bibfnamefont {Hrag}\ \bibnamefont
  {Karakachian}}, \bibinfo {author} {\bibfnamefont {Dmitry}\ \bibnamefont
  {Marchenko}}, \bibinfo {author} {\bibfnamefont {Kathrin}\ \bibnamefont
  {K\"uster}}, \ and\ \bibinfo {author} {\bibfnamefont {Ulrich}\ \bibnamefont
  {Starke}},\ }\bibfield  {title} {\enquote {\bibinfo {title} {Overdoping
  graphene beyond the van $\text{Hove}$ singularity},}\ }\href@noop {}
  {\bibfield  {journal} {\bibinfo  {journal} {Phys. Rev. Lett.}\ }\textbf
  {\bibinfo {volume} {125}},\ \bibinfo {pages} {176403} (\bibinfo {year}
  {2020})}\BibitemShut {NoStop}%
\bibitem [{\citenamefont {Giannozzi}\ \emph {et~al.}(2009)\citenamefont
  {Giannozzi}, \citenamefont {Andreussi}, \citenamefont {Brumme}, \citenamefont
  {Bunau}, \citenamefont {Nardelli}, \citenamefont {Calandra}, \citenamefont
  {Car}, \citenamefont {Cavazzoni}, \citenamefont {Ceresoli}, \citenamefont
  {Cococcioni}, \citenamefont {Colonna}, \citenamefont {Carnimeo},
  \citenamefont {Corso}, \citenamefont {de~Gironcoli}, \citenamefont {Delugas},
  \citenamefont {Jr}, \citenamefont {Ferretti}, \citenamefont {Floris},
  \citenamefont {Fratesi}, \citenamefont {Fugallo}, \citenamefont {Gebauer},
  \citenamefont {Gerstmann}, \citenamefont {Giustino}, \citenamefont {Gorni},
  \citenamefont {Jia}, \citenamefont {Kawamura}, \citenamefont {Ko},
  \citenamefont {Kokalj}, \citenamefont {K\"{u}c\"{u}kbenli}, \citenamefont
  {.Lazzeri}, \citenamefont {Marsili}, \citenamefont {Marzari}, \citenamefont
  {Mauri}, \citenamefont {Nguyen}, \citenamefont {Nguyen}, \citenamefont {de-la
  Roza}, \citenamefont {Paulatto}, \citenamefont {Ponc\'{e}}, \citenamefont
  {Rocca}, \citenamefont {Sabatini}, \citenamefont {Santra}, \citenamefont
  {Schlipf}, \citenamefont {Seitsonen}, \citenamefont {Smogunov}, \citenamefont
  {Timrov}, \citenamefont {Thonhauser}, \citenamefont {Umari}, \citenamefont
  {Vast}, \citenamefont {Wu},\ and\ \citenamefont
  {Baroni}}]{JPhys_CM_21_395502_2009}%
  \BibitemOpen
  \bibfield  {author} {\bibinfo {author} {\bibfnamefont {P.}~\bibnamefont
  {Giannozzi}}, \bibinfo {author} {\bibfnamefont {O.}~\bibnamefont
  {Andreussi}}, \bibinfo {author} {\bibfnamefont {T.}~\bibnamefont {Brumme}},
  \bibinfo {author} {\bibfnamefont {O.}~\bibnamefont {Bunau}}, \bibinfo
  {author} {\bibfnamefont {M.~Buongiorno}\ \bibnamefont {Nardelli}}, \bibinfo
  {author} {\bibfnamefont {M.}~\bibnamefont {Calandra}}, \bibinfo {author}
  {\bibfnamefont {R.}~\bibnamefont {Car}}, \bibinfo {author} {\bibfnamefont
  {C.}~\bibnamefont {Cavazzoni}}, \bibinfo {author} {\bibfnamefont
  {D.}~\bibnamefont {Ceresoli}}, \bibinfo {author} {\bibfnamefont
  {M.}~\bibnamefont {Cococcioni}}, \bibinfo {author} {\bibfnamefont
  {N.}~\bibnamefont {Colonna}}, \bibinfo {author} {\bibfnamefont
  {I.}~\bibnamefont {Carnimeo}}, \bibinfo {author} {\bibfnamefont {A.~Dal}\
  \bibnamefont {Corso}}, \bibinfo {author} {\bibfnamefont {S.}~\bibnamefont
  {de~Gironcoli}}, \bibinfo {author} {\bibfnamefont {P.}~\bibnamefont
  {Delugas}}, \bibinfo {author} {\bibfnamefont {R.~A.~DiStasio}\ \bibnamefont
  {Jr}}, \bibinfo {author} {\bibfnamefont {A.}~\bibnamefont {Ferretti}},
  \bibinfo {author} {\bibfnamefont {A.}~\bibnamefont {Floris}}, \bibinfo
  {author} {\bibfnamefont {G.}~\bibnamefont {Fratesi}}, \bibinfo {author}
  {\bibfnamefont {G.}~\bibnamefont {Fugallo}}, \bibinfo {author} {\bibfnamefont
  {R.}~\bibnamefont {Gebauer}}, \bibinfo {author} {\bibfnamefont
  {U.}~\bibnamefont {Gerstmann}}, \bibinfo {author} {\bibfnamefont
  {F.}~\bibnamefont {Giustino}}, \bibinfo {author} {\bibfnamefont
  {T.}~\bibnamefont {Gorni}}, \bibinfo {author} {\bibfnamefont {J}~\bibnamefont
  {Jia}}, \bibinfo {author} {\bibfnamefont {M.}~\bibnamefont {Kawamura}},
  \bibinfo {author} {\bibfnamefont {H.-Y.}\ \bibnamefont {Ko}}, \bibinfo
  {author} {\bibfnamefont {A.}~\bibnamefont {Kokalj}}, \bibinfo {author}
  {\bibfnamefont {E.}~\bibnamefont {K\"{u}c\"{u}kbenli}}, \bibinfo {author}
  {\bibfnamefont {M}~\bibnamefont {.Lazzeri}}, \bibinfo {author} {\bibfnamefont
  {M.}~\bibnamefont {Marsili}}, \bibinfo {author} {\bibfnamefont
  {N.}~\bibnamefont {Marzari}}, \bibinfo {author} {\bibfnamefont
  {F.}~\bibnamefont {Mauri}}, \bibinfo {author} {\bibfnamefont {N.~L.}\
  \bibnamefont {Nguyen}}, \bibinfo {author} {\bibfnamefont {H.-V.}\
  \bibnamefont {Nguyen}}, \bibinfo {author} {\bibfnamefont {A.~Otero}\
  \bibnamefont {de-la Roza}}, \bibinfo {author} {\bibfnamefont
  {L.}~\bibnamefont {Paulatto}}, \bibinfo {author} {\bibfnamefont
  {S.}~\bibnamefont {Ponc\'{e}}}, \bibinfo {author} {\bibfnamefont
  {D.}~\bibnamefont {Rocca}}, \bibinfo {author} {\bibfnamefont
  {R.}~\bibnamefont {Sabatini}}, \bibinfo {author} {\bibfnamefont
  {B.}~\bibnamefont {Santra}}, \bibinfo {author} {\bibfnamefont
  {M.}~\bibnamefont {Schlipf}}, \bibinfo {author} {\bibfnamefont {A.~P.}\
  \bibnamefont {Seitsonen}}, \bibinfo {author} {\bibfnamefont {A.}~\bibnamefont
  {Smogunov}}, \bibinfo {author} {\bibfnamefont {I.}~\bibnamefont {Timrov}},
  \bibinfo {author} {\bibfnamefont {T.}~\bibnamefont {Thonhauser}}, \bibinfo
  {author} {\bibfnamefont {P.}~\bibnamefont {Umari}}, \bibinfo {author}
  {\bibfnamefont {N.}~\bibnamefont {Vast}}, \bibinfo {author} {\bibfnamefont
  {X.}~\bibnamefont {Wu}}, \ and\ \bibinfo {author} {\bibfnamefont
  {S.}~\bibnamefont {Baroni}},\ }\bibfield  {title} {\enquote {\bibinfo {title}
  {$\text{Quantum ESPRESSO}$: a modular and open-source software project for
  quantum simulations of materials},}\ }\href@noop {} {\bibfield  {journal}
  {\bibinfo  {journal} {J. Phys.: Condens. Matter}\ }\textbf {\bibinfo {volume}
  {21}},\ \bibinfo {pages} {395502} (\bibinfo {year} {2009})}\BibitemShut
  {NoStop}%
\bibitem [{\citenamefont {Giannozzi}\ \emph {et~al.}(2017)\citenamefont
  {Giannozzi}, \citenamefont {Baroni}, \citenamefont {Bonini}, \citenamefont
  {Calandra}, \citenamefont {Car}, \citenamefont {Cavazzoni}, \citenamefont
  {Ceresoli}, \citenamefont {Chiarotti}, \citenamefont {Cococcioni},
  \citenamefont {amd A.~Dal~Corso}, \citenamefont {Fabris}, \citenamefont
  {Fratesi}, \citenamefont {de~Gironcoli}, \citenamefont {Gebauer},
  \citenamefont {Gerstmann}, \citenamefont {Gougoussis}, \citenamefont
  {Kokalj}, \citenamefont {Lazzeri}, \citenamefont {Martin-Samos},
  \citenamefont {Marzari}, \citenamefont {Mauri}, \citenamefont {Mazzarello},
  \citenamefont {Paolini}, \citenamefont {Pasquarello}, \citenamefont
  {Paulatto}, \citenamefont {Sbraccia}, \citenamefont {Scandolo}, \citenamefont
  {Sclauzero}, \citenamefont {Seitsonen}, \citenamefont {Smogunov},
  \citenamefont {Umari},\ and\ \citenamefont
  {Wentzcovitch}}]{JPhys_CM_29_465901_2017}%
  \BibitemOpen
  \bibfield  {author} {\bibinfo {author} {\bibfnamefont {P.}~\bibnamefont
  {Giannozzi}}, \bibinfo {author} {\bibfnamefont {S.}~\bibnamefont {Baroni}},
  \bibinfo {author} {\bibfnamefont {N.}~\bibnamefont {Bonini}}, \bibinfo
  {author} {\bibfnamefont {M.}~\bibnamefont {Calandra}}, \bibinfo {author}
  {\bibfnamefont {R.}~\bibnamefont {Car}}, \bibinfo {author} {\bibfnamefont
  {C.}~\bibnamefont {Cavazzoni}}, \bibinfo {author} {\bibfnamefont
  {D.}~\bibnamefont {Ceresoli}}, \bibinfo {author} {\bibfnamefont {G.~L.}\
  \bibnamefont {Chiarotti}}, \bibinfo {author} {\bibfnamefont {M.}~\bibnamefont
  {Cococcioni}}, \bibinfo {author} {\bibfnamefont {I.~Dabo}\ \bibnamefont {amd
  A.~Dal~Corso}}, \bibinfo {author} {\bibfnamefont {S.}~\bibnamefont {Fabris}},
  \bibinfo {author} {\bibfnamefont {G.}~\bibnamefont {Fratesi}}, \bibinfo
  {author} {\bibfnamefont {S.}~\bibnamefont {de~Gironcoli}}, \bibinfo {author}
  {\bibfnamefont {R.}~\bibnamefont {Gebauer}}, \bibinfo {author} {\bibfnamefont
  {U.}~\bibnamefont {Gerstmann}}, \bibinfo {author} {\bibfnamefont
  {C.}~\bibnamefont {Gougoussis}}, \bibinfo {author} {\bibfnamefont
  {A.}~\bibnamefont {Kokalj}}, \bibinfo {author} {\bibfnamefont
  {M.}~\bibnamefont {Lazzeri}}, \bibinfo {author} {\bibfnamefont
  {L.}~\bibnamefont {Martin-Samos}}, \bibinfo {author} {\bibfnamefont
  {N.}~\bibnamefont {Marzari}}, \bibinfo {author} {\bibfnamefont
  {F.}~\bibnamefont {Mauri}}, \bibinfo {author} {\bibfnamefont
  {R.}~\bibnamefont {Mazzarello}}, \bibinfo {author} {\bibfnamefont
  {S.}~\bibnamefont {Paolini}}, \bibinfo {author} {\bibfnamefont
  {A.}~\bibnamefont {Pasquarello}}, \bibinfo {author} {\bibfnamefont
  {L.}~\bibnamefont {Paulatto}}, \bibinfo {author} {\bibfnamefont
  {C.}~\bibnamefont {Sbraccia}}, \bibinfo {author} {\bibfnamefont
  {S.}~\bibnamefont {Scandolo}}, \bibinfo {author} {\bibfnamefont
  {G.}~\bibnamefont {Sclauzero}}, \bibinfo {author} {\bibfnamefont {A.~P.}\
  \bibnamefont {Seitsonen}}, \bibinfo {author} {\bibfnamefont {A.}~\bibnamefont
  {Smogunov}}, \bibinfo {author} {\bibfnamefont {P.}~\bibnamefont {Umari}}, \
  and\ \bibinfo {author} {\bibfnamefont {R.~M.}\ \bibnamefont {Wentzcovitch}},\
  }\bibfield  {title} {\enquote {\bibinfo {title} {Advanced capabilities for
  materials modelling with $\text{Quantum ESPRESSO}$},}\ }\href@noop {}
  {\bibfield  {journal} {\bibinfo  {journal} {J. Phys.: Condens. Matter}\
  }\textbf {\bibinfo {volume} {29}},\ \bibinfo {pages} {465901} (\bibinfo
  {year} {2017})}\BibitemShut {NoStop}%
\bibitem [{\citenamefont {Perdew}\ \emph {et~al.}(1996)\citenamefont {Perdew},
  \citenamefont {Burke},\ and\ \citenamefont
  {Ernzerhof}}]{PhysRevLett_77_3865_1996}%
  \BibitemOpen
  \bibfield  {author} {\bibinfo {author} {\bibfnamefont {J.~P.}\ \bibnamefont
  {Perdew}}, \bibinfo {author} {\bibfnamefont {K.}~\bibnamefont {Burke}}, \
  and\ \bibinfo {author} {\bibfnamefont {M.}~\bibnamefont {Ernzerhof}},\
  }\bibfield  {title} {\enquote {\bibinfo {title} {Generalized gradient
  approximation made simple},}\ }\href@noop {} {\bibfield  {journal} {\bibinfo
  {journal} {Phys. Rev. Lett.}\ }\textbf {\bibinfo {volume} {77}},\ \bibinfo
  {pages} {3865} (\bibinfo {year} {1996})}\BibitemShut {NoStop}%
\bibitem [{\citenamefont {Hamann}(2013)}]{PhysRevB_88_085117_2013}%
  \BibitemOpen
  \bibfield  {author} {\bibinfo {author} {\bibfnamefont {D.~R.}\ \bibnamefont
  {Hamann}},\ }\bibfield  {title} {\enquote {\bibinfo {title} {Optimized
  norm-conserving vanderbilt pseudopotentials},}\ }\href@noop {} {\bibfield
  {journal} {\bibinfo  {journal} {Phys. Rev. B}\ }\textbf {\bibinfo {volume}
  {88}},\ \bibinfo {pages} {085117} (\bibinfo {year} {2013})}\BibitemShut
  {NoStop}%
\bibitem [{\citenamefont {Schlipf}\ and\ \citenamefont
  {Gygi}(2015)}]{CompPhysComms_196_36_2015}%
  \BibitemOpen
  \bibfield  {author} {\bibinfo {author} {\bibfnamefont {M.}~\bibnamefont
  {Schlipf}}\ and\ \bibinfo {author} {\bibfnamefont {F.}~\bibnamefont {Gygi}},\
  }\bibfield  {title} {\enquote {\bibinfo {title} {Optimization algorithm for
  the generation of $\text{ONCV}$ pseudopotentials},}\ }\href@noop {}
  {\bibfield  {journal} {\bibinfo  {journal} {Comp. Phys. Comm.}\ }\textbf
  {\bibinfo {volume} {196}},\ \bibinfo {pages} {36--44} (\bibinfo {year}
  {2015})}\BibitemShut {NoStop}%
\bibitem [{\citenamefont {Scherpelz}\ \emph {et~al.}(2016)\citenamefont
  {Scherpelz}, \citenamefont {Govoni}, \citenamefont {Hamada},\ and\
  \citenamefont {Galli}}]{JChemTheoryComput_12_3523_2016}%
  \BibitemOpen
  \bibfield  {author} {\bibinfo {author} {\bibfnamefont {P.}~\bibnamefont
  {Scherpelz}}, \bibinfo {author} {\bibfnamefont {M.}~\bibnamefont {Govoni}},
  \bibinfo {author} {\bibfnamefont {I.}~\bibnamefont {Hamada}}, \ and\ \bibinfo
  {author} {\bibfnamefont {G.}~\bibnamefont {Galli}},\ }\bibfield  {title}
  {\enquote {\bibinfo {title} {Implementation and validation of fully
  relativistic $\text{GW}$ calculations: Spin–orbit coupling in molecules,
  nanocrystals, and solids},}\ }\href@noop {} {\bibfield  {journal} {\bibinfo
  {journal} {J. Chem. Theory Comput.}\ }\textbf {\bibinfo {volume} {12}},\
  \bibinfo {pages} {3523--3544} (\bibinfo {year} {2016})}\BibitemShut {NoStop}%
\bibitem [{\citenamefont {Marzari}\ \emph {et~al.}(1999)\citenamefont
  {Marzari}, \citenamefont {Vanderbilt}, \citenamefont {Vita},\ and\
  \citenamefont {Payne}}]{PhysRevLett_82_3296_1999_Marzari}%
  \BibitemOpen
  \bibfield  {author} {\bibinfo {author} {\bibfnamefont {Nicola}\ \bibnamefont
  {Marzari}}, \bibinfo {author} {\bibfnamefont {David}\ \bibnamefont
  {Vanderbilt}}, \bibinfo {author} {\bibfnamefont {Alessandro~De}\ \bibnamefont
  {Vita}}, \ and\ \bibinfo {author} {\bibfnamefont {M.~C.}\ \bibnamefont
  {Payne}},\ }\bibfield  {title} {\enquote {\bibinfo {title} {Thermal
  contraction and disordering of the $\text{Al(110)}$ surface},}\ }\href@noop
  {} {\bibfield  {journal} {\bibinfo  {journal} {Phys. Rev. Lett.}\ }\textbf
  {\bibinfo {volume} {82}},\ \bibinfo {pages} {3296} (\bibinfo {year}
  {1999})}\BibitemShut {NoStop}%
\bibitem [{\citenamefont {Brumme}\ \emph {et~al.}(2014)\citenamefont {Brumme},
  \citenamefont {Calandra},\ and\ \citenamefont
  {Mauri}}]{PhysRevB_89_245406_2014_Brumme}%
  \BibitemOpen
  \bibfield  {author} {\bibinfo {author} {\bibfnamefont {T.}~\bibnamefont
  {Brumme}}, \bibinfo {author} {\bibfnamefont {M.}~\bibnamefont {Calandra}}, \
  and\ \bibinfo {author} {\bibfnamefont {F.}~\bibnamefont {Mauri}},\ }\bibfield
   {title} {\enquote {\bibinfo {title} {Electrochemical doping of few-layer
  $\text{ZrNCl}$ from first principles: Electronic and structural properties in
  field-effect configuration},}\ }\href@noop {} {\bibfield  {journal} {\bibinfo
   {journal} {Phys. Rev. B}\ }\textbf {\bibinfo {volume} {89}},\ \bibinfo
  {pages} {245406} (\bibinfo {year} {2014})}\BibitemShut {NoStop}%
\bibitem [{\citenamefont {Brumme}\ \emph {et~al.}(2015)\citenamefont {Brumme},
  \citenamefont {Calandra},\ and\ \citenamefont
  {Mauri}}]{PhysRevB_91_155436_2015_Brumme}%
  \BibitemOpen
  \bibfield  {author} {\bibinfo {author} {\bibfnamefont {T.}~\bibnamefont
  {Brumme}}, \bibinfo {author} {\bibfnamefont {M.}~\bibnamefont {Calandra}}, \
  and\ \bibinfo {author} {\bibfnamefont {F.}~\bibnamefont {Mauri}},\ }\bibfield
   {title} {\enquote {\bibinfo {title} {First-principles theory of field-effect
  doping in transition-metal dichalcogenides: Structural properties, electronic
  structure, $\text{Hall}$ coefficient, and electrical conductivity},}\
  }\href@noop {} {\bibfield  {journal} {\bibinfo  {journal} {Phys. Rev. B}\
  }\textbf {\bibinfo {volume} {91}},\ \bibinfo {pages} {155436} (\bibinfo
  {year} {2015})}\BibitemShut {NoStop}%
\bibitem [{\citenamefont {Liu}\ \emph {et~al.}(2013)\citenamefont {Liu},
  \citenamefont {Shan}, \citenamefont {Yao}, \citenamefont {Yao},\ and\
  \citenamefont {Xiao}}]{PhysRevB_88_085433_Liu}%
  \BibitemOpen
  \bibfield  {author} {\bibinfo {author} {\bibfnamefont {Gui-Bin}\ \bibnamefont
  {Liu}}, \bibinfo {author} {\bibfnamefont {Wen-Yu}\ \bibnamefont {Shan}},
  \bibinfo {author} {\bibfnamefont {Yugui}\ \bibnamefont {Yao}}, \bibinfo
  {author} {\bibfnamefont {Wang}\ \bibnamefont {Yao}}, \ and\ \bibinfo {author}
  {\bibfnamefont {Di}~\bibnamefont {Xiao}},\ }\bibfield  {title} {\enquote
  {\bibinfo {title} {Three-band tight-binding model for monolayers of
  group-$\text{VIB}$ transition metal dichalcogenides},}\ }\href@noop {}
  {\bibfield  {journal} {\bibinfo  {journal} {Phys. Rev. B}\ }\textbf {\bibinfo
  {volume} {88}},\ \bibinfo {pages} {085433} (\bibinfo {year}
  {2013})}\BibitemShut {NoStop}%
\bibitem [{\citenamefont {Slater}\ and\ \citenamefont
  {Koster}(1954)}]{PhysRev_94_1498_1954_Slater_Koster}%
  \BibitemOpen
  \bibfield  {author} {\bibinfo {author} {\bibfnamefont {J.~C.}\ \bibnamefont
  {Slater}}\ and\ \bibinfo {author} {\bibfnamefont {G.~F.}\ \bibnamefont
  {Koster}},\ }\bibfield  {title} {\enquote {\bibinfo {title} {Simplified
  $\text{LCAO}$ method for the periodic potential problem},}\ }\href@noop {}
  {\bibfield  {journal} {\bibinfo  {journal} {Phys. Rev.}\ }\textbf {\bibinfo
  {volume} {94}},\ \bibinfo {pages} {1498} (\bibinfo {year}
  {1954})}\BibitemShut {NoStop}%
\bibitem [{\citenamefont {Dresselhaus}\ \emph {et~al.}(2008)\citenamefont
  {Dresselhaus}, \citenamefont {Dresselhaus},\ and\ \citenamefont
  {Jorio}}]{Group_Theory_Dresselhaus}%
  \BibitemOpen
  \bibfield  {author} {\bibinfo {author} {\bibfnamefont {M.~S.}\ \bibnamefont
  {Dresselhaus}}, \bibinfo {author} {\bibfnamefont {G.}~\bibnamefont
  {Dresselhaus}}, \ and\ \bibinfo {author} {\bibfnamefont {A.}~\bibnamefont
  {Jorio}},\ }\href@noop {} {\emph {\bibinfo {title} {Group Theory: Application
  to the Physics of Condensed Matter}}},\ \bibinfo {edition} {1st}\ ed.\
  (\bibinfo  {publisher} {Springer-Verlag Berlin Heidelberg},\ \bibinfo {year}
  {2008})\BibitemShut {NoStop}%
\bibitem [{\citenamefont {Winkler}(2003)}]{SOC_Winkler}%
  \BibitemOpen
  \bibfield  {author} {\bibinfo {author} {\bibfnamefont {R.}~\bibnamefont
  {Winkler}},\ }\href@noop {} {\emph {\bibinfo {title} {Spin-orbit Coupling
  Effects in Two-Dimensional Electron and Hole Systems}}},\ \bibinfo {edition}
  {1st}\ ed.,\ Vol.\ \bibinfo {volume} {191}\ (\bibinfo  {publisher}
  {Springer-Verlag Berlin Heidelberg},\ \bibinfo {year} {2003})\BibitemShut
  {NoStop}%
\bibitem [{\citenamefont {Sinova}\ \emph {et~al.}(2015)\citenamefont {Sinova},
  \citenamefont {Valenzuela}, \citenamefont {Wunderlich}, \citenamefont
  {Back},\ and\ \citenamefont {Jungwirth}}]{RevModPhys_87_1213_2015_Sinova}%
  \BibitemOpen
  \bibfield  {author} {\bibinfo {author} {\bibfnamefont {J.}~\bibnamefont
  {Sinova}}, \bibinfo {author} {\bibfnamefont {S.}~\bibnamefont {Valenzuela}},
  \bibinfo {author} {\bibfnamefont {J.}~\bibnamefont {Wunderlich}}, \bibinfo
  {author} {\bibfnamefont {C.~H.}\ \bibnamefont {Back}}, \ and\ \bibinfo
  {author} {\bibfnamefont {T.}~\bibnamefont {Jungwirth}},\ }\bibfield  {title}
  {\enquote {\bibinfo {title} {Spin $\text{Hall}$ effects},}\ }\href@noop {}
  {\bibfield  {journal} {\bibinfo  {journal} {Rev. Mod. Phys.}\ }\textbf
  {\bibinfo {volume} {87}},\ \bibinfo {pages} {1213} (\bibinfo {year}
  {2015})}\BibitemShut {NoStop}%
\bibitem [{\citenamefont {Sinova}\ \emph {et~al.}(2004)\citenamefont {Sinova},
  \citenamefont {Culcer}, \citenamefont {Niu}, \citenamefont {Sinitsyn},
  \citenamefont {Jungwirth},\ and\ \citenamefont
  {MacDonald}}]{PhysRevLett_92_126603_2004_Sinova}%
  \BibitemOpen
  \bibfield  {author} {\bibinfo {author} {\bibfnamefont {J.}~\bibnamefont
  {Sinova}}, \bibinfo {author} {\bibfnamefont {D.}~\bibnamefont {Culcer}},
  \bibinfo {author} {\bibfnamefont {Q.}~\bibnamefont {Niu}}, \bibinfo {author}
  {\bibfnamefont {N.~A.}\ \bibnamefont {Sinitsyn}}, \bibinfo {author}
  {\bibfnamefont {T.}~\bibnamefont {Jungwirth}}, \ and\ \bibinfo {author}
  {\bibfnamefont {A.~H.}\ \bibnamefont {MacDonald}},\ }\bibfield  {title}
  {\enquote {\bibinfo {title} {Universal intrinsic spin $\text{Hall}$
  effect},}\ }\href@noop {} {\bibfield  {journal} {\bibinfo  {journal} {Phys.
  Rev. Lett.}\ }\textbf {\bibinfo {volume} {92}},\ \bibinfo {pages} {126603}
  (\bibinfo {year} {2004})}\BibitemShut {NoStop}%
\bibitem [{\citenamefont {Guo}\ \emph {et~al.}(2005)\citenamefont {Guo},
  \citenamefont {Yao},\ and\ \citenamefont
  {Niu}}]{PhysRevLett_94_226601_2005_Guo}%
  \BibitemOpen
  \bibfield  {author} {\bibinfo {author} {\bibfnamefont {G.}~\bibnamefont
  {Guo}}, \bibinfo {author} {\bibfnamefont {Y.}~\bibnamefont {Yao}}, \ and\
  \bibinfo {author} {\bibfnamefont {Q.}~\bibnamefont {Niu}},\ }\bibfield
  {title} {\enquote {\bibinfo {title} {Ab initio calculation of the intrinsic
  spin $\text{Hall}$ effect in semiconductors},}\ }\href@noop {} {\bibfield
  {journal} {\bibinfo  {journal} {Phys. Rev. Lett.}\ }\textbf {\bibinfo
  {volume} {94}},\ \bibinfo {pages} {226601} (\bibinfo {year}
  {2005})}\BibitemShut {NoStop}%
\bibitem [{\citenamefont {Yao}\ and\ \citenamefont
  {Fang}(2005)}]{PhysRevLett_95_156601_2005_Yao}%
  \BibitemOpen
  \bibfield  {author} {\bibinfo {author} {\bibfnamefont {Y.}~\bibnamefont
  {Yao}}\ and\ \bibinfo {author} {\bibfnamefont {Z.}~\bibnamefont {Fang}},\
  }\bibfield  {title} {\enquote {\bibinfo {title} {Sign changes of intrinsic
  spin $\text{Hall}$ effect in semiconductors and simple metals :
  First-principles calculations},}\ }\href@noop {} {\bibfield  {journal}
  {\bibinfo  {journal} {Phys. Rev. Lett.}\ }\textbf {\bibinfo {volume} {95}},\
  \bibinfo {pages} {156601} (\bibinfo {year} {2005})}\BibitemShut {NoStop}%
\bibitem [{\citenamefont {Guo}\ \emph {et~al.}(2008)\citenamefont {Guo},
  \citenamefont {Murakami}, \citenamefont {Chen},\ and\ \citenamefont
  {Nagaosa}}]{PhysRevLett_100_096401_2008_Guo}%
  \BibitemOpen
  \bibfield  {author} {\bibinfo {author} {\bibfnamefont {G.~Y.}\ \bibnamefont
  {Guo}}, \bibinfo {author} {\bibfnamefont {S.}~\bibnamefont {Murakami}},
  \bibinfo {author} {\bibfnamefont {T.-W.}\ \bibnamefont {Chen}}, \ and\
  \bibinfo {author} {\bibfnamefont {N.}~\bibnamefont {Nagaosa}},\ }\bibfield
  {title} {\enquote {\bibinfo {title} {Intrinsic spin $\text{Hall}$ effect in
  platinum: First-principles calculations},}\ }\href@noop {} {\bibfield
  {journal} {\bibinfo  {journal} {Phys. Rev. Lett.}\ }\textbf {\bibinfo
  {volume} {100}},\ \bibinfo {pages} {096401} (\bibinfo {year}
  {2008})}\BibitemShut {NoStop}%
\bibitem [{\citenamefont {Matthes}\ \emph {et~al.}(2016)\citenamefont
  {Matthes}, \citenamefont {K\"{u}fner}, \citenamefont {Furthm\"{u}ller},\ and\
  \citenamefont {Bechstedt}}]{PhysRevB_94_085410_2016_Matthes}%
  \BibitemOpen
  \bibfield  {author} {\bibinfo {author} {\bibfnamefont {L.}~\bibnamefont
  {Matthes}}, \bibinfo {author} {\bibfnamefont {S.}~\bibnamefont {K\"{u}fner}},
  \bibinfo {author} {\bibfnamefont {J.}~\bibnamefont {Furthm\"{u}ller}}, \ and\
  \bibinfo {author} {\bibfnamefont {F.}~\bibnamefont {Bechstedt}},\ }\bibfield
  {title} {\enquote {\bibinfo {title} {Intrinsic spin $\text{Hall}$
  conductivity in one-, two-, and three-dimensional trivial and topological
  systems},}\ }\href@noop {} {\bibfield  {journal} {\bibinfo  {journal} {Phys.
  Rev. B}\ }\textbf {\bibinfo {volume} {94}},\ \bibinfo {pages} {085410}
  (\bibinfo {year} {2016})}\BibitemShut {NoStop}%
\bibitem [{\citenamefont {Feng}\ \emph {et~al.}(2012)\citenamefont {Feng},
  \citenamefont {Yao}, \citenamefont {Zhu}, \citenamefont {Zhou}, \citenamefont
  {Yao},\ and\ \citenamefont {Xiao}}]{PhysRevB_86_165108_2012_Feng}%
  \BibitemOpen
  \bibfield  {author} {\bibinfo {author} {\bibfnamefont {W.}~\bibnamefont
  {Feng}}, \bibinfo {author} {\bibfnamefont {Y.}~\bibnamefont {Yao}}, \bibinfo
  {author} {\bibfnamefont {W.}~\bibnamefont {Zhu}}, \bibinfo {author}
  {\bibfnamefont {J.}~\bibnamefont {Zhou}}, \bibinfo {author} {\bibfnamefont
  {W.}~\bibnamefont {Yao}}, \ and\ \bibinfo {author} {\bibfnamefont
  {D.}~\bibnamefont {Xiao}},\ }\bibfield  {title} {\enquote {\bibinfo {title}
  {Intrinsic spin $\text{Hall}$ effect in monolayers of group-$\text{VI}$
  dichalcogenides: A first-principles study},}\ }\href@noop {} {\bibfield
  {journal} {\bibinfo  {journal} {Phys. Rev. B}\ }\textbf {\bibinfo {volume}
  {86}},\ \bibinfo {pages} {165108} (\bibinfo {year} {2012})}\BibitemShut
  {NoStop}%
\bibitem [{\citenamefont {Zhou}\ \emph {et~al.}(2019)\citenamefont {Zhou},
  \citenamefont {Qiao}, \citenamefont {Bournel},\ and\ \citenamefont
  {Zhao}}]{PhysRevB_99_060408_2019_Zhou}%
  \BibitemOpen
  \bibfield  {author} {\bibinfo {author} {\bibfnamefont {J.}~\bibnamefont
  {Zhou}}, \bibinfo {author} {\bibfnamefont {J.}~\bibnamefont {Qiao}}, \bibinfo
  {author} {\bibfnamefont {A.}~\bibnamefont {Bournel}}, \ and\ \bibinfo
  {author} {\bibfnamefont {W.}~\bibnamefont {Zhao}},\ }\bibfield  {title}
  {\enquote {\bibinfo {title} {Intrinsic spin $\text{Hall}$ conductivity of the
  semimetals $\textrm{MoTe}_{2}$ and $\textrm{WTe}_{2}$},}\ }\href@noop {}
  {\bibfield  {journal} {\bibinfo  {journal} {Phys. Rev. B}\ }\textbf {\bibinfo
  {volume} {99}},\ \bibinfo {pages} {060408(R)} (\bibinfo {year}
  {2019})}\BibitemShut {NoStop}%
\bibitem [{\citenamefont {Ryoo}\ \emph {et~al.}(2019)\citenamefont {Ryoo},
  \citenamefont {Park},\ and\ \citenamefont
  {Souza}}]{PhysRevB_99_235113_2019_Ryoo}%
  \BibitemOpen
  \bibfield  {author} {\bibinfo {author} {\bibfnamefont {J.~H.}\ \bibnamefont
  {Ryoo}}, \bibinfo {author} {\bibfnamefont {C-H.}\ \bibnamefont {Park}}, \
  and\ \bibinfo {author} {\bibfnamefont {I.}~\bibnamefont {Souza}},\ }\bibfield
   {title} {\enquote {\bibinfo {title} {Computation of intrinsic spin
  $\text{Hall}$ conductivities from first principles using maximally localized
  $\text{Wannier}$ functions},}\ }\href@noop {} {\bibfield  {journal} {\bibinfo
   {journal} {Phys. Rev. B}\ }\textbf {\bibinfo {volume} {99}},\ \bibinfo
  {pages} {235113} (\bibinfo {year} {2019})}\BibitemShut {NoStop}%
\end{thebibliography}
\end{document}